\documentclass[iop]{emulateapj}

\usepackage{natbib,aas_macros}
\usepackage{multirow,color}

\begin{document}
\shorttitle{
ALMA Census of Faint 1.2 mm Sources
}
\shortauthors{Fujimoto et al.}
\slugcomment{ApJS in press}

\title{%
ALMA Census of Faint 1.2 \lowercase{mm} Sources Down to 0.02 \lowercase{m}J\lowercase{y}:\\
Extragalactic Background Light and Dust-Poor High-$z$ Galaxies
}

\author{%
Seiji Fujimoto\altaffilmark{1}, 
Masami Ouchi\altaffilmark{1,2}, 
Yoshiaki Ono\altaffilmark{1}, 
Takatoshi Shibuya\altaffilmark{1}, \\
Masafumi Ishigaki\altaffilmark{1}, 
Hiroshi Nagai \altaffilmark{3},
and 
Rieko Momose\altaffilmark{3} 
}

\email{sfseiji@icrr.u-tokyo.ac.jp}

\altaffiltext{1}{%
Institute for Cosmic Ray Research, The University of Tokyo,
Kashiwa, Chiba 277-8582, Japan
}
\altaffiltext{2}{%
Kavli Institute for the Physics andMathematics of the Universe 
(Kavli IPMU), WPI, The University of Tokyo, 
Kashiwa, Chiba 277-8583, Japan
}
\altaffiltext{3}{%
National Astronomical Observatory of Japan, Mitaka, 
Tokyo 181-8588, Japan
}

\def\aj{AJ}%
\def\actaa{Acta Astron.}%
\def\araa{ARA\&A}%
\def\apj{ApJ}%
\def\apjl{ApJ}%
\def\apjs{ApJS}%
\def\ao{Appl.~Opt.}%
\def\apss{Ap\&SS}%
\def\aap{A\&A}%
\def\aapr{A\&A~Rev.}%
\def\aaps{A\&AS}%
\def\azh{AZh}%
\def\baas{BAAS}%
\def\bac{Bull. astr. Inst. Czechosl.}%
\def\caa{Chinese Astron. Astrophys.}%
\def\cjaa{Chinese J. Astron. Astrophys.}%
\def\icarus{Icarus}%
\def\jcap{J. Cosmology Astropart. Phys.}%
\def\jrasc{JRASC}%
\def\mnras{MNRAS}%
\def\memras{MmRAS}%
\def\na{New A}%
\def\nar{New A Rev.}%
\def\pasa{PASA}%
\def\pra{Phys.~Rev.~A}%
\def\prb{Phys.~Rev.~B}%
\def\prc{Phys.~Rev.~C}%
\def\prd{Phys.~Rev.~D}%
\def\pre{Phys.~Rev.~E}%
\def\prl{Phys.~Rev.~Lett.}%
\def\pasp{PASP}%
\def\pasj{PASJ}%
\def\qjras{QJRAS}%
\def\rmxaa{Rev. Mexicana Astron. Astrofis.}%
\def\skytel{S\&T}%
\def\solphys{Sol.~Phys.}%
\def\sovast{Soviet~Ast.}%
\def\ssr{Space~Sci.~Rev.}%
\def\zap{ZAp}%
\def\nat{Nature}%
\def\iaucirc{IAU~Circ.}%
\def\aplett{Astrophys.~Lett.}%
\def\apspr{Astrophys.~Space~Phys.~Res.}%
\def\bain{Bull.~Astron.~Inst.~Netherlands}%
\def\fcp{Fund.~Cosmic~Phys.}%
\def\gca{Geochim.~Cosmochim.~Acta}%
\def\grl{Geophys.~Res.~Lett.}%
\def\jcp{J.~Chem.~Phys.}%
\def\jgr{J.~Geophys.~Res.}%
\def\jqsrt{J.~Quant.~Spec.~Radiat.~Transf.}%
\def\memsai{Mem.~Soc.~Astron.~Italiana}%
\def\nphysa{Nucl.~Phys.~A}%
\def\physrep{Phys.~Rep.}%
\def\physscr{Phys.~Scr}%
\def\planss{Planet.~Space~Sci.}%
\def\procspie{Proc.~SPIE}%

\begin{abstract}
We present statistics of 133 faint 1.2-mm continuum sources detected in about 120 deep ALMA pointing data 
that include all the archival deep data available by 2015 June. 
We derive number counts of 1.2 mm continuum sources down to 0.02 mJy partly with the assistance of 
gravitational lensing, 
and find that the total integrated 1.2 mm flux of the securely identified sources is 22.9$^{+6.7}_{-5.6}$ Jy deg$^{-2}$ 
that corresponds to 104$^{+31}_{-25}$\% of the extragalactic background light (EBL) measured by COBE observations. 
These results suggest that the major 1.2 mm EBL contributors are sources with 0.02 mJy, 
and that very faint 1.2 mm sources with $\lesssim$0.02 mJy contribute negligibly to the EBL with the possible 
flattening and/or truncation of number counts in this very faint flux regime. 
To understand the physical origin of our faint ALMA sources, we measure the galaxy bias $b_{g}$ by the counts-in-cells technique, 
and place a stringent upper limit of $b_{g}<3.5$ that is not similar to $b_{g}$ values of massive DRGs and SMGs 
but comparable to those of UV-bright sBzKs and LBGs. 
Moreover, in optical and near-infrared (NIR) deep fields, we identify optical-NIR counterparts for 59\% of our faint ALMA sources, 
majority of which have luminosities, colors, and the IRX-$\beta$ relation same as sBzKs and LBGs. 
We thus conclude that about a half of our faint ALMA sources are dust-poor high-$z$ galaxies as known as sBzKs and LBGs in optical studies, 
and that these faint ALMA sources are not miniature (U)LIRGs simply scaled down with the infrared brightness.
\end{abstract}


\section{Introduction}
\label{sec:intro}
Since the infrared (IR) extragalactic background light (EBL) was first identified by the {\it Cosmic Background Explorer } (COBE) satellite, it has been known that the total energy of IR EBL is comparable to that of the optical EBL \citep{puget1996,fixsen1998,hauser1998,hauser2001,dole2006}.  
Because dusty star-forming galaxies at high redshift should significantly contribute to the IR EBL,
the IR EBL is key to understanding the dusty-side of the cosmic star formation history and to
constraining physical parameters of galaxy formation models
\citep[e.g.,][]{granato2004,baugh2005,fontanot2007,shimizu2012,hayward2013}.

To reveal the origin of the IR EBL, we need reliable number counts from bright to faint flux limits.
The IR EBL of the low-frequency end, at submillimeter (submm) and millimeter (mm) wavelengths, is one of the most advantageous wavelength regimes, due to the negative k-correction.
Previous blank field observations with single dish telescopes have revealed that 
the bright dusty high-$z$ objects of submillimeter galaxies \citep[SMGs; ][]{lagache2005} are not major contributors of the EBL. 
SCUBA and LABOCA observations have resolved $20-40\%$ of the EBL at submm wavelength \citep[e.g.,][]{eales2000,smail2002,coppin2006,knudsen2008,weiss2009}.
Similar results have been obtained by AzTEC observations that they have resolved $10-20\%$ of the EBL at mm wavelength \citep[e.g.,][]{perera2008,hatsukade2011,scott2012}. 
These observations suggest that there should exist high-$z$ populations different from SMGs, 
and that such populations are major EBL contributors in submm and mm wavelengths.

There are two observational challenges for fully resolving the EBL. 
First is the spatial resolution.
The poor spatial resolution of the single-dish observations 
causes the serious source confusion \citep{condon1974} that skews the observed number counts.
Second is the sensitivity.  Most of the submm and mm telescopes reach
the limiting fluxes of $\sim 1$ mJy that correspond to the flux range of the SMGs. 
To detect submm sources fainter than $\sim 1$ mJy, 
there are many efforts \citep[e.g.,][]{smail2002,chapman2002,knudsen2008,johansson2011,chen2013}.
\cite{knudsen2008} and \cite{chen2013} observe
massive galaxy clusters at $850\ \mu$m
for gravitationally lensed sources,
and resolved 50\%$-$180\% of the EBL down to the intrinsic flux of $S_{\rm 850\ \mu m} > 0.1$ mJy.
These results indicate that there exist faint ($<1$ mJy) populations significantly contributing to the EBL \citep[e.g.,][]{smail2002,knudsen2008}. 
However, large uncertainties of the measurements still remain, due to
the source blending and cosmic variance. Moreover, the physical origin of these faint submm and mm sources 
is unknown.

The Atacama Large Millimeter $/$ submillimeter Array (ALMA) enables us to investigate the faint submm and mm sources with the negligibly small uncertainties of the source confusion and blending,
having its capabilities of high angular resolution and sensitivity.
The first ALMA mm search for faint sources below $S_{\rm 1mm} \sim 1.0$ mJy with no lensing effect
has been conducted by \cite{hatsukade2013}. 
The study reports that they resolve $\sim$80\% of the EBL down to $S_{\rm 1mm}=0.1$ mJy in 20 targets
residing in one blank field of the Subaru$/$XMM-Newton Deep Survey \citep[SXDS; ][]{furusawa2008}.
Due to the single field observations, this result would include unknown effects of cosmic variance.
Subsequently, using multi-field deep ALMA $1$mm maps of Bands 6 and 7
down to the flux limit of $S_{\rm 1mm}=0.1$ mJy,
\cite{ono2014} and \cite{carniani2015} have claimed an EBL resolved fraction of $\sim$60\% 
that is smaller than the one of \cite{hatsukade2013}. 
The uncertainties from cosmic variance are probably reduced with the multi-field data. 
Including the effects of the cosmic variance, only about a half of the EBL 
has been resolved with the present ALMA flux limit of $S_{\rm 1mm}=0.1$ mJy.

There is another remaining issue about the IR EBL. Although these studies newly identify
faint ALMA sources ($S_{\rm1mm}<1.0$ mJy) contributing to the half of the IR EBL, these studies do not
clearly answer to the question about the connection between the faint ALMA sources and 
optically selected high-$z$ galaxies.
Recent observations with {\it Herschel} have revealed that typical UV-selected galaxies such as Lyman-break galaxies (LBGs) 
have a median total (8-1000 $\mu$m) luminosity of 
$L_{\rm IR} \simeq 2.2\times10^{11}L_{\odot}$ \citep{reddy2012,lee2012,davies2013}. 
The stacking analysis of {\it Herschel} and ALMA data has also shown that $K$-selected galaxies, 
including star-forming BzK galaxies (sBzKs), have total IR luminosities of $L_{\rm IR}=(5-11)\times 10^{11}L_{\odot}$ \citep{decarli2014}. 
These ranges of the IR luminosities of LBGs and sBzKs correspond to the mm flux of $S_{\rm 1mm}\sim 0.1-1$ mJy if we assume a modified black body with $\beta_{\rm d}=1.8$, dust temperature $T_{\rm d}=35$K, and a source redshift $z=2.5$.
This mm flux range is similar to the one of the faint ALMA sources, implying that
these sources could be mm counterparts of the optically selected galaxies.

There is another approach for characterizing the faint ALMA sources.
Because the spatial distribution of galaxies is related to the underlying distribution of dark haloes
in the standard scenario of structure formation in $\Lambda$CDM universe,
the clustering analysis is a powerful tool for understanding the connection
between various galaxy populations. Clustering analyses have been carried out
over the past decade for bright SMGs observed with single-dish telescopes,
and concluded that the clustering amplitudes, or galaxy biases, are large,
and that the hosting dark halo masses of the bright SMGs are estimated
to be $\sim 10^{13}M_\odot$
(\citealt{webb2003,blain2004,weiss2009,williams2011,hickox2012}, cf. \citealt{miller2015}).
On the other hand, however,  the dark halo properties of the faint ALMA sources are poorly known, except
for the result of \cite{ono2014} who have obtained, for the first time, 
a meaningful constraint on a galaxy bias, $b_{\rm g}\leq4$, for faint ALMA sources.
In this way, the faint ALMA sources are not well studied.
To understand the physical origin of faint ALMA sources, 
one should study faint ALMA sources with a more complete data set
on the basis of individual sources as well as statistics.

In this paper, we analyze the large dataset of multi-field deep ALMA data including 
the complete deep datasets archived by 2015 June. 
Sixty-six independent-field and 
one cluster maps are taken
by  $\sim 120$ ALMA pointings 
in Band 6$/$7 to reveal the origins of the EBL and faint ALMA sources in mm wavelength. 
Note that this is the first step for statistically investigating faint sources behind a lensing cluster with ALMA.
The structure of this paper is as follows. In Section \ref{sec:alma_data}, we describe the observations and the data reduction. 
Section \ref{sec:data_analysis} outlines the method of the source extraction, our simulations for deriving the number counts, and the mass model development for the cluster. 
We compare the number counts from our and previous studies, 
and estimate the resolved fraction of the EBL at 1.2 mm in section \ref{sec:number_counts}. 
In Section \ref{sec:clustering}, we report the results of the clustering analysis for our faint ALMA sources. 
We investigate optical-NIR counterparts of our faint ALMA sources in Section \ref{sec:characterization}. 
The summary of this study is presented in Section \ref{sec:summary}.

Throughout this paper, we assume a flat universe with 
$\Omega_{\rm m} = 0.3$, 
$\Omega_\Lambda = 0.7$, 
$\sigma_8 = 0.8$, 
and $H_0 = 70$ km s$^{-1}$ Mpc$^{-1}$. 
We use magnitudes in the AB system (Oke \& Gunn 1983).

\section{Data and Reduction} 
\label{sec:alma_data}
We use 67 continuum maps obtained by $\sim 120$ pointing of 
the ALMA cycle 0-2 observations in Band 6$/$7 that accomplish high sensitivities and angular resolutions. 
Tables \ref{tab:sub_def} and \ref{tab:our_alma_maps} 
present the summary and the detailed properties, respectively, for
these 67 continuum maps. 
We define sub-datasets by the 
mapping modes and the depths 
because these two conditions would make different systematics.
For the
mapping mode definitions, there are 
maps targeting field regions by single pointing observations, referred to as 'field' data, 
4 and 62 out of which are taken from our programs
and the ALMA archive, respectively. 
One map for a galaxy cluster
taken by mapping observations, referred to as 'cluster' data,
is from the ALMA archive. 

For the depth definitions,
we divide these 66 maps 
of 'field' data
into two sub-datasets with low ($\leq 60\ \mu$Jy) and 
high ($>60\ \mu$Jy) noise levels, which we refer to as 'deep' and 'medium-deep' data, respectively. 
This is because the short-integration data would have more chances to contain systematic 
noise. As shown in Table \ref{tab:our_alma_maps}, we use the sub-dataset names
of A and B for the deep and medium-deep field data, respectively. 
The sub-dataset name of the 
deep cluster data is C.

In this section, we describe the details of these 67 continuum maps
of 'deep', 'medium-deep', and 'cluster' data,
and present our reduction procedures.

\begin{deluxetable}{ccc} 
\tablecolumns{2} 
\tablewidth{0pt} 
\tablecaption{Data Summary\label{tab:sub_def}}
\tablehead{ 
Sub-Dataset & Number of Maps \\
(1) &  (2)
}
\startdata 
Deep data (A) & 41 \\
Medium-deep data (B) & 25 & \\
Cluster data (C) & 1
\enddata 
\tablecomments{
The sub-dataset names of A, B, and C are written in parenthesis.
(1): ALMA maps with low ($\leq60\ \mu$Jy) and high ($>60\ \mu$Jy) noise levels 
are referred to as Deep and Medium-deep data, respectively.
(2): Number of the ALMA maps in each sub-dataset.
}
\end{deluxetable} 

\begin{deluxetable*}{clccccccc} 
\vspace{-1.0cm}
\tablecolumns{9} 
\tablewidth{0pt} 
\tablecaption{
ALMA Maps of this Study \label{tab:our_alma_maps}}
\tablehead{
\colhead{Map ID}
    & \colhead{Target} 
    & \colhead{$\lambda_{\rm obs}$}    
    & \colhead{$\nu_{\rm obs}$ (Band)} 
    & \colhead{$\sigma_{\rm cont}$} 
    & \colhead{beam size}
    & \colhead{$S_{1.2{\rm mm}} / S_{\rm obs}$}  
    & \colhead{Ref.}
    & \colhead{Project ID} \\
\colhead{ }
    & \colhead{  }    
    & \colhead{(mm)}
    & \colhead{(GHz)}
    & \colhead{($\mu$Jy beam$^{-1}$)}
    & \colhead{($''\times''$)}
    & \colhead{  }
    & \colhead{  }
    & \colhead{  }     \\
\colhead{ }
    & \colhead{ }
    & \colhead{(1)}
    & \colhead{(2)}
    & \colhead{(3)}    
    & \colhead{(4)}    
    & \colhead{(5)}
    & \colhead{(6)} 
    & \colhead{  }
    }
\startdata 
 &  \multicolumn{7}{c}{Deep Data (A)} &  \\ \hline 
1 & BDF-3299 &1.30 &227 (6) & 8.5 & 0.79$\times$0.58 & 1.28 & (a),(b) & 2012.A.00040.S\\
   &                  &         &            &       &                              &        &            & 2012.1.00719.S\\
2 & GRB090423 & 1.35 & 222 (6) & 10.3& 0.98$\times$0.75 & 1.45 &  (c),(d) & 2012.1.00953.S\\
3 & NB101-S-2904 & 1.26 & 238 (6) & 11.4 & 0.43$\times$0.33 &1.16 & (e)& 2012.1.00088.S\\
4 & NB816-S-61269 &1.03 &291 (7)& 13.0 & 0.45$\times$0.42 & 0.63& (f) & 2012.1.00602.S\\
5 & rxj0806 & 1.31 & 229 (6) & 16.6& 0.78$\times$0.63 & 1.31 &  (g) & 2012.1.00610.S\\
6 & Himiko & 1.16 & 259 (6) & 17.0 & 0.81$\times$0.56& 0.90& (h),(i) & 2011.0.00115.S\\
7 & BDF-521 & 1.30 & 230 (6) & 17.7 & 0.68$\times$0.51 & 1.28  & (a),(b) & 2012.1.00719.S\\
8 & IOK-1 & 1.29 & 232 (6) & 18.8 & 1.08$\times$0.78 & 1.25& (j),(k) & 2011.0.00767.S\\ 
9 & rxj2143 & 1.26 & 238 (6) & 20.8& 0.60$\times$0.43 & 1.16  & (l) & 2012.1.00610.S\\
10 & CFHQSJ2329-0301 &  1.20 & 250 (6) & 21.0 & 0.73$\times$0.63&1.00 &  (m),(n) & 2011.0.00243.S\\
11 & SDF-46975 & 1.22 & 244 (6) & 21.2 & 1.23$\times$0.98 & 1.05 & (o),(b) & 2012.1.00719.S\\
12 & ID239 & 1.29 & 232 (6) & 21.7&  1.88$\times$0.90 & 1.25 & $\cdots$  & 2012.1.00076.S\\
13 & GRB020819B& 1.18 & 254 (6) & 22.6 & 0.91$\times$0.72 & 0.95& (p) & 2011.0.00232.S\\
14 & CFHQSJ0055+0146 &1.09 & 273 (6) & 23.3 & 0.69$\times$0.55 &0.75 & (m) & 2012.1.00676.S\\
15 & ID182 &1.23 & 244 (6) & 23.8 & 1.33$\times$0.89 & 1.08& $\cdots$ & 2012.1.00076.S\\
16 & ID220 &1.23 & 244 (6) & 24.2 & 1.33$\times$0.88 & 1.08& $\cdots$ & 2012.1.00076.S\\
17 & ID244 & 1.29 & 232 (6) & 24.1& 1.87$\times$0.90 & 1.25 & $\cdots$ & 2012.1.00076.S\\
18 & CFHQSJ2229+1457 &1.12 & 266 (6) & 24.9 & 0.79$\times$0.70 &0.81 &  (m) & 2012.1.00676.S\\
19 & ID247 & 1.29 & 232 (6) & 25.1& 1.87$\times$0.90 & 1.25 & $\cdots$ & 2012.1.00076.S\\
20 & ID209 & 1.29 & 232 (6) & 25.8& 1.23$\times$1.01 & 1.25 & $\cdots$ & 2012.1.00076.S\\
21 & CFHQSJ0210-0456 & 1.20 & 249 (6) & 27.2 & 0.92$\times$0.61 & 1.00&  (m),(n) & 2011.0.00243.S\\ 
22 & MSDM80+3 & 1.06 & 283 (7) & 28.4& 0.84$\times$0.63 & 0.69 &  (q) & 2012.1.00536.S\\
23 & GRB051022 &1.15 & 262 (6) & 28.6 & 1.15$\times$0.77 & 0.88&  (p) & 2011.0.00232.S\\
24 & ID217 &1.23 & 244 (6) & 30.3 & 1.33$\times$0.93 &1.08 & $\cdots$ & 2012.1.00076.S\\
25 & ID225 &1.23 & 244 (6) & 31.6 & 1.34$\times$0.93 &1.08 & $\cdots$ & 2012.1.00076.S\\
26 & SXDS19723 &1.27 &236 (6) & 38.5 & 0.83$\times$0.68 &1.19 & (r),(s) & 2011.0.00648.S\\
27 & SXDS28019 &1.27 &236 (6) & 38.9 & 0.84$\times$0.68 &1.19 & (r),(s) & 2011.0.00648.S\\
28 & SXDS22198 &1.27 &236 (6) & 39.0 & 0.86$\times$0.68 &1.19 & (r),(s) & 2011.0.00648.S\\
29 & SXDS35572 &1.27 &236 (6) & 39.3 & 0.84$\times$0.68 &1.19 & (r),(s) & 2011.0.00648.S\\
30 & SXDS103139 &1.27 &236 (6) & 39.7 & 0.83$\times$0.68 &1.19 & (r),(s) & 2011.0.00648.S\\
31 & NB921-N-79144 & 1.22 & 245 (6) & 50.1 & 0.76$\times$0.63 & 1.05 &  (h),(f) & 2012.1.00602.S\\
32 & SXDS42087  &1.30 &231 (6) & 52.9& 0.80$\times$0.66&1.28 & (r),(s) & 2011.0.00648.S\\
33 & SXDS79307 & 1.30 &231 (6) & 53.7 & 0.80$\times$0.66 &1.28 & (r),(s) & 2011.0.00648.S\\
34 & SXDS31189 & 1.30 &231 (6) & 54.4 & 0.81$\times$0.66 &1.28 & (r),(s) & 2011.0.00648.S\\
35& MSDM29.5-5 & 1.06 & 283 (7) & 54.6& 0.71$\times$0.53 & 0.69  & (q) & 2012.1.00536.S\\
36 & HiZELS-UDS-NBK-8806 & 1.39 & 216 (6) & 55.0& 1.86$\times$0.89 & 1.58 & $\cdots$ & 2012.1.00934.S\\
37 & HiZELS-UDS-NBK-11473 & 1.39 & 216 (6) & 55.3& 1.86$\times$0.89 & 1.58 & $\cdots$ & 2012.1.00934.S\\
38 & MSDM71-5 & 1.06 & 283 (7) & 56.1& 0.72$\times$0.53 & 0.69   & (q) & 2012.1.00536.S\\
39 & HiZELS-UDS-NBK-13486 & 1.39 & 216 (6) & 56.4& 1.86$\times$0.89 & 1.58 & $\cdots$ & 2012.1.00934.S\\
40 & HiZELS-UDS-NBK-11961 & 1.39 & 216 (6) & 58.3& 1.86$\times$0.89 & 1.58 &  $\cdots$ & 2012.1.00934.S\\
41 & ID125 &1.09 & 273 (6) & 59.1 & 1.30$\times$0.73 &0.75 & $\cdots$ & 2012.1.00076.S\\ \hline
 &  \multicolumn{7}{c}{Medium-Deep Data (B)} & \\ \hline
42 & ID098 &1.09 & 273 (6) & 60.1 & 1.30$\times$0.72 &0.75 & $\cdots$ &  2012.1.00076.S\\
43 & ID093 &1.09 & 273 (6) & 60.3 & 1.30$\times$0.72 &0.75 & $\cdots$ &  2012.1.00076.S\\
44 & ID143 &1.09 & 273 (6) & 61.7 & 1.31$\times$0.72 &0.75 & $\cdots$ &  2012.1.00076.S\\ 
45 & SXDS33244 &1.25 &240 (6) & 67.5 & 1.04$\times$0.64 &1.13 & (r),(s) &  2011.0.00648.S\\
46 & ID163 & 1.19 &252 (6) & 68.2 & 1.36$\times$0.92 & 0.97& $\cdots$ &  2012.1.00076.S\\
47 & ID204 & 1.19 &252 (6) & 69.2 & 1.24$\times$0.86 & 0.97& $\cdots$ &  2012.1.00076.S\\ 
48 & SXDS13316 &1.23 &244 (6)& 69.3 &  0.87$\times$0.64&1.08 & (r),(s) &  2011.0.00648.S\\
49 & ID158 & 1.19 &252 (6)& 69.5 & 1.35$\times$0.92 &0.97 & $\cdots$ &  2012.1.00076.S\\
50 & ID192 & 1.19 &252 (6)& 69.7 & 1.24$\times$0.86 &0.97 & $\cdots$ &  2012.1.00076.S\\
51 & SXDS59863 &1.23 &244 (6) & 71.6 & 0.85$\times$0.65&1.08 & (r),(s)&  2011.0.00648.S\\
52 & ID177 & 1.13 &264 (6)& 73.0 & 1.24$\times$0.86 & 0.83& $\cdots$ &  2012.1.00076.S\\
53 & ID107 & 1.13 &264 (6)& 73.8 & 1.24$\times$0.86 & 0.83& $\cdots$ &  2012.1.00076.S\\
54 & SXDS67002 &1.23 &244 (6)& 74.0 & 0.88$\times$0.64 &1.08 & (r),(s)&  2011.0.00648.S\\
55 & ID112 & 1.13 &264 (6)& 75.4 & 1.24$\times$0.86 & 0.83& $\cdots$ &  2012.1.00076.S\\
56 & ID117 & 1.13 &264 (6)& 75.5 & 1.24$\times$0.86 & 0.83& $\cdots$ &  2012.1.00076.S\\
57 & SXDS9364 &1.23 &244 (6)& 76.7 & 0.87$\times$0.64 &1.08 &  (r),(s)&  2011.0.00648.S\\
58 & SXDS13015 &1.23 &244 (6)& 81.3 & 0.85$\times$0.65 & 1.08&  (r),(s)&  2011.0.00648.S\\
 59 &  113083 &  1.32 &  226 (6) &  82.1&  1.18$\times$0.72 &  1.36 & $\cdots$ &  2012.1.00323.S\\
60 & SXDS68849 &1.25 &240 (6)& 82.3 & 0.96$\times$0.65 & 1.13&  (r),(s)&  2011.0.00648.S\\
61 & SXDS79518 &1.25 &240 (6)& 83.0 & 1.02$\times$0.65 & 1.13&  (r),(s)&  2011.0.00648.S\\
 62 &  1374240 &  1.32 &  226 (6) &  84.6&  1.20$\times$0.72 &  1.36 & $\cdots$ &  2012.1.00323.S\\
63 & SXDS101746 &1.25 &240 (6)& 86.7 & 1.13$\times$0.65 & 1.13& (r),(s)&  2011.0.00648.S\\
64 & SXDS110465 &1.25 &240 (6)& 87.6 & 1.13$\times$0.64 &  1.13& (r),(s)&  2011.0.00648.S\\
65 & SXDS1723 &1.25 &240 (6)& 94.0 & 1.29$\times$0.64 & 1.13& (r),(s) &  2011.0.00648.S\\
66 & SXDS59914 &1.25 &240 (6)& 95.2 & 1.20$\times$0.65 & 1.13& (r),(s) &  2011.0.00648.S\\ \hline
 &  \multicolumn{7}{c}{Cluster Data (C)} & \\ \hline
67 & A1689 & 1.32 &231 (6)& 38.4-40.6$^{\dagger}$ &  0.96$\times$0.74 & 1.36 &(t) & 2011.0.00319.S \\
  & & & & & & & & 2012.1.00261.S 
\enddata 
\tablecomments{
\footnotesize{
(1): Wavelength in the observed frame. 
(2): Frequency in the observed frame. 
(3): One sigma noise measured in each map 
before primary beam correction.
(4): Synthesized beam size of our ALMA maps (weighting $=$ 'natural').
(5): Ratio of the flux density at $1.2$ mm, $S_{1.2{\rm mm}}$, 
to the one at the observed wavelength, $S_{\rm obs}$, 
that is estimated with the modified blackbody spectrum of
$\beta_{\rm d}=1.8$, a dust temperature of $T_{\rm d}=35$K, and a redshift of $z=2.5$.
(6): Reference. 
(a) \cite{vanzella2011};
(b) \cite{maiolino2015};
(c) \cite{tanvir2012};
(d) \cite{berger2014};
(e) \cite{konno2014};
(f) \cite{ouchi2010};
(g)\cite{haberl1998}; 
(h) \cite{ouchi2013}; 
(i) \cite{ono2014};
(j) \cite{iye2006};
(k) \cite{ota2014};
(l) \cite{zampieri2001};
(m) \cite{willott2010};
(n) \cite{willott2013}; 
(o) \cite{ouchi2009}; 
(p) \cite{hatsukade2014};
(q) \cite{martin2006};
(r) \cite{yabe2012};
(s) \cite{hatsukade2013};
(t) \cite{watson2015};
}}
\tablenotetext{$\dagger$}{%
Noise levels are measured in the 9 regions (see text for the details). 
}
\end{deluxetable*} 

\subsection{Our Data}
\label{sec:our_data}
The 4 maps of the field data were obtained by our ALMA 
single pointing observations.
Two out of the four maps were taken in the ALMA cycle-0 Band 6 observations
for the spectroscopically confirmed Ly$\alpha$ emitter (LAE) at $z=6.595$, Himiko \citep{ouchi2013} and 
an LAE at $z=6.511$, NB921-N-79144 \citep[][PI R. Momose]{ouchi2010}. 
See \cite{ono2014} for the summary of these observations. 
We also utilize two maps of newly obtained (ALMA cycle 2) field data of Bands 6 and 7
taken for spectroscopically confirmed LAEs 
at $z=5.7$ (NB816-S-61269; \citealt{ouchi2008}) and
at $z=7.3$ (NB101-S-2904; \citealt{konno2014}, Ouchi et al. in preparation). 
The NB816-S-61269 observations were carried out in the ALMA program of \#2012.1.00602.S (PI: R. Momose) on 2014 May 20, Jun 19, and July 7 with 43 12-m antennae array in the range of $18-650$ m baseline. 
The full width at half maximum (FWHM) of the primary beam was $20\farcs6$.
The available 7.5 GHz bandwidth with four spectral windows was centered at an observed frequency of 290.8 GHz (i.e. $\sim1.03$ mm). 
J0217+0144 and J2258-279 were observed as a flux calibrator, while J0006-0623 was used for bandpass calibrator.  
Phase calibration was generally performed by using observations of J0217+0144 and J0215-0222.
The total on-source observed time was $\sim$ 2.3 hours. 
The NB101-S-2904 data were taken in the ALMA program of \#2012.1.00088.S (PI: M. Ouchi) on 2014 July 22, August 6, 7, 14, and 18, with 55 12-m antennae array in the extended configuration of 18-1300 m baseline. 
The full width at half maximum (FWHM) of the primary beam was $25\farcs3$. The center of the observed frequency is 237.9 GHz (i.e. $\sim1.26$ mm). 
J0238+166, J2258-279, and J0334-401 were used for primary flux calibrators.
Bandpass and phase calibrations were performed with
J0241-0815/J0238+1636 and J0215-0222/J0217+0144, respectively.
The total on-source integration time was $\sim$ 5.2 hours.

\subsection{Archival Data}
\label{sec:archival_data}
To increase the number of ALMA  sources,
we make full use of ALMA archival data of cycles 0 and 1 that became public  by 2015 June.
We select all of the available ALMA Band 6$/$7 data sets fulfilling the following three criteria: 
1) The 1$\sigma$ noise level is $\lesssim 100$ $\mu$Jy/beam for the continuum maps, 
2) No bright ($\geq 50\sigma$) sources like AGN, QSO and SMG are included in the maps, and 
3) The galactic latitude is high enough to avoid Galactic objects.
The second criterion is important to reduce the chance for selecting 
residual side lobes of the bright sources that might remain even after the CLEAN algorithm on Common Astronomy Software Applications \citep[{\sc casa};][]{mcmullin2007} package.  
Moreover, the second criterion significantly reduces the systematic uncertainties
from galaxy clustering because it is reported that there is a clear excess of submm number counts
around bright sources such as AGNs and SMGs \citep{silva2015,simpson2015b}

\subsubsection{Field Data}
\label{sec:field_data}
On the basis of our complete archival data search up to 2015 June 
with these three criteria, we collect the 62 maps of the field data
taken by single-pointing observations:  
2 from \cite{willott2013}, 20 from \cite{hatsukade2013}, 1 from \cite{ota2014},
2 from \cite{hatsukade2014}, 1 from \cite{berger2014}, 3 from \cite{carniani2015}, 20 from ALMA \#2012.1.00076, 2 from ALMA\#2012.1.00676.S,
 2 from ALMA \#2012.1.00323, 3 from ALMA \#2012.1.00536, 4 from ALMA \#2012.1.00934, and 2 from ALMA \#2012.1.00610. 
The ALMA \#2012.1.00076 data were taken
by PI K. Scott for 20 IR galaxies at $z=0.25-0.65$ 
in January 2014 with Band 6. The total on-source integration is $\sim 20-40$ min for each map. 
The ALMA \#2012.1.00676.S observations were carried out by PI C. Willott for 2 quasars at $z\sim6$ 
in the end of November 2013 with Band 6. 
The total on-source integration is $\sim40$ min.
The ALMA \#2012.1.00323.S observations were performed by PI G. Popping for 2 star-forming galaxies in March 2014 with Band 6. 
The total on-source integration is $\sim$7 min for each source.
The ALMA \#2012.1.00536.S data were taken by PI C. Martin for 3 LAEs at $z=5.7$ with Band 7.
The total on-source integration is $\sim 20-40$ min for each source.
The ALMA \#2012.1.00934.S observations were conducted by PI B. Phillip for 4 normal star-forming galaxies at $z=2.23$ in January 2014 with Band 6.
The total on-source integration is $\sim 6$ min for each source. 
The ALMA \#2012.1.00610.S observations were carried out by PI B. Posselt for 2 nearby neutron stars in April and Jun 2014 with Band 6.
The total on-source integrations are $\sim 30$ and 60 mins for  each source.

\subsubsection{Cluster Data}
\label{sec:cluster_data}
One-cluster data were taken for Abell 1689 (A1689) in ALMA cycle 0 and 1 observations (PI: J. Richard)
of ALMA \#2011.0.00319.S and \#2012.1.00261.S. 
These two programs of A1689 included two different frequency settings, high and low tuning modes.
The central-observed frequencies were chosen at 231 and 222 GHz (i.e. $\sim$ 1.30 and 1.35 mm) for the high and the low tuning modes, respectively.
Thus, a total of four observing sets were conducted for A1689 through ALMA cycle 0 and 1.
Each observing set comprises $2-4$ sub-blocks of observations in which the 
source integration time is $\sim10-30$ minutes to make a mosaic map covering 
the highly magnified area of A1689 by $\sim$50 pointings.
The first observation set was obtained on 2012 June 17-18, and July 3-4 with
20 antennas in the high-tuning mode. 
 
Mars and Titan were observed as flux calibrators. 
The bandpass and phase calibrations were performed using observations of 3C 279.
The second observation set was taken on 2012 July 14,15, 28, August 9, and 2013 January 1 with 
24 antennas in the low-tuning mode.
The available 7.5 GHz bandwidth with four spectral windows was centered at an observed frequency of 222 GHz (i.e. $\sim$1.35 mm). 
Titan was observed as a flux calibrator. Bandpass and phase calibrations were performed with 3C 279.  
The third observation set was obtained on 2014 March 11 and April 4 with 34 antennas in the low-tuning mode. 
Titan was observed as a flux calibrator. Bandpass and phase calibrations were carried out with J1256-0547. 
The fourth observing set was performed on 2014 March 22 and April 4 with 37 antennas in the high-tuning mode.
Ceres was observed as a flux calibrator. Bandpass and phase calibrations were done with J1256-0547.
Throughout these four observing sets, the total on-source integration is $\sim5.5$ hour.

\subsection{Data Reduction}
\label{sec:data_reduction}

Basically, the data are reduced with CASA version 4.3.0 in a standard manner
with the scripts used for the data reduction provided by the ALMA observatory.
In this process, we also use previous CASA versions from 3.4 to 4.2.2, if we find problems in the final image
that the noise level is significantly higher than the calibrated products provided by the ALMA observatory,
or that there remain striped patterns.
Exceptionally for the cycle 0 data with the high noise level or the patterned noise,
we use the calibrated data produced by the ALMA observatory.
Similarly for the data of ALMA \#2011.0.00648.S (PI: Ohta),  
we use a re-calibrated data provided by Seko et al. because they find that the coordinate of 
a phase calibrator of \#2011.0.00648.S is wrong ($\sim0\farcs3$), which causes 
positional offsets \citep[Seko et al. in preparation;][]{hatsukade2015b}.

Our CASA reduction has three major steps: bad data flagging, 
bandpass calibration, and 
gain calibration with flux scaling. 
In the first step, we remove the data of the shadowed antennas
and the edge channels of spectral windows. 
We also do not use the unreliable data such with jumps 
or a low phase/amplitude gain.
We apply the flaggings that are shown in the scripts,
but no additional flaggings.
This is because we find that the flaggings in the scripts 
are good enough for our scientific goals. 
The second step is the bandpass calibration.
After we calibrate the phase time variation on the bandpass calibrator scan,
we obtain the bandpass calibration in the phase and amplitude.
The final step is 
the gain calibration. 
We estimate the time variations of the phase and amplitude on the phase calibrators, 
and then transfer the calibration to the target source by applying the linear interpolation 
with the results of the phase calibrators. 
The water vapor radiometer is used for the correction of the short-time phase variations 
that cannot be traced by the phase calibrators. 
We estimate flux scaling factors, using solar-system objects and bright QSOs 
with the flux models of 'Bulter-JPL-Horizons 2012' and ALMA Calibrator Source Catalog, respectively. 
The difference of the opacity between the flux calibrators and the target sources is corrected with the system noise temperature (T$_{\rm sys}$) measurements. 
The systematic flux uncertainty is typically 10\% in bands 6 and 7 (ALMA proposer's guide
\footnote{Section A7: https://almascience.nao.ac.jp/documents-and-tools/cycle-1/alma-proposers-guide}).

We perform Fourier transformation for the original {\it uv}-data to create ''dirty'' maps. 
The continuum maps are made using all the line free channels of the four spectral windows. 
We process the maps with the auto CLEAN algorithm down to the depth of 3$\sigma$ noise levels of 
the dirty maps, using natural weighting. 
The final cleaned maps achieve angular resolutions of $0\farcs45\times0\farcs33$ $- 1\farcs88\times0\farcs90$. 
The sensitivities of these ALMA maps range from 8.5 to 95.2 $\mu$Jy beam$^{-1}$
before the primary beam corrections.

\begin{figure}
\begin{center}
\includegraphics[trim=0.0cm 0cm 0.0cm 0cm, clip, angle=0,width=0.5\textwidth]{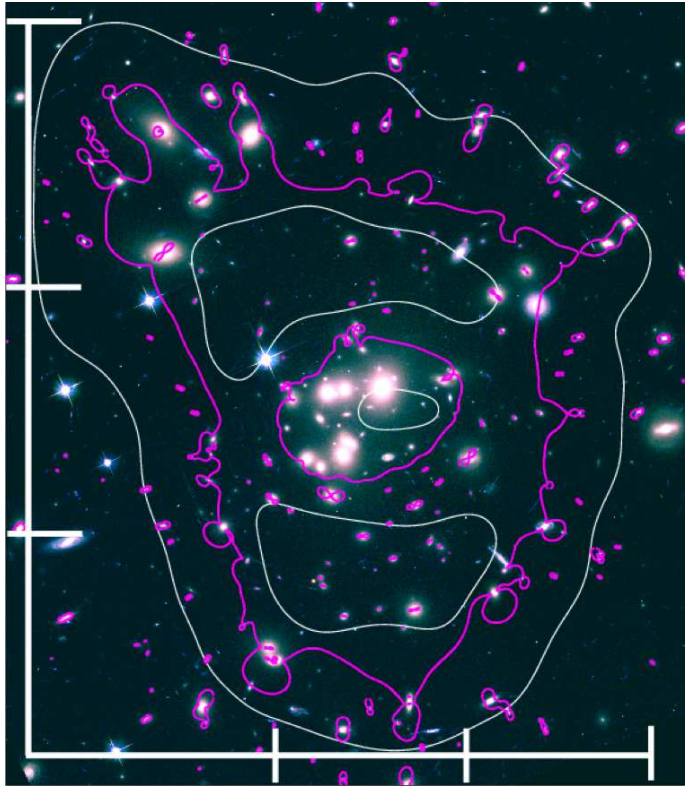}
\caption[]
{False-color image of A1689 cluster (red: $i_{775}$, green: $V_{625}$, red: $B_{475}$). The critical lines for background sources at $z=2.5$ are shown with the magenta lines. 
The white curves are the contours indicating 
the $50$\% sensitivity of the deepest part in the mosaic map.
The white thick lines and ticks indicate the 9 regions of our noise estimates.
\label{fig:a1689}}
\end{center}
\end{figure}

Because the observations of the cluster data were taken place in 4 different epochs, 
we recalculate the data weights with the $statwt$ task in CASA based on its visibility scatters 
which include the effects of integration time, channel width, and systematic temperature. 
After applying the data weights to the datasets, we combine these two datasets with the $concat$ task in CASA.
Since the mosaic mapping observations were conducted for the cluster, the noise levels vary by positions.
To estimate the noise levels of the cluster data, we divide the cluster's mosaic map in 9 regions 
as shown in Figure \ref{fig:a1689}.
The data of the 9 regions achieve the 1$\sigma$ noise levels in the range of 38.4-40.6 $\mu$Jy.


\section{Data Analysis} 
\label{sec:data_analysis}

We show our data analyses in the following subsections. 
The wavelengths of our mm maps fall in the range of $1.03-1.32$ mm. 
Our data set is composed of 63 Band 6 maps and 4 Band 7 maps (Table 2),
and the majority of our data are taken with Band 6.
Because our major data of the ALMA Band 6 maps have the average wavelength of 1.23 mm,
we derive the number counts with a flux density at 1.2 mm, $S_{\rm 1.2mm}$. 
We scale the flux densities of our mm maps to $S_{\rm 1.2mm}$ with the
flux density ratios summarized in Table \ref{tab:our_alma_maps}.
For the data of the previous studies defined with the wavelengths different from 1.2 mm, 
we estimate $S_{\rm 1.2mm}$ with the ratios of
$S_{\rm 1.2mm} / S_{850\mu{\rm m}} = 0.37$,  
$S_{\rm 1.2mm} / S_{870\mu{\rm m}} = 0.39$,  
$S_{\rm 1.2mm} / S_{\rm 1.1mm} = 0.77$, 
and 
$S_{\rm 1.2mm} / S_{\rm 1.3mm} = 1.28$.
All of these flux density ratios are calculated based on a modified blackbody 
whose spectral index $\beta_{\rm d}$ and dust temperature $T_{\rm d}$ are
similar to those of typical SMGs; $\beta_{\rm d} = 1.8$ \citep[e.g., ][; Planck Collaboration \citeyear{planck2011}]{chapin2009}
and $T_{\rm d}=35$ K \citep[e.g.,][]{kovacs2006,coppin2008}.
Here we assume a source redshift of $z=2.5$ that is  
a median redshift
value of SMGs \citep[e.g.,][]{chapman2005,yun2012,simpson2014}.

\subsection{Source Detection}
\label{sec:source_extraction}

We conduct source extraction for our ALMA maps before primary beam corrections 
with SExtractor version 2.5.0 \citep{bertin1996}. 
textcolor{blue}
{
The source extraction is carried out in high sensitivity regions.
For the single-pointing maps presented in Sections \ref{sec:our_data} and \ref{sec:field_data},
we use the regions with the primary beam sensitivity greater than 50\%.
For the mosaic map shown in Section \ref{sec:cluster_data}, we perform source extraction
where the relative sensitivity to the deepest part of the mosaic map is greater than 50\%.
}

We identify sources with a positive peak count above the 3.0$\sigma$ level,
assuming the sources are not resolved out in our ALMA maps. 
The catalog of these sources is referred to as the 3.0$\sigma$-detection catalog. 
From our 3.0$\sigma$-detection catalog,
we remove the objects located at the map centers 
that are main science targets of the archived-data ALMA observations.

\begin{figure}
\begin{center}
\center{\includegraphics[trim=0.0cm 0.2cm 1.0cm 0cm, clip,angle=0,width=0.5 \textwidth]{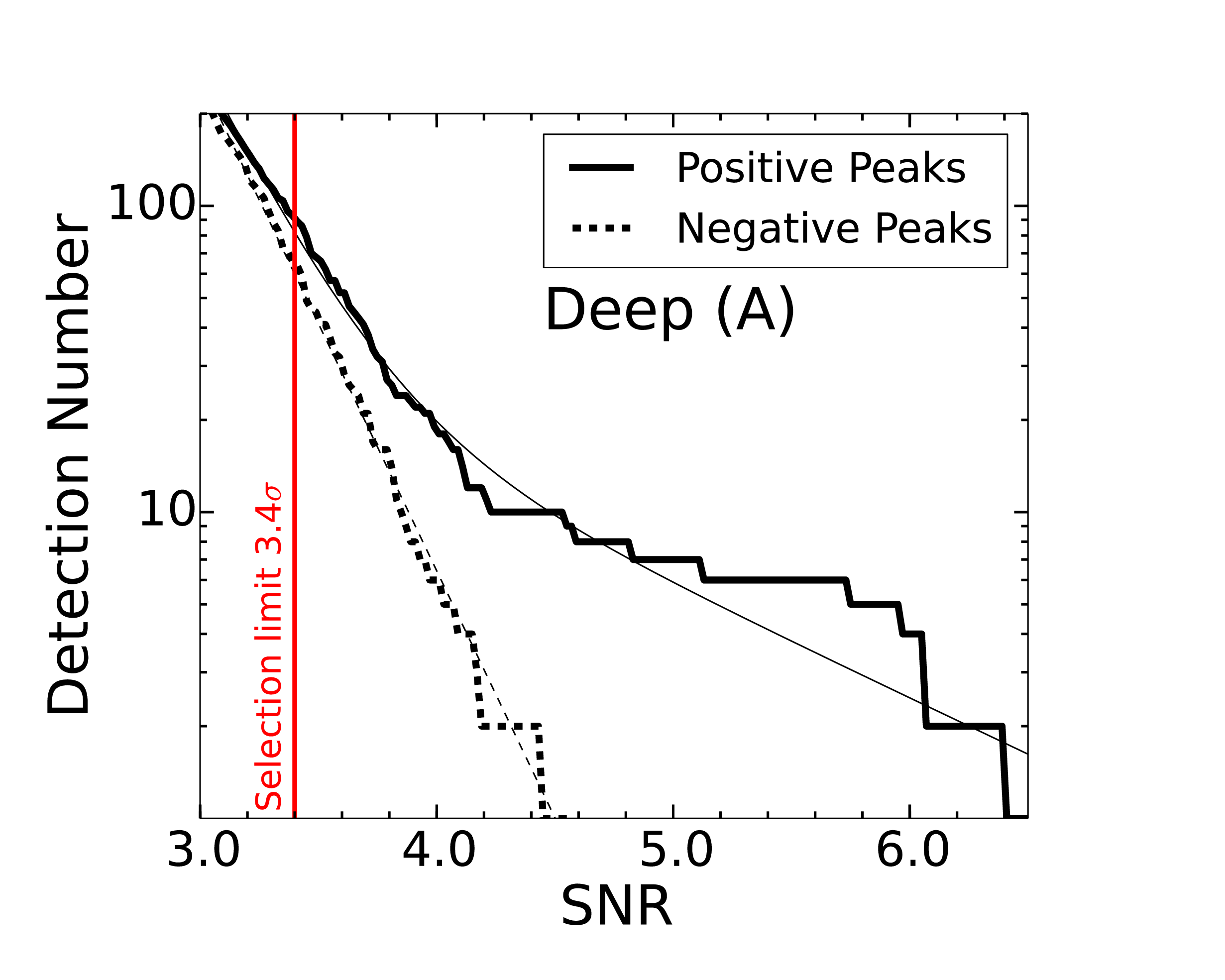}}
\vspace{-1.2cm}
\center{\includegraphics[trim=0.0cm 0.2cm 1.0cm 0cm, clip,angle=0,width=0.5\textwidth]{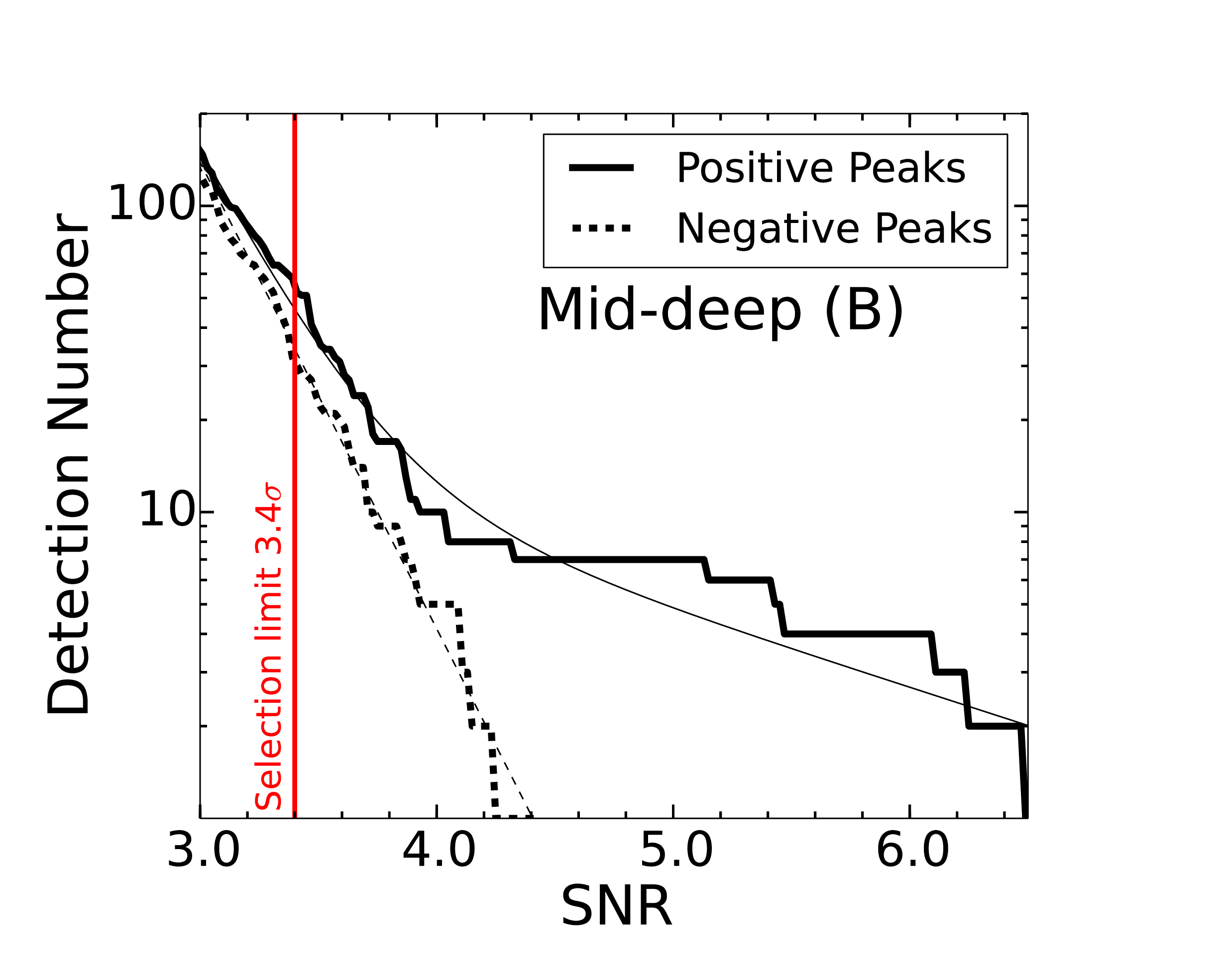}}
\vspace{-1.2cm}
\center{\includegraphics[trim=0.0cm 0.2cm 1.0cm 0cm, clip,angle=0,width=0.5\textwidth]{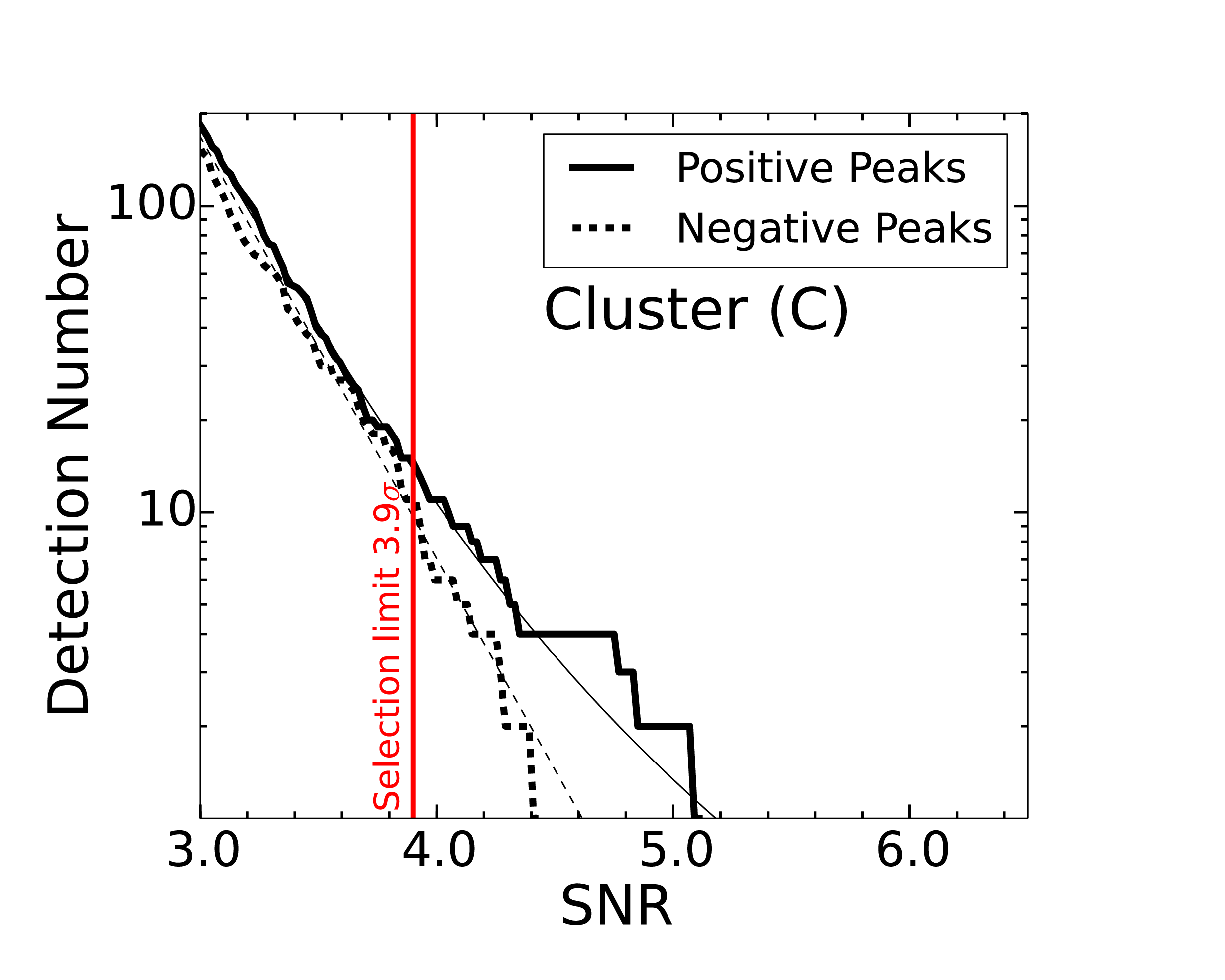}}
\caption{
Total numbers of positive and negative peaks in the deep (A),  medium-deep (B), and cluster (C) sub-datasets as a function of SNR.
The thick solid and dashed lines show the total numbers of the positive and negative peaks, respectively. 
The thin solid and dashed curves denote the best-fit model functions. 
The red lines are selection limits for our source catalog. 
See text for more details.
\label{fig:pm_count}}
\end{center}
\end{figure}

\subsection{Spurious Sources and Source Catalog}
\label{sec:spurious_rate}
The 3.0$\sigma$-detection catalog should include many spurious sources,
because the detection is simply determined with a peak pixel count. 
To evaluate the number of the spurious sources, we conduct negative peak analysis 
\citep{hatsukade2013, ono2014, carniani2015}
for the deep (A), medium-deep (B), and cluster (C) datasets.
We multiply $-1$ to each pixel value of our ALMA maps,
and perform the source extraction 
for negative peaks
in the same manner as 
those in Section \ref{sec:source_extraction}.
The numbers of negative and positive peaks in our ALMA maps are shown in Figure \ref{fig:pm_count}. 
The excess of the positive to negative peak numbers is regarded as the real source numbers.
We model the negative and positive peak distributions of Figure \ref{fig:pm_count} with the functions of
signal-to-noise ratio SNR,
\begin{eqnarray}
\label{eq:peak_distribution_deep}
&{N_{\rm np}^{\rm c}}&\ =\ a_{\rm n}\times 10^{-b_{\rm n}\times {\rm SNR}},  \\
\label{eq:peak_distribution_mdeep}
&{N_{\rm pp}^{\rm c}}&\ =\ a_{\rm n}\times 10^{-b_{\rm n}\times {\rm SNR}} + a_{\rm p}\times 10^{-b_{\rm p}\times {\rm SNR}}, 
\end{eqnarray} 
where $N_{\rm np}^{\rm c}$ and $N_{\rm pp}^{\rm c}$ are the cumulative numbers of negative and positive peaks, respectively, 
and $a_{\rm n}$, $b_{\rm n}$, $a_{\rm p}$, and $b_{\rm p}$ are free parameters. 

\begin{figure}
\begin{center}
\includegraphics[trim=0.0cm 0.2cm 1.0cm 0cm, clip,angle=0,width=0.5\textwidth]{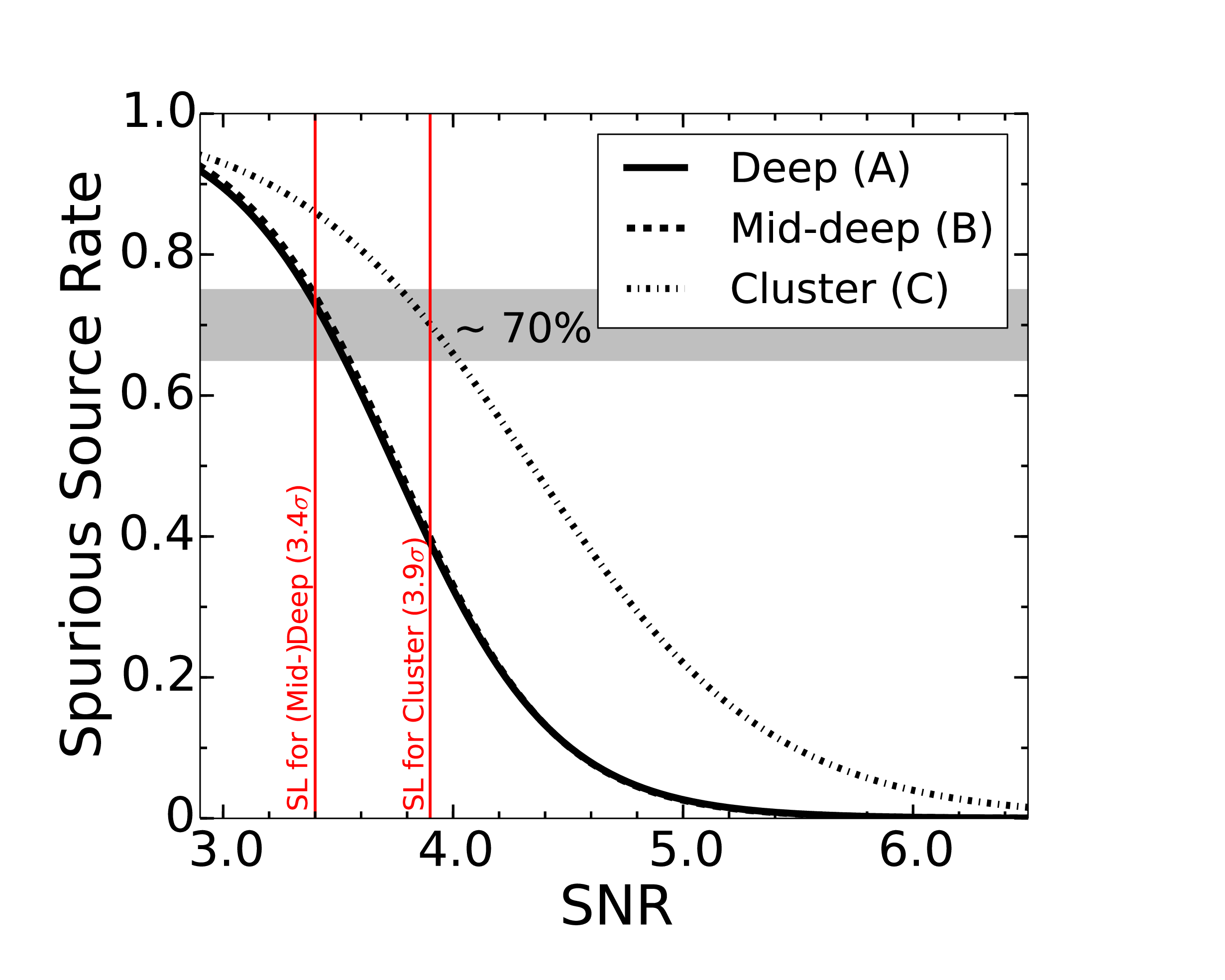}
\caption[]
{Cumulative spurious source rates as a function of SNR.
The solid, dashed, and dotted lines indicate the deep (A), medium-deep (B), and cluster (C) sub-datasets, respectively.
The red lines at SNRs of 3.4 and 3.9 are the selection limits (SL) of our source catalog.
\label{fig:sp_rate}}
\end{center}
\end{figure}

We estimate spurious source rates $f_{\rm sp}$ defined by the ratio of negative to positive peak numbers,
\begin{eqnarray}
\label{eq:sp_rate}
&f_{\rm sp}({\rm SNR})&\ =\ \frac{N_{\rm np}^{\rm c}({\rm SNR})}{N_{\rm pp}^{\rm c}({\rm SNR})}.
\end{eqnarray}
We evaluate $f_{\rm sp}$ with the best-fit functions of eqs. 
(\ref{eq:peak_distribution_deep}) and (\ref{eq:peak_distribution_mdeep}), 
and present $f_{\rm sp}$ in Figure \ref{fig:sp_rate}. 

Figures \ref{fig:sp_rate} indicates that the field data of deep (A) and medium-deep (B)
have an almost identical distribution of the spurious source rates, while the cluster data (C) 
shows spurious source rates higher than (A) and (B) at a given SNR.
This result would suggest that spurious source rates do not depend on
the data depth, but the mapping modes. The complex distribution of the
depths in the mapping data of (C) would have the relation of SNR and spurious source rate
that is different from the smooth distribution of  the depth in the single pointing data of (A) and (B).
We use sources down to an SNR whose spurious source rate is $\lesssim 70$\%,
and adopt 3.4, 3.4, and $3.9\sigma$
levels for our selection limits of  
the data (A), (B), and (C), respectively.
Applying these selection limits to our 3.0$\sigma$-detection catalog,
we obtain a source catalog consisting of 133 sources: 122 from the field data (A and B) and 11 from the cluster data (C).
The source catalog of our 133 faint ALMA sources is presented in
Table \ref{tab:our_catalog}.

Note that A1689-zD1, an LBG at $z=7.5$ behind A1689 \citep{watson2015}, is not included in our source catalog because 
A1689-zD1 is located outside of the primary beam with a $>50\%$ sensitivity that is one of our selection criteria. 
Nevertheless, we have looked at the position of A1689-zD1 in our data, and found that there is a source 
with an SNR of 4 whose flux is the same as the one derived by \cite{watson2015} within the $\sim1\sigma$ uncertainty.
We confirm the ALMA source detection at the position of A1689-zD1.
%

\subsection{Completeness and Flux Boosting} 
\label{sec:completeness}

\begin{figure}
\begin{center}
\includegraphics[trim=0.0cm 0.2cm 1.0cm 0cm, clip, angle=0,width=0.5\textwidth]{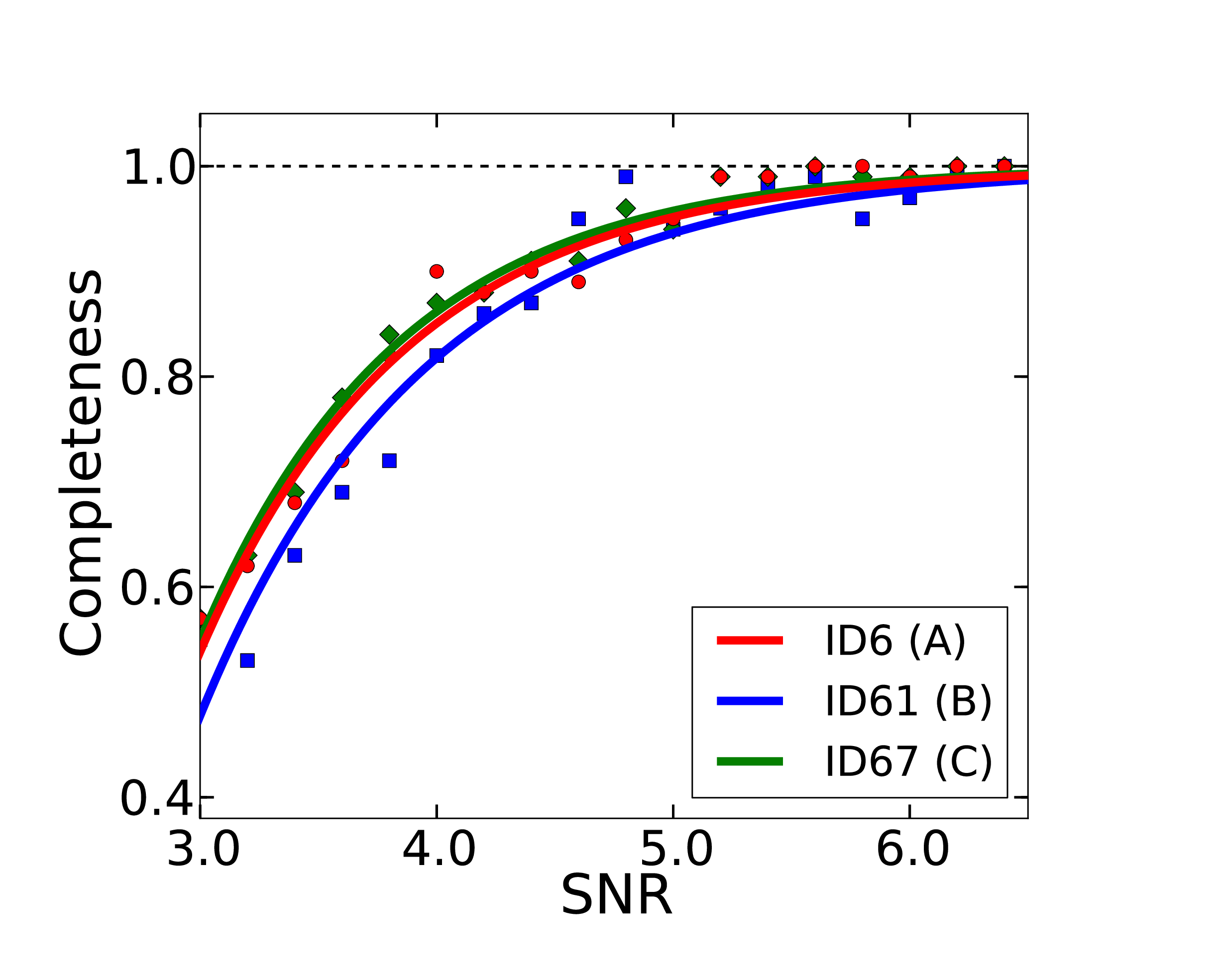}
\caption[]
{Completeness as functions of SNR for three sub-datasets.  The circles (red line), squares (blue line), 
and diamonds (green line) represent the completeness values (the best-fit functions) of 
the ID 6, 61 and 67 maps, respectively.
\label{fig:comp}}
\end{center}
\end{figure}

We estimate the detection completeness by Monte-Carlo simulations.
We put a flux-scaled synthesized beam into a map as an artificial source on a random position. 
These artificial sources have SNRs ranging from 3.0 to 7.0 with an SNR step of 0.2 dex.
If an artificial source is extracted within a distance of
the synthesized beam size from the input position, 
we regard that the source is recovered.
These completeness estimations are conducted 
in the data before primary beam corrections,
because we perform the source extractions in the real data (Section \ref{sec:source_extraction})
uncorrected for primary beam attenuations.
For all of our ALMA maps, we iterate this process 100 times at an SNR bin. 
Figure \ref{fig:comp} displays our completeness estimates for three sub-datasets. 
We fit the function of $1-\exp(a {\rm SNR} -b)$ to our completeness estimates,
where $a$ and $b$ are free parameters. 
The best-fit functions are also shown in Figure \ref{fig:comp}. 

We examine the flux boosting that is caused by the confusions of the undetected faint sources
\citep{austermann2009, austermann2010}. 
We use the detected objects of the Monte-Carlo simulations conducted for the completeness estimates,
and measure the output flux densities at the input positions.
We find that the ratios of the input to output flux densities are 
almost unity with a variance of only 5\% 
over the wide SNR range of our ALMA maps.
This small flux boosting is probably found, because 
the angular resolution of ALMA is high.
We thus conclude that the effect of flux boosting is negligibly small.

\subsection{Flux Measurement}
\label{sec:flux_measure}

We estimate the source fluxes of our 133 objects
from the integrated flux densities 
measured with the 2D Gaussian-fitting routine of $imfit$ in CASA.
Because the integrated flux density values for low-SNR sources would include 
large systematic uncertainties originated from the spatially-correlated positive noise 
at around the source position, we use the integrated flux densities of $imfit$ 
only for the sources 
which are identified as 'resolved' 
with an SNR of $\geq$ 5.
For the rest of the sources, we use the peak fluxes of the $imfit$ best-fit Gaussian measurements. 
We confirm that the peak flux values of $imfit$, $S_{\rm imfit}$, agree 
with those of SExtractor, $S_{\rm SEx}$, within the 1$\sigma$ error,
$S_{\rm imfit}/S_{\rm SEx}=0.93\pm0.15$. 

Because the sources in the cluster data of A1689 would be distorted by the gravitational lensing effects, 
we make a low-resolution map of A1689 produced with a $1''$ circular gaussian uv-taper that is 
the same as the one produced by \cite{watson2015}.
The taped beam size is $1\farcs31\times1\farcs12$. 
In this map, we perform the $imfit$ flux measurements (as described above)
at the source positions determined in the original map.
Here we define flux measurements in the tapered and non-tapered maps 
as $S_{\rm taper}$ and $S_{\rm no\,taper}$, respectively. 
We then obtain $S_{\rm taper} / S_{\rm no\,taper} = 1.01 \pm 0.19$ on average, 
if we omit two sources showing signatures of source confusions by neighboring objects. 
However, the strongly-lensed source of A1 shows a systematically high fraction of $S_{\rm taper}$ / $S_{\rm no\,taper} =1.23$. 
Thus, we decide to use $S_{\rm taper}$ for the cluster data of A1689, except for those two confusion sources.
We also make low-resolution maps for some of the field data, and perform the same test of tapering. 
Because we find that $S_{\rm taper}$ and $S_{\rm no\,taper}$ are almost identical in these data, 
we use $S_{\rm no\,taper}$ for the field data.

We test the reliability of our flux measurements, performing flux recovery simulations.
First, we select the typical data sets of ID22 and ID55 that have
beam sizes of $\sim0\farcs8$ and $\sim1\farcs2$ corresponding to
the two peaks of bimodal size distribution for our data.
We then create 200 input sources of elliptical Gaussian model profiles with the uniform distribution of SNRs 
in $3-7$ and major-axis sizes of $0\farcs2$ and $0\farcs$5. Here, the input source sizes are chosen 
from recent ALMA high-resolution studies. \citet{ikarashi2014} and \citet{simpson2015a}
claim that the median sizes of individual submm sources are $\sim0\farcs2-0\farcs3$ in FWHM. 
\cite{simpson2015a} also find that the stacked image of submm sources with a low SNR of $4-5$ 
has the size of FWHM $0\farcs35^{+0.17}_{-0.10}$. 
Thus, we investigate the model sources of $0\farcs2$ and $0\farcs5$ 
that are taken from the median individual source size and the 1$\sigma$ upper limit stacked source size.
We add these model sources to ID22 and ID55 maps at random positions.
Finally, we obtain flux densities of the model sources in the same manner as our flux measurements for the real sources,
and compare these flux measurements with the input source fluxes.
In the case of the $0\farcs2$ input source size, we find that the flux recovery ratio (defined by the ratio of the output to input fluxes) is 
$1.0\pm0.1$ ($1.0\pm0.2$) for ID22 (ID55) data. Similarly, in the case of the $0\farcs5$ input source size, we obtain $0.8\pm 0.1$ ($0.9\pm 0.2$) 
for ID22 (ID55) data. These results suggest that the flux recovery ratios are nearly unity 
within the $1-2\sigma$ uncertainties, and that our flux measurements are reliable.

Note that missing fluxes of interferometric observations are 
negligible.
ALMA proposer's guide 
\footnote{Table A2 of the ALMA proposers' guide: https://almascience.nao.ac.jp/documents-and-tools/cycle-1/alma-proposers-guide}
shows that there exist missing flux effects for
sources with $\simeq 2\farcs0-3\farcs0$ at Band 6/7
for the shortest-baseline configuration that we use.
The source sizes of $\simeq 2\farcs0-3\farcs0$ are
significantly larger than our ALMA sources whose angular sizes 
are $\lesssim 0\farcs5$, and 
the missing fluxes should be negligible.

\begin{longtable*}{cccccccc}
\tablecolumns{8}
\tabletypesize{\scriptsize} 
\setlength{\tabcolsep}{0.03in}
\tablewidth{0pt} 
\tablecaption{Source Catalog \label{tab:our_catalog}}
\tablehead{
\colhead{ID}
    & \colhead{Map ID}
    & \colhead{R.A.(J2000)}    
    & \colhead{Decl.(J2000)}    
    & \colhead{$S^{\rm obs}_{\lambda_{\rm obs}}$}    
    & \colhead{$S^{\rm corr}_{1.2{\rm mm}}$}    
    & \colhead{SNR}    
    & \colhead{$f_{\rm sp}$}    
\\
\colhead{ }
&\colhead{ }
    & \colhead{ }    
    & \colhead{ }
    & \colhead{(mJy)}        
    & \colhead{(mJy)}        
    & \colhead{ }
    & \colhead{ }
\\
\colhead{(1)}
& \colhead{(2)}
    & \colhead{ }
    & \colhead{ }
    & \colhead{(3)}
    & \colhead{(4)}    
    & \colhead{(5)}    
    & \colhead{(6)}
}
1 & 1 & 337.049805 & -35.169643 & 0.06 $\pm$ 0.02 & 0.07 $\pm$ 0.02 & 3.6 & 0.57 \\ 
2 & 1 & 337.050232 & -35.164864 & 0.41 $\pm$ 0.01 & 0.68 $\pm$ 0.07 & 39.3 & 0.00 \\ 
3 & 1 & 337.050293 & -35.170044 & 0.07 $\pm$ 0.02 & 0.09 $\pm$ 0.02 & 4.0 & 0.34 \\ 
4 & 1 & 337.053314 & -35.165119 & 0.06 $\pm$ 0.01 & 0.12 $\pm$ 0.02 & 5.7 & 0.00 \\ 
5 & 1 & 337.054413 & -35.167576 & 0.05 $\pm$ 0.01 & 0.07 $\pm$ 0.02 & 4.1 & 0.27 \\ 
6 & 1 & 337.054718 & -35.166332 & 0.05 $\pm$ 0.01 & 0.06 $\pm$ 0.02 & 3.5 & 0.66 \\ 
7 & 2 & 148.887314 & 18.150942 & 0.05 $\pm$ 0.01 & 0.04 $\pm$ 0.02 & 3.6 & 0.63 \\ 
8 & 2 & 148.891403 & 18.151211 & 0.06 $\pm$ 0.02 & 0.06 $\pm$ 0.02 & 3.4 & 0.71 \\ 
9 & 3 & 34.446007 & -5.008159 & 0.05 $\pm$ 0.01 & 0.05 $\pm$ 0.02 & 3.5 & 0.70 \\ 
10 & 3 & 34.446648 & -5.007715 & 0.05 $\pm$ 0.01 & 0.04 $\pm$ 0.02 & 3.4 & 0.70 \\ 
11\tablenotemark{$\dagger$} & 4 & 34.437332 & -5.492185 & 0.06 $\pm$ 0.02 & 0.04 $\pm$ 0.01 & 3.7 & 0.53 \\ 
12\tablenotemark{$\dagger$} & 4 & 34.437649 & -5.494778 & 0.06 $\pm$ 0.02 & 0.03 $\pm$ 0.01 & 3.5 & 0.67 \\ 
13\tablenotemark{$\dagger$} & 4 & 34.437805 & -5.494339 & 0.06 $\pm$ 0.01 & 0.03 $\pm$ 0.01 & 3.8 & 0.48 \\ 
14\tablenotemark{$\dagger$} & 4 & 34.439056 & -5.494986 & 0.06 $\pm$ 0.02 & 0.03 $\pm$ 0.01 & 3.6 & 0.60 \\ 
15\tablenotemark{$\dagger$} & 4 & 34.440075 & -5.494945 & 0.07 $\pm$ 0.02 & 0.04 $\pm$ 0.01 & 3.6 & 0.63 \\ 
16 & 5 & 121.593948 & -41.376675 & 0.10 $\pm$ 0.03 & 0.08 $\pm$ 0.04 & 3.6 & 0.60 \\ 
17 & 5 & 121.596870 & -41.375671 & 0.06 $\pm$ 0.02 & 0.08 $\pm$ 0.02 & 3.6 & 0.63 \\ 
18 & 5 & 121.599480 & -41.373417 & 0.08 $\pm$ 0.02 & 0.10 $\pm$ 0.03 & 3.7 & 0.55 \\ 
19 & 5 & 121.599564 & -41.375568 & 0.07 $\pm$ 0.02 & 0.08 $\pm$ 0.03 & 3.6 & 0.61 \\ 
20 & 5 & 121.600471 & -41.375629 & 0.08 $\pm$ 0.02 & 0.09 $\pm$ 0.03 & 3.6 & 0.60 \\ 
21 & 7 & 336.940674 & -35.119549 & 0.11 $\pm$ 0.03 & 0.13 $\pm$ 0.04 & 3.8 & 0.48 \\ 
22 & 8 & 201.001526 & 27.413574 & 0.10 $\pm$ 0.03 & 0.11 $\pm$ 0.04 & 3.5 & 0.70 \\ 
23 & 8 & 201.001526 & 27.413574 & 0.10 $\pm$ 0.03 & 0.11 $\pm$ 0.04 & 3.5 & 0.70 \\ 
24 & 8 & 201.001526 & 27.413574 & 0.10 $\pm$ 0.03 & 0.11 $\pm$ 0.04 & 3.5 & 0.70 \\ 
25 & 9 & 325.760986 & 6.904217 & 0.13 $\pm$ 0.04 & 0.17 $\pm$ 0.05 & 3.5 & 0.68 \\ 
26 & 9 & 325.762177 & 6.902022 & 0.14 $\pm$ 0.04 & 0.17 $\pm$ 0.05 & 3.6 & 0.62 \\ 
27 & 9 & 325.763336 & 6.905892 & 0.08 $\pm$ 0.02 & 0.10 $\pm$ 0.03 & 3.4 & 0.72 \\ 
28 & 9 & 325.767029 & 6.902850 & 0.15 $\pm$ 0.04 & 0.16 $\pm$ 0.05 & 3.6 & 0.60 \\ 
29 & 10 & 352.281616 & -3.033988 & 0.13 $\pm$ 0.04 & 0.12 $\pm$ 0.04 & 3.4 & 0.71 \\ 
30 & 10 & 352.284851 & -3.031080 & 0.16 $\pm$ 0.03 & 0.16 $\pm$ 0.03 & 5.9 & 0.00 \\ 
31 & 10 & 352.285248 & -3.030138 & 0.17 $\pm$ 0.04 & 0.17 $\pm$ 0.04 & 4.6 & 0.08 \\ 
32 & 10 & 352.285309 & -3.035150 & 0.11 $\pm$ 0.03 & 0.10 $\pm$ 0.03 & 3.8 & 0.47 \\ 
33 & 11 & 200.929321 & 27.340736 & 0.15 $\pm$ 0.04 & 0.13 $\pm$ 0.04 & 3.9 & 0.40 \\ 
34 & 12 & 150.547440 & 1.910628 & 0.13 $\pm$ 0.04 & 0.13 $\pm$ 0.05 & 3.7 & 0.55 \\ 
35 & 13 & 351.831116 & 6.267834 & 0.13 $\pm$ 0.03 & 0.12 $\pm$ 0.03 & 4.2 & 0.22 \\ 
36 & 13 & 351.833496 & 6.267574 & 0.16 $\pm$ 0.04 & 0.15 $\pm$ 0.04 & 3.7 & 0.53 \\ 
37 & 14 & 13.760888 & 1.771644 & 0.10 $\pm$ 0.03 & 0.08 $\pm$ 0.02 & 4.0 & 0.35 \\ 
38 & 14 & 13.761566 & 1.768838 & 0.16 $\pm$ 0.04 & 0.12 $\pm$ 0.04 & 3.7 & 0.52 \\ 
39 & 14 & 13.763390 & 1.770106 & 0.11 $\pm$ 0.03 & 0.08 $\pm$ 0.03 & 3.7 & 0.52 \\ 
40 & 14 & 13.764014 & 1.770272 & 0.13 $\pm$ 0.03 & 0.09 $\pm$ 0.03 & 3.7 & 0.50 \\ 
41 & 15 & 149.775772 & 1.736944 & 0.16 $\pm$ 0.03 & 0.18 $\pm$ 0.04 & 5.1 & 0.02 \\ 
42 & 15 & 149.775955 & 1.738510 & 0.15 $\pm$ 0.04 & 0.17 $\pm$ 0.05 & 3.9 & 0.41 \\ 
43 & 15 & 149.776794 & 1.736661 & 0.09 $\pm$ 0.03 & 0.07 $\pm$ 0.03 & 3.6 & 0.63 \\ 
44 & 18 & 337.254547 & 14.950516 & 0.17 $\pm$ 0.05 & 0.14 $\pm$ 0.04 & 3.8 & 0.48 \\ 
45 & 18 & 337.256073 & 14.954236 & 0.19 $\pm$ 0.03 & 0.21 $\pm$ 0.03 & 6.1 & 0.00 \\ 
46 & 19 & 149.722565 & 2.606706 & 0.11 $\pm$ 0.03 & 0.10 $\pm$ 0.04 & 3.4 & 0.70 \\ 
47 & 19 & 149.724045 & 2.602218 & 0.21 $\pm$ 0.05 & 0.28 $\pm$ 0.07 & 4.1 & 0.26 \\ 
48 & 20 & 149.792572 & 2.103762 & 0.12 $\pm$ 0.03 & 0.10 $\pm$ 0.04 & 4.0 & 0.30 \\ 
49 & 21 & 32.553406 & -4.939113 & 0.12 $\pm$ 0.03 & 0.11 $\pm$ 0.03 & 3.9 & 0.37 \\ 
50 & 21 & 32.553787 & -4.939454 & 0.11 $\pm$ 0.03 & 0.15 $\pm$ 0.03 & 3.7 & 0.56 \\ 
51 & 21 & 32.555397 & -4.937203 & 0.13 $\pm$ 0.03 & 0.13 $\pm$ 0.04 & 3.7 & 0.54 \\ 
52\tablenotemark{$\dagger$} & 22 & 150.127991 & 2.288839 & 0.13 $\pm$ 0.04 & 0.08 $\pm$ 0.03 & 3.4 & 0.72 \\ 
53 & 23 & 359.017120 & 19.607481 & 0.10 $\pm$ 0.03 & 0.08 $\pm$ 0.03 & 3.4 & 0.71 \\ 
54 & 24 & 150.406784 & 2.034714 & 0.14 $\pm$ 0.04 & 0.12 $\pm$ 0.04 & 3.5 & 0.64 \\ 
55 & 25 & 150.393936 & 1.998264 & 0.15 $\pm$ 0.04 & 0.16 $\pm$ 0.05 & 3.7 & 0.56 \\ 
56 & 26 & 34.100346 & -5.154171 & 0.15 $\pm$ 0.04 & 0.14 $\pm$ 0.05 & 3.5 & 0.64 \\ 
57 & 26 & 34.100658 & -5.156616 & 0.16 $\pm$ 0.05 & 0.17 $\pm$ 0.06 & 3.5 & 0.66 \\ 
58 & 26 & 34.102978 & -5.155593 & 0.15 $\pm$ 0.04 & 0.15 $\pm$ 0.05 & 3.4 & 0.70 \\ 
59 & 27 & 34.034069 & -5.104988 & 0.17 $\pm$ 0.04 & 0.18 $\pm$ 0.06 & 3.8 & 0.45 \\ 
60 & 28 & 34.472076 & -4.711596 & 0.26 $\pm$ 0.07 & 0.27 $\pm$ 0.09 & 3.5 & 0.65 \\ 
61 & 28 & 34.473747 & -4.713997 & 0.16 $\pm$ 0.04 & 0.20 $\pm$ 0.06 & 3.6 & 0.62 \\ 
62 & 29 & 34.392979 & -5.043759 & 0.17 $\pm$ 0.04 & 0.19 $\pm$ 0.06 & 3.7 & 0.51 \\ 
63 & 29 & 34.396526 & -5.042582 & 0.21 $\pm$ 0.06 & 0.23 $\pm$ 0.07 & 3.6 & 0.59 \\ 
64 & 29 & 34.396526 & -5.046781 & 0.25 $\pm$ 0.07 & 0.26 $\pm$ 0.09 & 3.4 & 0.72 \\ 
65 & 31 & 34.612003 & -4.588018 & 0.25 $\pm$ 0.07 & 0.28 $\pm$ 0.08 & 3.5 & 0.69 \\ 
66 & 31 & 34.612686 & -4.583063 & 0.36 $\pm$ 0.07 & 0.43 $\pm$ 0.09 & 4.9 & 0.04 \\ 
67 & 31 & 34.613739 & -4.588505 & 0.31 $\pm$ 0.09 & 0.31 $\pm$ 0.10 & 3.5 & 0.68 \\ 
68 & 31 & 34.613930 & -4.585264 & 0.19 $\pm$ 0.06 & 0.18 $\pm$ 0.06 & 3.5 & 0.70 \\ 
69 & 33 & 34.272991 & -4.859830 & 0.29 $\pm$ 0.08 & 0.34 $\pm$ 0.11 & 3.5 & 0.68 \\ 
70 & 33 & 34.274231 & -4.860201 & 0.32 $\pm$ 0.09 & 0.43 $\pm$ 0.12 & 3.6 & 0.59 \\ 
71 & 33 & 34.274761 & -4.859237 & 0.26 $\pm$ 0.07 & 0.32 $\pm$ 0.09 & 3.8 & 0.46 \\ 
72 & 34 & 34.305080 & -5.067344 & 0.29 $\pm$ 0.07 & 0.39 $\pm$ 0.10 & 3.9 & 0.41 \\ 
73 & 34 & 34.305538 & -5.069685 & 0.23 $\pm$ 0.06 & 0.33 $\pm$ 0.09 & 3.6 & 0.61 \\ 
74 & 34 & 34.305553 & -5.067293 & 0.29 $\pm$ 0.07 & 0.22 $\pm$ 0.09 & 4.2 & 0.19 \\ 
75 & 34 & 34.307556 & -5.071889 & 0.33 $\pm$ 0.09 & 0.48 $\pm$ 0.12 & 3.6 & 0.61 \\ 
76 & 34 & 34.309822 & -5.069085 & 0.29 $\pm$ 0.08 & 0.23 $\pm$ 0.11 & 3.5 & 0.65 \\ 
77\tablenotemark{$\dagger$} & 35 & 230.738800 & -0.125928 & 0.35 $\pm$ 0.10 & 0.23 $\pm$ 0.07 & 3.5 & 0.69 \\ 
78\tablenotemark{$\dagger$} & 35 & 230.739777 & -0.125576 & 0.28 $\pm$ 0.08 & 0.20 $\pm$ 0.06 & 3.6 & 0.59 \\ 
79\tablenotemark{$\dagger$} & 35 & 230.743118 & -0.127994 & 0.28 $\pm$ 0.08 & 0.18 $\pm$ 0.06 & 3.5 & 0.66 \\ 
80 & 36 & 34.327038 & -5.131583 & 0.21 $\pm$ 0.06 & 0.27 $\pm$ 0.09 & 3.8 & 0.48 \\ 
81 & 37 & 34.401154 & -5.152167 & 0.33 $\pm$ 0.10 & 0.37 $\pm$ 0.15 & 3.4 & 0.72 \\ 
82 & 37 & 34.401531 & -5.154098 & 0.30 $\pm$ 0.08 & 0.34 $\pm$ 0.13 & 3.7 & 0.55 \\ 
83 & 37 & 34.403275 & -5.149917 & 0.36 $\pm$ 0.11 & 0.61 $\pm$ 0.18 & 3.4 & 0.71 \\ 
84\tablenotemark{$\dagger$} & 38 & 231.036255 & -0.176025 & 0.39 $\pm$ 0.11 & 0.22 $\pm$ 0.08 & 3.5 & 0.67 \\ 
85\tablenotemark{$\dagger$} & 38 & 231.037949 & -0.178142 & 0.22 $\pm$ 0.06 & 0.15 $\pm$ 0.04 & 3.7 & 0.54 \\ 
86 & 40 & 34.414261 & -5.201880 & 0.34 $\pm$ 0.10 & 0.56 $\pm$ 0.16 & 3.5 & 0.65 \\ 
87 & 40 & 34.419544 & -5.199385 & 0.31 $\pm$ 0.08 & 0.52 $\pm$ 0.14 & 3.8 & 0.46 \\ 
88 & 41 & 149.590530 & 2.807018 & 0.28 $\pm$ 0.07 & 0.23 $\pm$ 0.05 & 4.2 & 0.21 \\ 
89 & 41 & 149.591949 & 2.805352 & 0.26 $\pm$ 0.07 & 0.21 $\pm$ 0.05 & 4.0 & 0.31 \\ 
90 & 42 & 150.688171 & 2.681118 & 0.37 $\pm$ 0.07 & 0.26 $\pm$ 0.06 & 5.4 & 0.01 \\ 
91 & 45 & 34.194702 & -5.056259 & 0.43 $\pm$ 0.12 & 0.39 $\pm$ 0.15 & 3.5 & 0.71 \\ 
92 & 45 & 34.197121 & -5.059778 & 0.31 $\pm$ 0.08 & 0.37 $\pm$ 0.10 & 3.7 & 0.55 \\ 
93 & 46 & 150.409714 & 2.358208 & 0.30 $\pm$ 0.09 & 0.32 $\pm$ 0.09 & 3.5 & 0.71 \\ 
94 & 47 & 149.770050 & 1.804996 & 0.32 $\pm$ 0.09 & 0.31 $\pm$ 0.10 & 3.4 & 0.72 \\ 
95 & 48 & 34.411190 & -4.745051 & 0.30 $\pm$ 0.08 & 0.14 $\pm$ 0.09 & 3.8 & 0.44 \\ 
96 & 48 & 34.411701 & -4.745893 & 0.31 $\pm$ 0.08 & 0.34 $\pm$ 0.09 & 4.0 & 0.31 \\ 
97 & 49 & 150.600494 & 2.444365 & 0.47 $\pm$ 0.13 & 0.53 $\pm$ 0.14 & 3.6 & 0.59 \\ 
98 & 51 & 34.440269 & -4.912953 & 0.39 $\pm$ 0.11 & 0.40 $\pm$ 0.12 & 3.6 & 0.60 \\ 
99 & 51 & 34.441357 & -4.912045 & 0.32 $\pm$ 0.08 & 0.29 $\pm$ 0.09 & 3.9 & 0.43 \\ 
100 & 51 & 34.441635 & -4.913539 & 0.46 $\pm$ 0.13 & 0.45 $\pm$ 0.14 & 3.6 & 0.62 \\ 
101 & 51 & 34.442795 & -4.911066 & 0.53 $\pm$ 0.08 & 0.56 $\pm$ 0.11 & 6.2 & 0.00 \\ 
102 & 51 & 34.443920 & -4.909602 & 0.41 $\pm$ 0.12 & 0.35 $\pm$ 0.13 & 3.5 & 0.65 \\ 
103 & 52 & 150.151459 & 1.751314 & 0.49 $\pm$ 0.14 & 0.50 $\pm$ 0.13 & 3.6 & 0.64 \\ 
104 & 54 & 34.759140 & -4.832398 & 0.32 $\pm$ 0.09 & 0.37 $\pm$ 0.11 & 3.5 & 0.71 \\ 
105 & 55 & 150.022736 & 2.027032 & 0.53 $\pm$ 0.08 & 0.46 $\pm$ 0.08 & 6.8 & 0.00 \\ 
106 & 57 & 34.141144 & -5.232085 & 0.47 $\pm$ 0.13 & 0.42 $\pm$ 0.15 & 3.6 & 0.60 \\ 
107 & 58 & 34.303791 & -5.161064 & 0.47 $\pm$ 0.14 & 0.58 $\pm$ 0.16 & 3.4 & 0.73 \\ 
108 & 59 & 150.201645 & 2.021740 & 0.57 $\pm$ 0.15 & 0.72 $\pm$ 0.22 & 3.7 & 0.53 \\ 
109 & 60 & 34.250214 & -4.804325 & 0.34 $\pm$ 0.09 & 0.34 $\pm$ 0.10 & 3.9 & 0.39 \\ 
110 & 60 & 34.250927 & -4.802006 & 0.37 $\pm$ 0.11 & 0.39 $\pm$ 0.13 & 3.5 & 0.68 \\ 
111 & 61 & 34.744495 & -4.855921 & 0.35 $\pm$ 0.10 & 0.27 $\pm$ 0.12 & 3.5 & 0.70 \\ 
112 & 61 & 34.747295 & -4.858103 & 0.37 $\pm$ 0.10 & 0.26 $\pm$ 0.11 & 3.8 & 0.45 \\ 
113 & 62 & 150.757401 & 1.700786 & 0.50 $\pm$ 0.14 & 0.57 $\pm$ 0.21 & 3.5 & 0.70 \\ 
114 & 62 & 150.758331 & 1.703298 & 0.42 $\pm$ 0.12 & 0.52 $\pm$ 0.17 & 3.6 & 0.60 \\ 
115 & 62 & 150.760971 & 1.703856 & 1.01 $\pm$ 0.10 & 1.17 $\pm$ 0.18 & 9.7 & 0.00 \\ 
116 & 62 & 150.762024 & 1.698676 & 0.57 $\pm$ 0.16 & 0.85 $\pm$ 0.24 & 3.5 & 0.69 \\ 
117 & 62 & 150.762161 & 1.701653 & 0.34 $\pm$ 0.10 & 0.37 $\pm$ 0.14 & 3.5 & 0.68 \\ 
118 & 62 & 150.762741 & 1.700690 & 0.43 $\pm$ 0.12 & 0.61 $\pm$ 0.17 & 3.6 & 0.62 \\ 
119 & 62 & 150.763138 & 1.702965 & 0.47 $\pm$ 0.13 & 0.70 $\pm$ 0.19 & 3.6 & 0.62 \\ 
120 & 64 & 34.587593 & -5.318766 & 0.32 $\pm$ 0.09 & 0.21 $\pm$ 0.10 & 3.6 & 0.59 \\ 
121 & 64 & 34.589180 & -5.315989 & 0.61 $\pm$ 0.17 & 0.69 $\pm$ 0.21 & 3.5 & 0.65 \\ 
122 & 65 & 34.387825 & -5.220457 & 0.41 $\pm$ 0.11 & 0.42 $\pm$ 0.13 & 3.7 & 0.51 \\ 
123 & 67 & 197.860764 & -1.337225 & 0.25 $\pm$ 0.06 & 0.03\tablenotemark{$\ddag$} $\pm$ 0.01 & 4.3 & 0.53 \\ 
124 & 67 & 197.862061 & -1.342226 & 0.24 $\pm$ 0.06 & 0.00\tablenotemark{$\ddag$} $\pm$ 0.00 & 4.3 & 0.54 \\ 
125 & 67 & 197.864731 & -1.343368 & 0.27 $\pm$ 0.06 & 0.03\tablenotemark{$\ddag$} $\pm$ 0.01 & 4.7 & 0.32 \\ 
126 & 67 & 197.869086 & -1.337143 & 0.31 $\pm$ 0.08 & 0.08\tablenotemark{$\ddag$} $\pm$ 0.03 & 4.1 & 0.60 \\ 
127 & 67 & 197.871590 & -1.345783 & 0.30 $\pm$ 0.06 & 0.02\tablenotemark{$\ddag$} $\pm$ 0.01 & 5.1 & 0.20 \\ 
128 & 67 & 197.874725 & -1.326138 & 0.22 $\pm$ 0.05 & 0.02\tablenotemark{$\ddag$} $\pm$ 0.01 & 4.3 & 0.51 \\ 
129 & 67 & 197.881638 & -1.323127 & 0.26 $\pm$ 0.05 & 0.05\tablenotemark{$\ddag$} $\pm$ 0.02 & 4.8 & 0.29 \\ 
130 & 67 & 197.882812 & -1.356609 & 0.28 $\pm$ 0.07 & 0.05\tablenotemark{$\ddag$} $\pm$ 0.02 & 4.1 & 0.64 \\ 
131 & 67 & 197.883881 & -1.355241 & 0.30 $\pm$ 0.08 & 0.06\tablenotemark{$\ddag$} $\pm$ 0.03 & 4.0 & 0.65 \\ 
132 & 67 & 197.884125 & -1.334885 & 0.30 $\pm$ 0.07 & 0.13\tablenotemark{$\ddag$} $\pm$ 0.05 & 4.2 & 0.58 \\ 
133 & 67 & 197.886185 & -1.328756 & 0.24 $\pm$ 0.05 & 0.04\tablenotemark{$\ddag$} $\pm$ 0.01 & 5.1 & 0.19 \\ 
\end{longtable*}
\tablecomments{
\footnotesize{
(1): Source ID.
(2): Map ID that corresponds to the one in Table 2.
(3): 
Peak flux density of the SExtractor measurement (Section \ref{sec:source_extraction}) 
with the primary beam correction at the observed wavelength. 
(4): 
The best-estimate source flux density at 1.2 mm. 
The source flux is estimated with the 2D Gaussian-fitting routine of {\it imfit} in CASA (Section \ref{sec:flux_measure})
with the primary beam correction and the flux scaling to 1.2 mm (Section \ref{sec:data_analysis}). 
For the sources found in the cluster data, the lensing magnification corrections are also applied (Section \ref{sec:mass_model}). 
The lensing magnification factors are estimated at the ALMA flux peak positions. 
}
The error bar includes the random noise, 10\% of ALMA system's flux measurement uncertainty, 
and the lensing magnification uncertainty (Section \ref{sec:derivation_nc}).
(5): Signal-to-noise ratio of the peak flux density.  
(6): Spurious source rate that is defined by eq. \ref{eq:sp_rate}.
$^{\dagger}$ Source identified in the Band 7 map. Unless otherwise specified,
the source is found in the Band 6 map.
$^{\ddag}$ The lensing magnification is corrected. The ID 124 source has
$0.0020\pm0.0010$ mJy for the best-estimate source flux density at 1.2 mm with the primary-beam and lensing-magnification corrections.
}

\subsection{Mass Model}
\label{sec:mass_model}

We construct a mass model for the cluster data (C) of A1689 at $z = 0.183$. 
We make the mass model with the parametric gravitational lensing package 
GLAFIC \citep{oguri2010} in the same manner as \cite{ishigaki2015}.
Our mass model consists of three types of mass distributions: cluster-scale halos, 
cluster member galaxy halos, and external perturbation. 
For evaluating the mass distributions, we make a galaxy catalog of A1689, 
conducting source extractions in the optical images of $g_{475},\ r_{625},$ and $i_{775}$ bands 
taken with $Hubble\ Space\ Telescope$ (HST). 
The cluster-scale halos are estimated with three brightest member galaxies in the core of the cluster.
The cluster member galaxies are selected 
by the color criteria of 
\begin{eqnarray}
&r_{625}&\ <\ 24, \nonumber \\
&g_{475}-r_{625}&\ <\ -\frac{1}{18}(r_{625}-24)+1.3, \nonumber \\ 
&g_{475}-r_{625}&\ >\ -\frac{1}{18}(r_{625}-24)+0.7,
\end{eqnarray}
where the source extraction and photometry are carried out with SExtractor.
The external perturbation is calculated with the theoretical model 
under the assumption that the perturbation is weak \citep[e.g.,][]{kochanek1991}.

Using the positions of the multiple images presented in the literature \citep{coe2010,diego2015}, 
we optimize free parameters of the mass profiles based on 
the standard $\chi^{2}$ minimization to determine the best-fit mass model. 
Figure \ref{fig:a1689} presents the best-fit mass model.
The best-fit mass model achieves that all of the offsets between the model 
and observed positions of multiple images are within $1\farcs0$ in the image plane.
We then calculate magnification factors $\mu$ at the positions of our sources.
For the $\mu$ estimates, we assume our source redshift of $z = 2.5$
that is the same as the one used in the flux scaling shown in Section \ref{sec:data_analysis}. 
This assumption does not change our statistical results,
because most of
mm sources in the target fields at the high-galactic latitude 
should reside at $z>1$ where the $\mu$ values 
do not depend much on a redshift. 
Here we calculate the magnification factors for sources at $z=7.5$ that correspond to the redshift of A1689-zD1,
and evaluate the differences between the magnification factors for sources at $z=2.5$  ($\mu_{2.5}$) and 
$z=7.5$ ($\mu_{7.5}$).
We find that the average difference, $\Delta\mu\equiv(\mu_{7.5}-\mu_{2.5})/\mu_{2.5}$, is $0.36\pm0.25$ for 
our 11 sources in A1689. Although the difference is not large, 
we include the lensing magnification difference of $0.36$
into the errors of the intrinsic flux estimates for our 11 sources in A1689. Accordingly,
we propagate these uncertainties to our major results such as number counts
shown in Section \ref{sec:number_counts}.

\subsection{Survey Area}
\label{sec:ef_area}

\begin{figure}
\begin{center}
 \includegraphics[trim=0.0cm 0.2cm 1.0cm 0cm, clip, angle=0,width=0.5\textwidth]{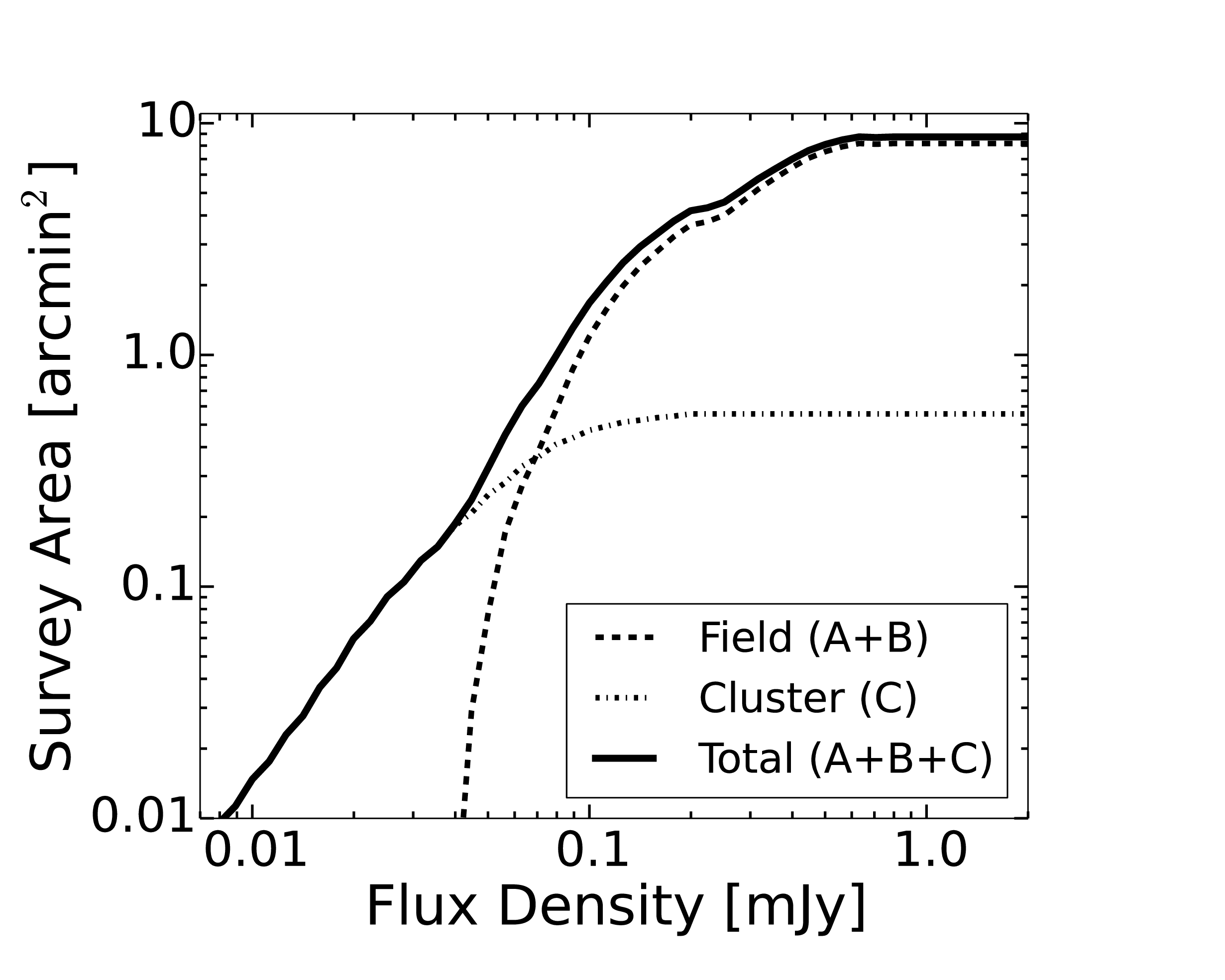}
 \caption[]
{Survey areas as a function of intrinsic flux densities. The dashed, dotted, and solid lines represent the survey areas of the field (A and B), cluster (C), and total (A, B, and C) datasets, respectively. 
\label{fig:sum_ef}}

\end{center}
\end{figure}

We estimate survey areas of our data of A, B, and C,
and 
present these estimates in Figure \ref{fig:sum_ef}.
The survey areas 
are defined by the high sensitivity regions that are detailed in
Section \ref{sec:source_extraction}.
Because the sensitivities of our ALMA maps are not spatially uniform, 
the survey areas depend on the flux densities.

In the field data of A and B, the sensitivity of the primary beam decreases 
with increasing radius from a map center.
In other words, the detectable intrinsic flux densities (corrected for the primary beam attenuation)
increase from the center to the edge of the data.
First, we find the radius where sources with the intrinsic flux densities 
can be detected at our selection limit SNR.
Then, we calculate areas with the radius
given in each map, and sum up these areas
to obtain the total survey areas.

In the cluster data of C, 
the spatial distribution of 
the sensitivity is not expressed by a simple function of radius, 
because the cluster data are taken by mosaic mapping.
Moreover, there are cluster-lensing magnification effects
that allow us to detect intrinsically faint sources.
We make an effective magnification map, multiplying, at each position, 
the sensitivity and the magnification factor estimated with the mass model of Section \ref{sec:mass_model}.
We then calculate the survey area of the C data where a source with an intrinsic flux density 
can be observed above the selection limit.

\section{Number Counts and EBL}
\label{sec:number_counts}

\subsection{Number Counts at 1.2 mm}
\label{sec:derivation_nc}  

We derive differential 
number counts at 1.2 mm.
For the number counts, 
we use the 122 sources identified in the ALMA Band 6 maps 
covering the 1.2 mm band, because there is a possibility
that the 11 sources in the Band 7 would include unknown systematics 
in the flux scaling to 1.2 mm fluxes (Section 3).
To derive the number counts, we basically follow
the methods 
used in the previous ALMA studies \citep{hatsukade2013,ono2014,carniani2015}. 
A contribution from an identified source with an intrinsic flux density of $S$ to the number counts, $\xi$, is determined by
\begin{eqnarray}
\label{eq:ef_num}
\xi(S) = \frac{1-f_{\rm sp}(S)}{C(S)A_{\rm eff}(S)},
\end{eqnarray}
where $C$ is the completeness, and $A_{\rm eff}$ is the survey area. 
For C, $A_{\rm eff}$, and $f_{\rm sp}$ (eq.\ref{eq:sp_rate}),  we use the values derived in Section \ref{sec:data_analysis}.
Then, we calculate a sum of the contributions for each flux bin,
\begin{eqnarray}
n(S) = \frac{\Sigma\xi(S)}{\Delta \log S},
\end{eqnarray}
where $\Delta \log S$ is the scaling factor of $0.25$ for
the 1-dex width logarithmic differential number counts. 
We estimate the errors that include both Poisson statistical errors of the source numbers
and flux uncertainties of the identified sources. 
Because the numbers of our sources are small, 
we use the Poisson uncertainty values presented in \cite{gehrels1986}
that are applicable for the small number statistics. 
The flux uncertainties are composed of the random noise 
and the ALMA system's flux measurement uncertainty whose typical value
is $\sim$10\% (Section \ref{sec:data_reduction}).
Moreover, for our 11 sources in the A1689 data,
the intrinsic flux estimates include
the lensing magnification uncertainties 
(see Section \ref{sec:mass_model}).
To evaluate these flux uncertainties, we perform Monte-Carlo simulations. 
We make a mock catalog of the faint ALMA sources whose flux densities 
follow the Gaussian probability distributions whose standard deviations
are given by the combination of 
the random noise and the system (+ the lensing magnification) uncertainties.
We obtain the number counts for each mock catalog 
in the same manner as those for our real sources. 
We repeat this processes 1000 times, and calculate 
the standard deviation of the number counts per flux bin caused by the flux uncertainties. 
Combining the standard deviation of the flux uncertainties with the Poisson errors,
we finally obtain the $1\sigma$ uncertainties of the number counts.

\begin{figure*}
\begin{center}
\includegraphics[trim=0.0cm 0cm 0.0cm 0cm, clip, angle=0,width=0.9\textwidth]{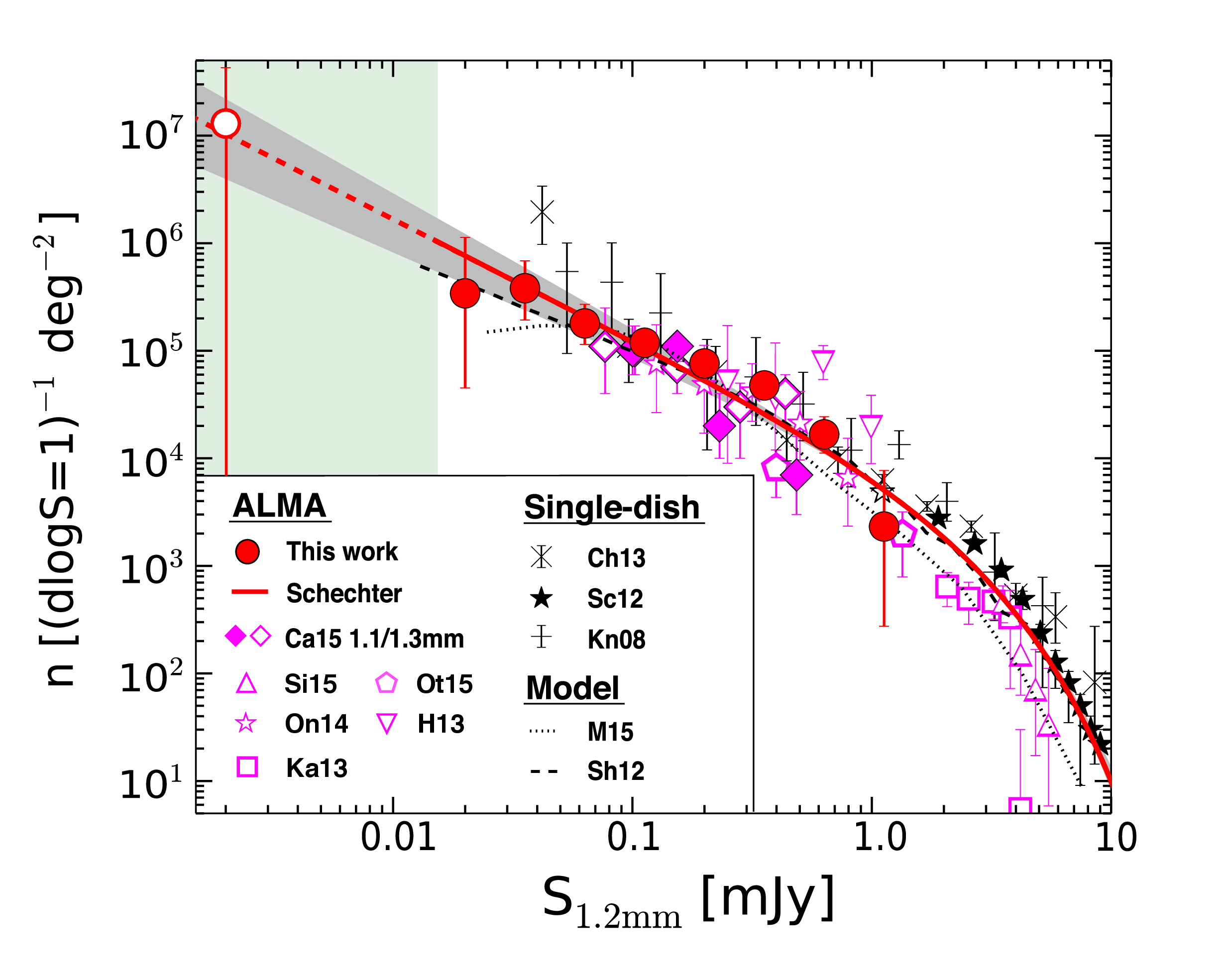}
\caption[]
{
Differential number counts at 1.2 mm.
The red filled circles
are our number counts derived from our faint ALMA sources.
The red curve and the gray region denote the best-fit Schechter function and the associated
$1\sigma$ error. 
In the Schechter function fitting, we use our number counts 
and the previous measurements shown with 
the black filled stars (Sc12; \citealt{scott2012}) and the magenta filled diamonds (Ca15 1.1 mm; \citealt{carniani2015}).
Note that the 1.1 mm results of Ca15 are obtained from ALMA sources, none of which are used in our number count measurements. 
Because the rest of the existing ALMA studies include 
some ALMA data covered by our study,
for our Schechter function fitting we do not include the previous ALMA measurements shown with 
the magenta open squares (Ka13; \citealt{karim2013}), 
open inverse triangles (H13; \citealt{hatsukade2013}), 
open stars (O14; \citealt{ono2014}), 
open diamonds (Ca15 1.3 mm; \citealt{carniani2015}), 
open triangles (Si15; \citealt{simpson2015b}), 
and open pentagons (Ot15; \citealt{oteo2015}).
Note that the faintest data point of our study
(red open circle) is also removed from the sample of our Schether function fitting,
due to the possible large systematic uncertainties.
The pale-green region presents a flux range
where no reliable number-count data bin exists (see text). 
We do not use the faintest data point of Sc12 (black open star) 
for our fitting, because this data point has unrealisticly small errors.
The dashed and dotted curves show the model predictions for number counts based on cosmological hydrodynamic 
simulations with {\sc GADGET-3} (Sh12; \citealt{shimizu2012}) and semi-analytic models (M15; \citealt{makiya2015}), respectively. 
These model predictions are roughly consistent with our results.
For the previous measurements of the number counts at 1.1 mm and 1.3 mm
different from our 1.2 mm band,
we scale the flux densities with the methods shown in Section \ref{sec:data_analysis}.
Since the flux scaling factors for $<1$ mm wavelengths would contain relatively large systematic uncertainties, 
we do not use the $<1$ mm band data, for our Schechter function fitting, that are presented with the open triangles (Kn08; \citealt{knudsen2008}) and 
open hexagons (Ch13; \citealt{chen2013}).
\label{fig:nc_fig}}
\end{center}
\end{figure*}

The differential number counts and the associated $1 \sigma$ uncertainties are shown in Figure \ref{fig:nc_fig}
and Table \ref{tab:nc_tab}. With the technique same as those deriving the differential number counts,
we estimate the cumulative number counts that are summarized in 
Table \ref{tab:nc_tab}.

In Figure \ref{fig:nc_fig}, 
we find that the faintest data point
at $\sim0.002$ mJy (red open circle) is composed of a source
with a very high magnification factor of $\mu>100$.
The highly magnified sources would potentially include 
large systematic uncertainties originated from the mass modeling. 
Moreover, the faintest data point has the large error bar
that does not provide important constraints on the number count
measurements. Thus, we do not use the faintest data point in the following
discussion. 
Except for this very high magnification factor object, the faintest and brightest
sources in our sample have intrinsic fluxes of 0.018 and 1.2 mJy, respectively. Thus,
our study covers the flux density range of  $\sim0.02-1$ mJy.

\begin{deluxetable}{ccc} 
\tablecolumns{3} 
\tablewidth{0pt} 
\tablecaption{Differential and Cumulative Number Counts at $1.2$ mm \label{tab:nc_tab}}
\tablehead{
\colhead{$S$}
    & \colhead {$\log(n)$}
    & \colhead {$N_{\rm diff}$} \\
 \colhead{(mJy)}
    & \colhead{([$\Delta \log S=1$]$^{-1}$deg$^{-2}$ )}
    &
    }
\startdata
 (0.002) &  $7.1_{-7.1}^{+0.5}$ &  1 \\
 0.020   &  $5.5_{-0.5}^{+0.3}$ &  2 \\
 0.036   &  $5.6_{-0.4}^{+0.3}$ & 6 \\
 0.063   &  $5.3_{-0.4}^{+0.2}$ &  15 \\
 0.112   &  $5.1_{-0.2}^{+0.2}$ &  25 \\
 0.200   &  $4.9_{-0.2}^{+0.1}$ &  27 \\
 0.356   &  $4.7_{-0.2}^{+0.1}$ &  29 \\
 0.633   &  $4.2_{-0.2}^{+0.2}$ &  15 \\
 1.124   &  $3.4_{-1.3}^{+0.5}$ &  2 \\ \hline
\rule[0pt]{0pt}{5pt}
$S$     &  $\log(n(>S))$ & $N_{\rm cum}$ \\
(mJy)  &   (deg$^{-2}$) &                     \\ \hline
 (0.002) &  $6.6_{-1.1}^{+0.5}$ &  122 \\
 0.015 &    $5.5_{-0.4}^{+0.3}$ &   121 \\
 0.027 &    $5.3_{-0.3}^{+0.2}$ &  119 \\
 0.047 &    $5.0_{-0.2 }^{+0.2}$ & 113 \\
 0.084 &    $4.8_{-0.2}^{+0.1}$ &   98 \\
 0.150 &    $4.6_{-0.1}^{+0.1}$ &   73 \\
 0.267 &    $4.2_{-0.2}^{+0.2}$ &   46 \\
 0.474 &    $3.7_{-0.3}^{+0.2}$ &   17 \\
 0.843 &    $2.8_{-0.5}^{+0.6}$ &   2
\enddata 
\tablecomments{
The $1\sigma$ uncertainties are estimated from the combination of the number-count Poisson statistical errors 
and the flux uncertainties (see text).
The column of $N_{\rm diff}$ presents the numbers of our 1.2 mm sources in each flux bin, 
while the one of $N_{\rm cum}$ denotes the cumulative source numbers down to the flux densities.
For our analysis, we do not use the faintest bin data indicated with the parentheses (see text).
}
\end{deluxetable} 

\subsection{Comparison with Previous Number Count Measurements}
\label{sec:comp_prenc}

We compare our number counts to previous measurements at the $\sim 1.2$ mm band.
In the flux range close to our study,
there are two types of previous studies,
blank-field observations with ALMA and cluster observations for lensed sources 
with single-dish telescopes.
For the measurements of 1.1 and 1.3 mm bands that are different from 1.2 mm band,
we scale the flux densities with the methods described in Section \ref{sec:data_analysis}. 

\subsubsection{Blank-Field Observations}
\label{sec:field_nc}

To derive number counts,
\cite{hatsukade2013} and \cite{ono2014} use
20 ALMA maps in one blank field of SXDS
and 10 ALMA maps in 10 independent fields,
respectively.
\cite{carniani2015} obtain number counts 
with the 18 ALMA maps composed of 9 independent fields at 1.1 mm band
and the 9 ALMA maps in one blank field of Cosmic Evolution Survey (COSMOS; \citealt{scoville2007}) 
at 1.2 mm band.
After the flux density scaling, all of these previous number counts 
agree with our measurements within the $1\sigma$ uncertainties. 
Because the amount of our ALMA data is much larger than
the previous studies, the statistical and systematic uncertainties of 
the number counts measurements are significantly smaller than 
those of previous studies.

\subsubsection{Cluster Observations for Lensed Sources}
\label{sec:lense_nc}

Single-dish observations of SCUBA and SCUBA-2 
for lensed sources behind galaxy clusters
successfully reach the intrinsic flux limits 
comparable to the previous ALMA blank field observations. 
\cite{knudsen2008} conduct  SCUBA observations for 12 clusters, and derive the number counts 
at 850 $\mu$m with the sample of 15 gravitationally lensed faint sources 
with flux densities below the SCUBA's blank-field confusion limit ($\sim2$ mJy at 850 $\mu$m). 
\cite{chen2013} observe 2 clusters as well as 3 blank fields with SCUBA-2, and 
obtain the number counts similar to those of \cite{knudsen2008}.
The number counts from these previous studies are also shown in Figure \ref{fig:nc_fig}.
These previous results generally agree with our results within the $1\sigma$ errors.
Although the number counts of the previous studies are slightly higher than 
ours systematically at $\lesssim 0.1$ mJy, these small systematic differences 
would be explained by the uncertainties of the flux scaling from $850 \mu$m to 1.2 mm.

\subsection{Contributions to the EBL}
\label{sec:ebl}

In Section \ref{sec:derivation_nc}, we obtain the 1.2-mm number counts with our faint ALMA source.
Because our sample lacks sources brighter than
1.2 mJy, we use previous 
studies of bright sources in the literature 
to investigate the source contributions to the IR EBL. 
There are a number of studies that investigate the bright sources (Figure \ref{fig:nc_fig}), but
we use the number counts at $2-10$ mJy obtained from a survey for six blank fields 
with a single-dish telescope instrument of AzTEC \citep{scott2012}, which shows
reliable number counts at 1.1 mm that is close to 1.2 mm band .
Although there is another systematic survey for bright sources from the single-dish
observations with LABOCA \citep{weiss2009}, we do not use their data for our analysis,
due to their observation wavelength of $870\ \mu$m far from our 1.2 mm band.
It should be noted that the number counts of bright sources are still under
debate \citep[e.g.,][]{karim2013,chen2013,simpson2015b}, because the recent high-resolution
ALMA observations reveal multiple sources in a beam of single-dish observations
\citep{hodge2013}. 
Nevertheless, our results below do not depend on
the bright-source number counts, because
the contribution from the bright sources is not large (Section \ref{sec:intro}).

The source contributions to the IR EBL can be calculated by integrating the number counts down to a flux density limit.  
To characterize the shape of the number counts, we use the Schechter function 
form \citep{schechter1976} in the same manner as the previous studies, 
\begin{eqnarray}
\phi(S)dS=\phi_{*}\left(\frac{S}{S_{*}}\right)^{\alpha} \rm exp \it \left(-\frac{S}{S_{*}}\right)d\left(\frac{S}{S_{*}}\right),
\end{eqnarray}
where $\phi_{*}\rm$, $S_{*}$, and $\alpha$ are the normalization, characteristic flux density, and faint-end slope 
power-law index, respectively. 
The logarithmic form of $n(S)d(\rm log \it S)=\phi(S)dS$
is given by
\begin{eqnarray}
\label{eq:log_schechter}
n(S)=(\rm ln\ 10)\it \phi_{*}\left(\frac{S}{S_{*}}\right)^{(\alpha+1)} \rm \exp \it \left(-\frac{S}{S_{*}}\right).
\end{eqnarray}
Using eq. (\ref{eq:log_schechter}), we conduct $\chi^{2}$-fitting to the differential number counts derived from our and the previous observations 
shown with the filled symbols in Figure \ref{fig:nc_fig}
that are our measurements and the AzTEC \citep{scott2012} and ALMA 1.1 mm \citep{carniani2015} results.
Here we do not use the measurements of \cite{hatsukade2013}, \cite{ono2014}, \cite{oteo2015},
and the 1.3 mm results of \cite{carniani2015} that are presented with the open symbols of Figure \ref{fig:nc_fig}, 
because these studies use ALMA data covered by our study
or the sample that does not meet our selection criteria (Section \ref{sec:archival_data}).
We confirm that the entire ALMA 1.1 mm data of \cite{carniani2015} are different from ours.
Including the bright number counts at $2-10$ mJy (see above), we 
perform Schechter function fitting with
our number counts and those from the 1.1-mm observations
with AzTEC \citep{scott2012} and ALMA \citep{carniani2015}.  
We vary three Schechter parameters, and search for the best-fit parameter set of ($\phi_{*}$,$S_{*}$,$\alpha$) 
that minimizes $\chi^{2}$. The best-fit function and parameter set are presented
in Figure \ref{fig:nc_fig} and Table \ref{tab:ncfit_param}, respectively.
Figure \ref{fig:nc_fig} and the $\chi^{2}/dof$ of 16.9/19  (Table \ref{tab:ncfit_param}) indicate that 
the number counts are well represented by the Schechter function down to $\sim0.02$ mJy. 

We also carry out fitting of the double-power law (DPL) function in the same manner as
the Schechter function fitting. The DPL function has four free parameters of
$\phi_{*}^{\rm DPL}$, $S_{*}^{\rm DPL}$, $\alpha^{\rm DPL}$, and $\beta^{\rm DPL}$  
\citep[see, e.g.,][]{ono2014}.
The best-fit parameters are presented in Table \ref{tab:ncfit_param}.
The $\chi^{2}/dof$ value for the DPL fitting is nearly unity (12.8/18), and the DPL also well represents 
the number counts. Because the Schechter and DPL functions show similarly good fitting results,
we use the Schechter function fitting results for simplicity in the following discussion.

\begin{deluxetable}{ccccc} 
\tablecolumns{5} 
\tablewidth{0pt} 
\tablecaption{Best-fit Parameters \label{tab:ncfit_param}}
\tablehead{
\multicolumn{5}{c}{Schechter} \\ \hline
$S_{*}$ & $\phi_{*}$ & \multicolumn{2}{c}{$\alpha$} & $\chi^{2}/dof$ \\
(mJy) & ($10^{3}$ deg$^{-2}$) & & \\ 
(1)  & (2) & \multicolumn{2}{c}{(3)} & (4) 
}
\startdata 
$2.35^{+0.16}_{-0.16}$  & 1.54$^{+0.29}_{-0.26}$ & \multicolumn{2}{c}{$-2.12^{+0.07}_{-0.06}$}  & 16.6/19 \\ \hline \hline
\multicolumn{5}{c}{DPL} \\ \hline
$S_\ast^{\rm DPL}$ & $\phi_\ast^{\rm DPL}$ & $\alpha^{\rm DPL}$ & $\beta^{\rm DPL}$ & $\chi^2/dof$ \\ 
 (mJy) & ($10^3$ deg$^{-2}$) & & \\  
(5)  & (6)  & (7) & (8) &  \\ \hline
$4.43^{+0.34}_{-0.49}$  & 0.26$^{+0.05}_{-0.07}$ & $2.39^{+0.05}_{-0.06}$ & $6.20^{+0.48}_{-1.11}$ & 12.8/18
\enddata 
\tablecomments{
(1)-(3): Best-fit parameter set for the Schechter function.
(4): $\chi^{2}$ over the degree of freedom.
(5)-(8): Best-fit parameter set for the DPL function.
}
\end{deluxetable} 

\begin{figure}
\begin{center}
\includegraphics[trim=0.0cm 0.2cm 0.0cm 0.2cm, clip, angle=0,width=0.5\textwidth]{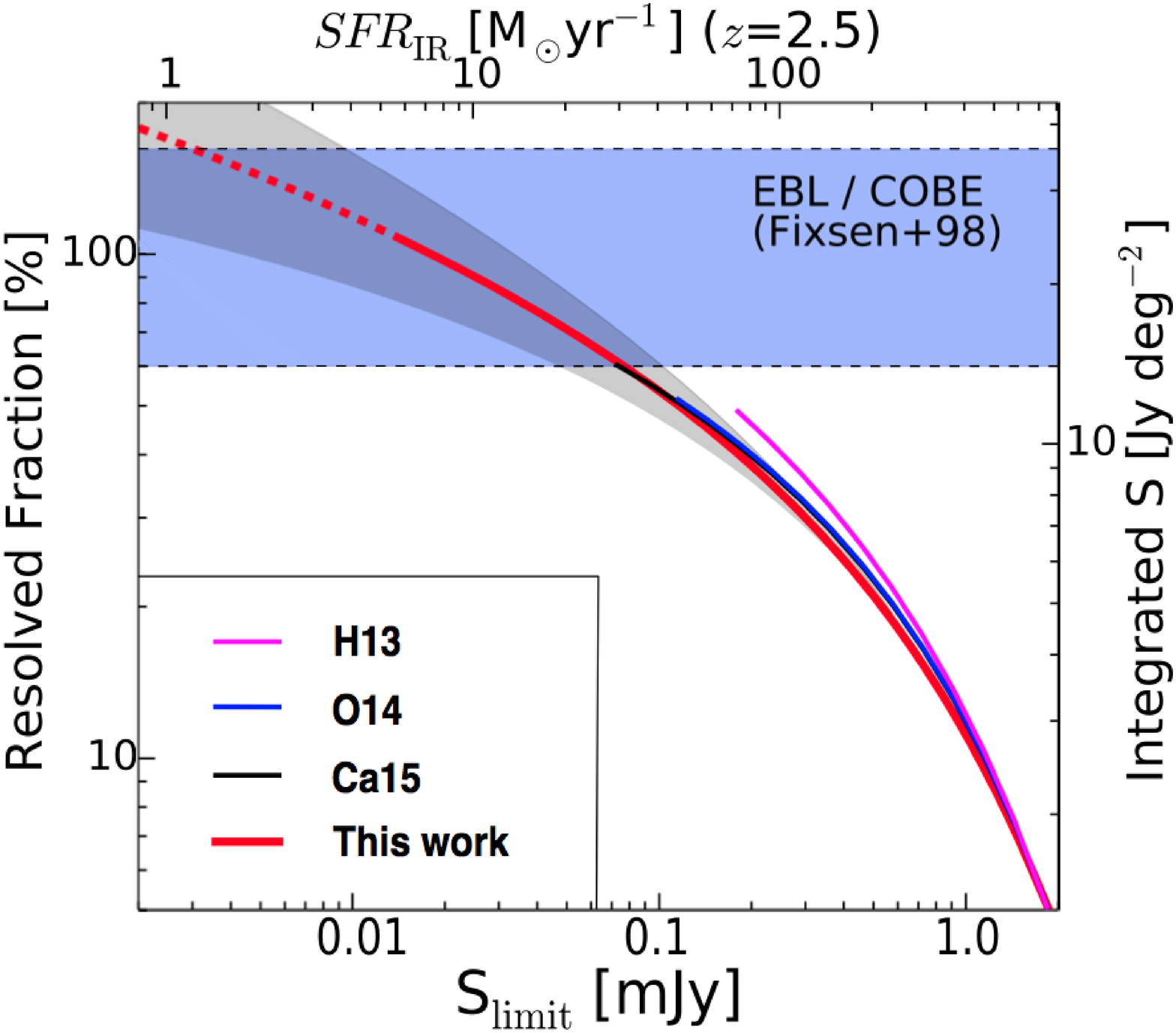}
\caption[]
{Fraction of resolved EBL at 1.2 mm as a function of the detection limit, $S_{\rm limit}$.
The red line and the gray shade indicate the best estimate from our study 
and the associated errors, respectively. The magenta, blue, and black lines
present the estimates from the previous ALMA studies 
\citep{hatsukade2013,ono2014,carniani2015}. 
Here we re-calculate the resolved fractions of the previous studies 
with the bright source counts of \cite{scott2012}.  
The blue shade denotes the IR EBL measurements from COBE \citep{fixsen1998}.
The right axis represents the absolute values of the integrated flux densities, $\int^{\infty}_{S_{\rm limit}} S\phi(S)dS$,
that correspond to the resolved EBL fraction.
The top axis ticks IR star-formation rates $SFR_{\rm IR}$ corresponding to the 1.2 mm flux \citep{kennicutt1998}
for the case that a source resides at $z=2.5$.
\label{fig:ebl}}

\end{center}
\end{figure}


With the best-fit Schechter parameter set, we calculate the integrated flux densities, 
$\int^{\infty}_{S_{\rm limit}}\it S\phi(S)dS$, down to the flux limit of  
$S_{\rm limit}$. Because the faintest bin of our number count data covers
down to 0.015 mJy (Figure \ref{fig:nc_fig}), we choose $S_{\rm limit}=0.015$ mJy.
We calculate the integrated flux density to be
$22.9^{+6.7}_{-5.6}$
Jy deg$^{-2}$ at 1.2 mm.
Figure \ref{fig:ebl} shows the integrated flux density.
We evaluate the fraction contributed to the total IR EBL measurement at 1.2 mm
($22^{+14}_{-8}$ Jy deg$^{-2}$; \citealt{fixsen1998}) that is obtained by the observations of 
the COBE Far Infrared Absolute Spectrophotometer (FIRAS),
and find that the resolved fraction in our study is $104^{+31}_{-25}\%$. 
This result suggests that our study fully resolves the 1.2 mm EBL 
into the individual sources within the uncertainties. 

Our result also indicates that the EBL contribution of sources with $\lesssim 0.02$ mJy are
negligibly small. It would suggest that there is 1) a flattening of number counts slope $\alpha$, or
2) a truncation of the number counts at a flux $S_{\rm trunc}$ below $\simeq 0.02$ mJy.
Using our resolved fraction of $f_{\rm res}=104^{+31}_{-25}\%$, we constrain $\alpha$ and $S_{\rm trunc}$
in the scenarios of 1) and 2). 
For the scenario of 1), we 
extrapolate the Schechter function from $S_{\rm limit}$ to 0 mJy 
with a power law of $n(S) \propto {S^{(\alpha_{\rm 0-S_{\rm limit}}}+1)}$,
where $\alpha_{\rm 0-S_{\rm limit}}$ is a power law slope at $0$ mJy to $S_{\rm limit}$,
and integrate the Schechter function with the power law extrapolation.
We constrain the $\alpha_{0-S_{\rm limit}}$ value
for the condition that the integrated flux density does not exceed the $1\sigma$ upper limit of the EBL measurement. 
We thus determine that the lower limit of $\alpha_{\rm 0-S_{\rm limit}}\gtrsim-1.0$.
For the scenario of 2), we extrapolate our best-fit Schechter function
below $S_{\rm limit}$. We find $S_{\rm trunc}\simeq 0.004$ mJy (corresponding to 
$SFR_{\rm IR}\sim 1.6$ $M_{\odot}{\rm yr^{-1}}$; \citealt{kennicutt1998}) 
where the $1\sigma$ lower error of $f_{\rm res}$ reaches 100\%
of the EBL contribution fraction.
Because these 1) and 2) scenarios are two extreme cases, 
our results would suggest that the faint-end slope at $\lesssim 0.02$ mJy is flatter than
the best-fit Schechter function
slope of $\alpha= -2.12^{+0.07}_{-0.06}$ (Table \ref{tab:ncfit_param}), 
and that there is a number-counts truncation, if any, at $\lesssim 0.02$ mJy.

\section{Clustering Analysis}
\label{sec:clustering}

\subsection{Galaxy Bias of our Faint ALMA Sources}
\label{sec:galaxy_bias}

We estimate a galaxy bias of our faint ALMA sources
by the counts-in-cells technique (e.g. \citealt{adelberger1998,robertson2010})
\begin{eqnarray}
b_{g}^{2} \approx \frac{\sigma_{N}^{2}-\bar{N}}{\bar{N}^{2}\sigma_{V}^{2}(z)},
\end{eqnarray} 
where $\sigma_{N}^{2}$ is the standard deviation of the detected source counts per field, $\sigma_{V}^{2}$ is the matter variance averaged over the survey volumes $V$, 
and $\bar{N}$ is the average source counts per field.

In this counts-in-cells analysis, 
we use 66 maps of the field data (A) and (B).
Because the cluster data (C) includes lensing effects of magnification and survey volume distortion
as well as the different spurious source rate (Section \ref{sec:spurious_rate}),
we do not include the cluster data (C) in the sample of our counts-in-cells analysis.
The survey volume is calculated by using the average effective area per field (Section \ref{sec:ef_area}). 
For the redshift distribution to calculate the volume, we again assume a median redshift of $z=2.5$ (Section \ref{sec:data_analysis})
and the top hat function covering $z=1-4$.
For $\sigma_{N}^{2}$ and $\bar{N}$,
we do not use the simple detection numbers, but the corrected detection numbers 
obtained with equation (\ref{eq:ef_num}), 
which allows us to remove the effects of spurious sources.
We calculate $\sigma_{V}^{2}$ with the analytic structure formation model \citep[e.g.,][]{mo2002}.
To estimate the $1\sigma$ uncertainty of $b_{g}$, we carry out Monte-Carlo simulations.
We first make 1000 mock catalogs of faint ALMA sources
consisting of source counts that 
follow the Poission probability distribution function
whose average values agree with
our observed source counts.
Then, we calculate galaxy bias values with each mock catalog,
and repeat this process for the 1000 mock catalogs.
We define a $1\sigma$ error of our observational $b_{g}$ estimate
by the 68 percentile of the $b_{g}$ distribution obtained from the 1000 mock catalog results.
Because the $1\sigma$ error is large, we cannot obtain the measurement
of $b_{g}$ but place a $1\sigma$ upper limit of 
$b_{g}<3.5$
for our faint ALMA sources.
This $1\sigma$ upper limit agrees with the previous
$b_{g}$ estimate for faint ALMA sources, $b_{g}<4$, 
given by \cite{ono2014}. 

We estimate the dark halo masses of our faint ALMA sources 
with our result of $b_{g}<3.5$ with the analytics 
structure formation model of
\cite{sheth1999}, 
assuming the one-to-one correspondence
between galaxies and dark halos.
We obtain the $1\sigma$ upper limit dark halo mass of 
$\lesssim 5\times10^{12}\ M_{\odot}$.

\begin{figure}
\begin{center}
\includegraphics[trim=0cm 0cm 1.0cm 1.0cm, clip, angle=0,width=0.5\textwidth]{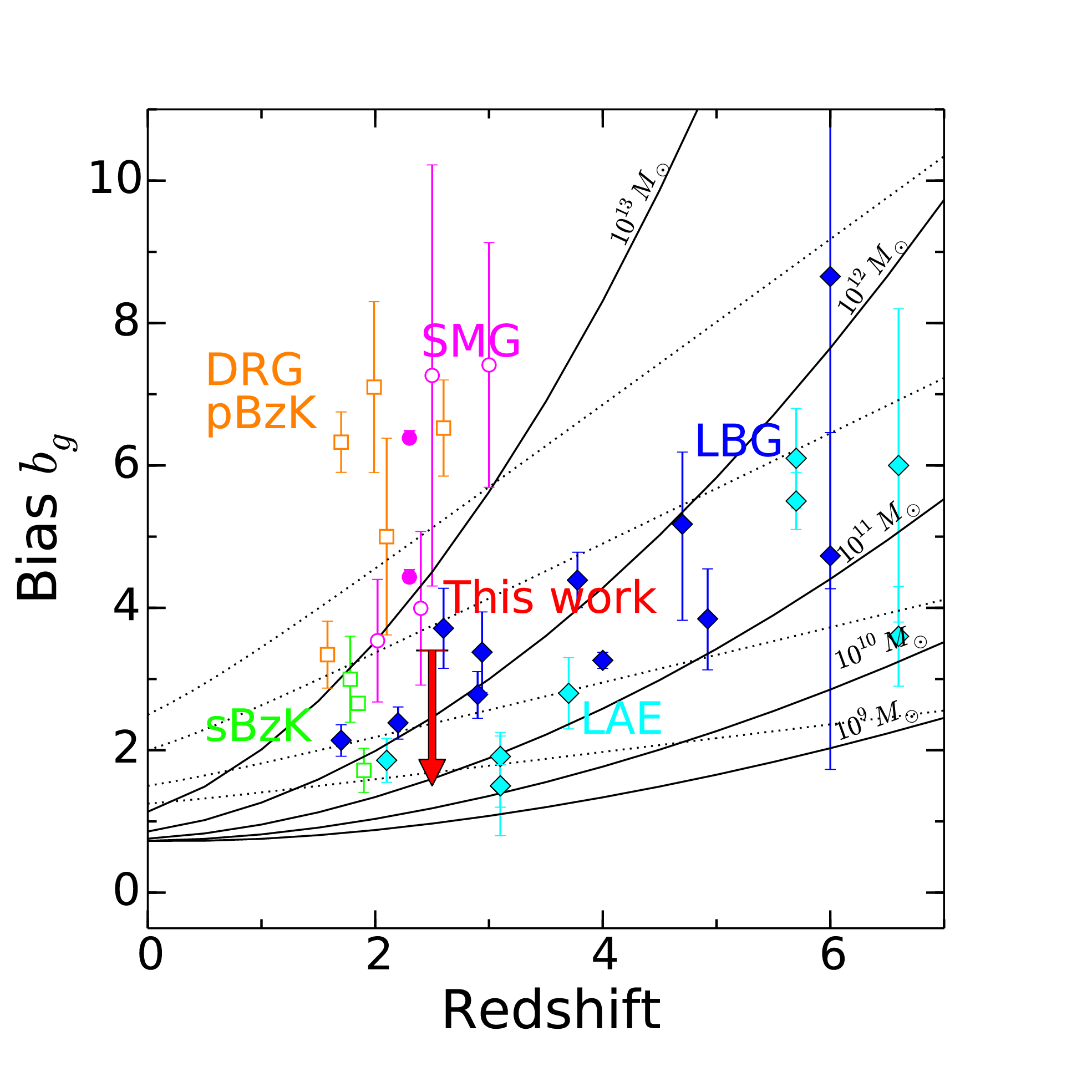}
 \caption[]
{Galaxy bias of our faint ALMA sources and the other
high-$z$ galaxies.
The red arrow shows the $1\sigma$ upper limit of our faint ALMA sources.
The magenta arrows are the measurements and the upper limit 
for bright SMGs, which are calculated from the clustering results of 
\citet{webb2003,blain2004,weiss2009,williams2011,hickox2012}. 
The orange and green squares denote
the $K$-selected galaxies of DRGs/pBzK and sBzK, respectively \citep{grazian2006,hayashi2007,quadri2007,blanc2008,furusawa2011,lin2012}. 
The blue and cyan diamonds represent the UV-selected galaxies of LBGs 
(including BX/BM) and LAEs, respectively 
\citep{ouchi2004,adelberger2005,overzier2006,lee2006,gawiser2007,ouchi2010}.
The solid curves indicate the galaxy biases of the dark halos 
with masses of $10^{9}$, $10^{10}$, $10^{11}$, $10^{12}$, and $10^{13}$ $M_{\odot}$ from bottom to top,
which are predicted by the model of \cite{sheth1999}
under the assumption of the one-to-one relation between galaxies and dark halos.
The dotted curves present evolutionary tracks of dark halos for the galaxy conserving case, 
which assume that the motion of galaxy is purely caused by gravity with no galaxy merges \citep{fry1996}. 
\label{fig:bg}}
\end{center}
\end{figure}

\subsection{Comparison with Other Populations}
\label{sec:comp_bias}

Figure \ref{fig:bg} shows the galaxy biases of our faint ALMA sources and 
a variety of high-$z$ galaxy populations; bright SMGs, $K$-selected galaxies including passively evolving BzK galaxies (pBzKs) and star-forming BzK galaxies (sBzKs), distant red galaxies (DRGs), and UV-selected galaxies composed of BX$/$BM galaxies, LBGs, and LAEs. 
Some of these high-$z$ galaxy studies shows no galaxy bias estimates, but only
the galaxy correlation lengths $r_{0}$, and the power-law indices of the two point spatial correlation function $\gamma$.
For these results, we calculate the galaxy biases in the same manner 
as \cite{ono2014} from $b_{g}=\sigma_{\rm 8,gal}/\sigma_{8}(z)$, 
where $\sigma_{8}(z)$ is a matter fluctuation in spheres of comoving radius of 8 $h^{-1}$Mpc, 
and $\sigma_{\rm 8,gal}$ is a galaxy fluctuation. 
The value of $\sigma_{\rm 8,gal}$ is derived with \citep[see eq. 7.72 in][]{peebles1993}
\begin{eqnarray}
\sigma_{8,\rm gal}^{2}=\frac{72}{(3-\gamma)(4-\gamma)(6-\gamma)2^{\gamma}}\left(\frac{r_{0}}{8\ h^{-1}\rm{Mpc}}\right)^{\gamma}.
\end{eqnarray}

In Figure \ref{fig:bg},
the galaxy bias estimates for the most of SMGs, DRGs, and pBzKs are higher than the upper limit of 
our faint ALMA sources. 
On the other hand, $K$-selected star-forming galaxies of sBzKs have the halo masses comparable 
with those of our faint ALMA sources. 
The UV-selected LBGs and LAEs also have the galaxy bias values similar to our faint ALMA sources.
Thus, we find that our faint ALMA sources are not similar to 
SMGs, DRGs, and pBzK with masses of $\gtrsim 10^{13} M_\odot$
but sBzKs, LBGs, and LAEs with masses $\lesssim 10^{13} M_\odot$ (Figure \ref{fig:bg}).
These clustering results suggest a possibility in a statistical sense 
that a large fraction of our faint ALMA sources 
could be mm counterparts of sBzKs, LBGs, and LAEs.

\section{Multi-Wavelength Properties of the Faint ALMA Sources}
\label{sec:characterization}

\subsection{Optical-NIR Counterparts}
\label{sec:counterpart_id}
We investigate optical-NIR counterparts of our faint ALMA sources, and 
characterize the sources with the multi-wavelength properties,
especially the optical-NIR colors.
We search for the optical-NIR counterparts in the regions of 
SXDS and A1689 that have rich multi-wavelength data. 
In these two regions, there are a total of 65 faint ALMA sources;
54 and 11 sources in SXDS and A1689 regions, respectively.
For the optical-NIR counterpart studies so far conducted, 
this is the largest sample of faint mm sources down to $\sim 0.02$ mJy, while there exist
studies, such as ALESS and COSMOS, that use a comparably large but
relatively bright sample \citep{hodge2013,scoville2015}.

Table \ref{tab:data_properties} summarizes the multi-wavelength data of SXDS and A1689
used in this study. The multi-wavelength data cover from X-ray to far-infrared (FIR) radiation.
In the SXDS region, we use published photometry catalogs 
in 0.5-10 keV \citep{ueda2008}, 
$u^{*}$ (Foucaud et al. in preparation),
$B$, $V$, $R$, $i^\prime$, $z^\prime$ \citep{furusawa2008},
$J$, $H$, $K$ (UKIDSS DR10; Almaini et al. in preparation), 
$3.6-24\ \mu$m (Spitzer UKIDSS Ultra Deep Survey; SpUDS),
$250-500\ \mu$m (Herschel Multi-tiered Extragalactic Survey; HerMES, \citealt{oliver2012}),
and 1.4 GHz \citep{simpson2006}
as well as the photometric redshift ($z_{\rm phot}$) catalog \citep{williams2009}.  
In the A1689 region, we investigate published photometry catalogs
of $g_{475}$, $r_{625}$, $i_{775}$, $z_{850}$, $J_{110}$, $H_{160}$ (Hubble Source Catalog; HSC), 
$3.6-24\ \mu$m ({\it Spitzer} Enhanced Imaging Products; SEIP), 
$250-500\ \mu$m (HerMES), and
the photometric redshift $z_{\rm photo}$ and spectroscopic redshift $z_{\rm spec}$ catalogs 
\citep{limousin2007,coe2010,diego2015}.  

We use the $u^{*}$ to $K$ ($g'$ to $H$) band images in the SXDS (A1689) region 
for the optical-NIR ($0.4-2\ \mu$m) counterpart identification, because these data
have a good spatial resolution.
The optical-NIR counterparts are defined by the criterion that
the distance between optical-NIR and ALMA source centers
is within a $1\farcs0$. If an ALMA source meets this criterion 
in any of optical-NIR bands, we regard that the ALMA source
has an optical-NIR counterpart.
The $1\farcs0$ radius is chosen, because the absolute positional 
accuracy (APA) of $\simeq 1\farcs0$ is guaranteed in our ALMA data (cf. optical-NIR data
for $\lesssim 0\farcs3$; see, e.g., \citealt{furusawa2008}).
Regarding the ALMA's APA,
the knowledge base page of ALMA Science portal presents 
that the APA is smaller than the synthesized beam width\footnote{
The APA is larger than $0\farcs1$ that is originated from the
uncertainties of phase calibrator positions and 
the baseline lengths of antennas.}.
Because the synthesized beam widths of our data are $\lesssim 1\farcs0$,
the $1\farcs0$ radius is applied.

Using the SXDS+A1689 source catalog of the
65 faint ALMA sources, we identify a total of 17 optical-NIR counterparts. 
Fifteen and two sources are 
identified in SXDS and A1689 and referred to as S1-15 and A1-2, respectively. 
Figure \ref{fig:postage_stamps} shows the postage stamps of these identified sources,
and Table \ref{tab:multiband_photometry} summarizes 
their photometric properties. 
The photometry values in Table \ref{tab:multiband_photometry}
are total magnitudes.
The $u^*$ magnitudes are {\tt MAG\_AUTO} values given by our {\sc SExtractor} photometry.
Similarly, $BVR_c i^\prime z^\prime$ magnitudes are {\tt MAG\_AUTO} values listed in the \citet{furusawa2008}'s catalogs.
The {\it Spitzer} photometry values are estimated from the SEIP aperture flux densities with
the aperture correction 
\footnote{http://irsa.ipac.caltech.edu/data/SPITZER/docs/irac/
iracinstrumenthandbook/29/}, 
while the {\it HST} photometry values (for A1 and A2) are the total magnitudes calculated 
from {\tt MAG\_APER} values of our {\sc SExtractor} measurements and the aperture correction
\footnote{ http://archive.stsci.edu/hst/hsc/help/FAQ/aperture
\_corrections.txt}.
Note that there is an offset of $1\farcs2$ ($1\farcs06$) between the flux peaks of the optical-NIR counterpart and S9 (S15) ALMA source.
Because S9 and S15 have an extended optical-NIR profile whose major component falls in the ALMA emitting region,
the optical-NIR source is selected as a counterpart of the ALMA source.

Because spurious sources are included in the 65 faint ALMA sources,
we calculate the real ALMA source number in the SXDS (A1689) region.
In the SXDS (A1689) region,
we subtract the expected spurious-source number of 31 (5) from
the SXDS (A1689) source number of 54 (11), and estimate the real ALMA source number
to be 23 (6). The total real ALMA source number in these two regions 
is 29 $(=23+6)$.
Thus, the counterpart-identification fraction that is the ratio of 
the total optical-NIR counterpart ALMA sources to the real ALMA sources 
is estimated to be 59\% (=17/29). 
We estimate the probability of the chance projection 
that a foreground or background object is located 
by chance in the $1\farcs0$-radius circle of an ALMA source, and obtain
$\sim10\%$ in the $P$-value \citep{downes1986}.
Multiplying this probability with the number of optical-NIR counterparts, 17,
we expect that $\sim 2$ objects would be the foreground/background contamination.
Even if we include this contamination effect, a half of our faint ALMA sources, 52\% ($=[17-2]/29$), have an optical-NIR counterpart.

Interestingly, there is a difference of the counterpart-identification fractions
between SXDS and A1689 regions, although the depths of the optical-NIR data
are comparable in these two regions ($27-28$ mag in optical and $25-26$ mag in NIR;
see Table \ref{tab:data_properties}).
The counterpart-identification fraction
in the SXDS region is 65\% $(=15/23)$, while the one in the A1689 region
is only 33\% $(=2/6)$. Including the $1\sigma$ statistical uncertainties,
these fractions are $65\pm 22$\% and $33\pm 30$\% for the SXDS and A1689 regions,
respectively. 
We carefully examine whether this difference is just an artifact originated from, 
for examples, a systematically high spurious rate and cluster member galaxy obscurations
in the cluster region. However, there are no hints of notable artificial effects in the data.
\citet{chen2014} also report a similarly low-identification rate of MIR/radio counterparts 
in A1689 based on their observations with SMA.
It should be noted that the intrinsic mm flux densities of our sources in the A1689 region are $\simeq 0.01-0.1$ mJy 
(magnification corrected) that is about an order of magnitude 
fainter than those of our sources in the SXDS region, $\simeq 0.1-1$ mJy.

There are two possible explanations for the difference between SXDS and A1689.
The first explanation is made by the negative k-correction.
Due to the faint flux densities,
the mm sources of the A1689 would reside at a redshift higher than
the SXDS sources on average, since
the abundance of the dusty population decreases towards 
the epoch of the first-generation galaxies.
Because the effect of the negative k-correction is significant, 
the ratio of mm to optical-NIR fluxes is higher for high-$z$ sources than low-$z$ sources.
In this case, the counterpart-identification fraction would be small in the A1689 region.
The second explanation is the A1689's lensing distortion effects
in conjunction with the intrinsic offsets between mm and optical-NIR emitting regions within a galaxy.
\cite{chen2015} report that there is an offset between submm and optical-NIR
emitting regions typically by $\sim 0\farcs4$. If there is an intrinsic $\sim 0\farcs4$ offset
in the source plane of A1689, the image distortion effects of lensing would magnify
the offset to $>1\farcs0$.
We thus estimate the probability that the offsets between mm and optical-NIR sources 
in the source plane is magnified to $>1\farcs0$ in the image plane 
by Monte-Carlo simulations.
We place 100 artificial sources in the source plane of A1689
at around an ALMA source in A1689. The positions of these artificial sources
follow the Gaussian distributions whose standard deviation is $0\farcs4$.
Using our mass model, we calculate the offsets between the ALMA source
and the artificial sources in the image plane. We repeat this process
for 11 ALMA sources in A1689.  We find that $\sim 50$\% of the artificial sources show 
$>1\farcs0$ offsets from the ALMA source in the image plane. 
Because the number of optical counterparts in A1689 is 2, our simulation results
suggest that a total of 4 ($=2/[50\%]$) optical counterparts should exist in the case of no lensing distortion effects (+the intrinsic offsets).
Thus, the true counterpart-identification fraction of A1689 is 67\%($=4/6$)
that is very similar to the counterpart-identification fraction of SXDS ($65\pm 22$\%).
If the lensing distortion effects (+the intrinsic offsets) would provide 2 ($=4-2$) sources whose optical counterparts 
have an offset from the ALMA sources by $>1\farcs0$ in the image plane, we would find such ALMA sources 
whose optical counterparts are located at the position just beyond $1\farcs0$ in our data.
We refer the distribution of the artificial optical counterparts from our simulations,
and search for potential optical-NIR counterparts whose distances from the ALMA sources are $>1\farcs0$.
We find such potential optical counterparts in 4 sources of A1689, some of which may be real optical counterparts, 
and confirm the hypothesis that the combination of the lensing distortion effects and the intrinsic offsets 
can be the major reason for the difference of 
the counterpart-identification fractions between the cluster of A1689 and the blank field of SXDS.%

Table \ref{tab:multiband_photometry} lists the photometry of the {\it Spitzer} ($3.6-24\ \mu$m) 
and {\it Herschel} ($250-500\ \mu$m) data. Because the spatial resolutions of
the {\it Spitzer} and {\it Herschel} data are poor, we determine the {\it Spitzer} and {\it Herschel} 
counterparts by visual inspection that allows sources residing at a position
slightly beyond the $1''$ search radius.
Although no sources are identified in the {\it Herschel} bands,
about a half of our faint ALMA sources are detected in the {\it Spitzer} data.
These results would indicate that our faint ALMA sources are very high redshift sources ($z > 4$).
However, it should be noted that the {\it Herschel} data can detect dusty starbursts whose
SFRs are very high, $>$ a few $100 M_{\odot}{\rm yr}^{-1}$ even at $z\sim 2-3$.

Here we examine whether we simply miss red spectral energy distribution (SED) objects that are 
not detected in optical-NIR ($0.4-2\ \mu$m) bands but
only in {\it Spitzer} and {\it Herschel} bands.
We search for {\it Spitzer} and {\it Herschel} counterparts of 
our faint ALMA sources with no optical-NIR counterparts,
adopting the visual inspection technique described above.
We find no {\it Spitzer} and {\it Herschel} counterparts that
have undetectable optical-NIR fluxes.
Again, these results do not suggest that our faint ALMA sources
are dusty starbursts.

We cross-match the catalogs of our faint ALMA sources and 
the radio 1.4 GHz and X-ray 0.5-10 keV sources. We find no counterparts
within a $1\farcs0-3\farcs0$
radius around our faint ALMA sources. Thus, our faint ALMA sources
have no signatures of AGN.

To summarize these counterpart identification results with
the {\it Spitzer} and {\it Herschel} as well as radio and X-ray data,
there are signs of neither dusty starbursts nor AGNs
in our faint ALMA sources.
Note that these cross-match analysis is completed only in SXDS because there are 
no radio and X-ray data available in A1689.

\begin{deluxetable*}{lllc} 
\tablecolumns{4} 
\tablewidth{0pt} 
\tablecaption{Multi-wavelength Data \label{tab:data_properties}}
\tablehead{
Region ($N_{counter}$)
 & Telescope/Instrument & Band (5$\sigma$ limit in mag) & Reference\\
 (1) &  & (2) & (3)
 }   
\startdata 
SXDS (15) 
& {\it XMM-Newton}/EPICs, pn, MOS & 0.5-10 keV ($6-50\times10^{-16}$ erg cm$^{-2}$ s$^{-1}$) &  (a)\\
& CFHT/MegaCam         & $u^{*} (27.7)$ &  (b) \\
& Subaru/Suprime-Cam & $B$ (28.4) & (c)\\ 
& Subaru/Suprime-Cam & $V$ (28.0)  & (c)\\
& Subaru/Suprime-Cam & $R_c$ (27.8)  & (c)\\
& Subaru/Suprime-Cam & $i^\prime$ (27.7) & (c)\\
& Subaru/Suprime-Cam & $z^\prime$ (26.7)  & (c)\\
& UKIRT/WFCAM & $J$ (25.6)  & (d)\\
& UKIRT/WFCAM & $H$ (24.8) & (d)\\
& UKIRT/WFCAM & $K$ (25.4)  & (d)\\
& {\it Spitzer}/IRAC & 3.6$\mu$m (24.7)  &(e)\\ 
& {\it Spitzer}/IRAC & 4.5$\mu$m (24.6)  &(e)\\
& {\it Spitzer}/IRAC & 5.8$\mu$m (22.3)  &(f)\\
& {\it Spitzer}/IRAC & 8.0$\mu$m (22.3)  &(f)\\
& {\it Spitzer}/MIPS & 24$\mu$m (18.0)  & (f)\\        
& {\it Herschel}/SPIRE & 250$\mu$m (13.8)  & (g)\\ 
& {\it Herschel}/SPIRE & 350$\mu$m (14.0)  & (g)\\ 
& {\it Herschel}/SPIRE & 500$\mu$m (13.6)  & (g)\\ 
& VLA & 1.4 GHz (20.0) & (h) \\ \hline
A1689 (2) 
& {\it HST}/ACS & $g_{475}$ (28.6)    & (i)\\
& {\it HST}/ACS & $r_{625}$ (28.3)     & (i)\\ 
& {\it HST}/ACS & $i_{775}$ (28.2)     & (i)\\
& {\it HST}/ACS & $z_{850}$ (27.8)    & (i)\\
& {\it HST}/NICMOS & $J_{110}$ (26.2)  & (j) \\  
& {\it HST}/WFC3 & $H_{160}$ ($26.3-28.1$)     &  (k)\\                  
& {\it Spitzer}/IRAC & 3.6$\mu$m (24.8) & (l)\\ 
& {\it Spitzer}/IRAC & 4.5$\mu$m (24.3) & (l)\\
& {\it Spitzer}/IRAC & 5.8$\mu$m (23.4) & (l)\\
& {\it Spitzer}/IRAC & 8.0$\mu$m (22.9) & (l)\\
& {\it Spitzer}/MIPS & 24$\mu$m (19.5) & (l)\\  
& {\it Herschel}/SPIRE & 250$\mu$m (14.0) &   (g)\\ 
& {\it Herschel}/SPIRE & 350$\mu$m (14.2) &   (g)\\ 
& {\it Herschel}/SPIRE & 500$\mu$m (13.8) &   (g)   
\enddata 
\tablecomments{Columns: 
(1) Name of the region.  
The number, $N_{\rm counter}$, in the parenseses
indicates the number of the optical-NIR counterparts.
(2) Limiting magnitude defined by 5$\sigma$ sky noise in a circular aperture with a radius of PSF FWHM.
(3) Reference or source catalogue: 
(a) \cite{ueda2008};
(b) Foucaud et al. (in preparation);
(c) \cite{furusawa2008};
(d) UKIDSS (O. Almaini et al. in preparation);
(e) SEDS \citep{ashby2013}; 
(f) SpUDS; 
(g) HerMES \citep{oliver2012}; 
(h) \cite{simpson2006};  
(i) Alavi et al. (2014); 
(j) \cite{bouwens2009}; 
(k) \cite{bouwens2015}; 
(l) SEIP.\footnote{http://irsa.ipac.caltech.edu/data/SPITZER/Enhanced/SEIP/overview.html}
}
\end{deluxetable*}

\begin{turnpage}
\begin{deluxetable*}{llllllllllllllllll} 
\tablecolumns{18} 
\tablewidth{0pt} 
\tablecaption{Photometry of the Optical-NIR Counterparts \label{tab:multiband_photometry}}
\tablehead{
\multicolumn{18}{c}{SXDS}  \\
ID & $u^{*}$ & $B$ & $V$ & $R_c$ & $i^\prime$ & $z^\prime$ & $J$ & $H$ & $K$ & 3.6$\mu$m & 4.5$\mu$m & 5.8$\mu$m & 8.0$\mu$m & 24$\mu$m & 250$\mu$m & 350$\mu$m & 500$\mu$m
} 
\startdata
S1 & $\cdots$ & 24.46 & 24.24 & 23.85 & 23.52 & 22.88 & $\cdots$ & $\cdots$ & $\cdots$ & 20.02 & 19.94 & 20.11 & 20.12 & $>18.56$ & $>14.35$ & $>14.55$ & $>14.15$ \\ 
S2 & $\cdots$ & 25.91 & 26.18 & 26.59 & 26.71 & 25.01 & $\cdots$ & $\cdots$ & $\cdots$ & $>25.27$ & $>25.16$ & $>22.85$ & $>22.81$ & $>18.56$ & $>14.35$ & $>14.55$ & $>14.15$ \\ 
S3 & 27.24 & 26.69 & 26.35 & 25.63 & 25.29 & 25.25 & 24.80 & 24.43 & 24.41 & $>25.27$ & $>25.16$ & $>22.85$ & $>22.81$ & $>18.56$ & $>14.35$ & $>14.55$ & $>14.15$ \\ 
S4 & 25.76 & 25.61 & 25.30 & 25.00 & 24.47 & 24.05 & 23.21 & 22.92 & 22.66 & 21.68 & 21.84 & $>22.85$ & $>22.81$ & $>18.56$ & $>14.35$ & $>14.55$ & $>14.15$ \\ 
S5 & 26.31 & 25.11 & 24.70 & 24.76 & 24.58 & 24.75 & 24.53 & 24.05 & 24.22 & $>25.27$ & $>25.16$ & $>22.85$ & $>22.81$ & $>18.56$ & $>14.35$ & $>14.55$ & $>14.15$ \\ 
S6 & 25.69 & 24.84 & 23.76 & 22.79 & 21.96 & 21.55 & 20.81 & 20.42 & 20.08 & 19.61 & 20.06 & 19.95 & 20.24 & $>18.56$ & $>14.35$ & $>14.55$ & $>14.15$ \\ 
S7 & $>28.23$ & 27.14 & 27.42 & 27.30 & 27.32 & 26.62 & $>26.18$ & $>25.31$ & $>25.94$ & $>25.27$ & $>25.16$ & $>22.85$ & $>22.81$ & $>18.56$ & $>14.35$ & $>14.55$ & $>14.15$ \\ 
S8\tablenotemark{$\dagger$} & 24.57 & $>28.93$ & $>28.56$ & $>28.33$ & $>28.24$ & $>27.22$ & $>26.18$ & $>25.31$ & $>25.94$ & $>25.27$ & $>25.16$ & $>22.85$ & $>22.81$ & $>18.56$ & $>14.35$ & $>14.55$ & $>14.15$ \\ 
S9\tablenotemark{$\dagger\dagger$} & $>28.23$ & 25.30 & 23.86 & 22.96 & 21.63 & 20.93 & 20.24 & 20.22 & 20.45 & 20.40 & 20.78 & 20.46 & $>22.81$ & $>18.56$ & $>14.35$ & $>14.55$ & $>14.15$ \\ 
S10 & $>28.23$ & 26.68 & 26.25 & 26.25 & 25.65 & 25.07 & 25.90 & 24.60 & 25.67 & $>25.27$ & $>25.16$ & $>22.85$ & $>22.81$ & $>18.56$ & $>14.35$ & $>14.55$ & $>14.15$ \\ 
S11 & 26.00 & 25.14 & 24.66 & 24.53 & 24.02 & 23.62 & 22.36 & 21.76 & 21.26 & 19.91 & 19.69 & 19.31 & 19.51 & 15.95 & $>14.35$ & $>14.55$ & $>14.15$ \\ 
S12 & 23.45 & 23.22 & 22.96 & 22.94 & 22.65 & 22.45 & 21.55 & 21.18 & 20.88 & 19.89 & 19.66 & 19.35 & 19.62 & 16.66 & $>14.35$ & $>14.55$ & $>14.15$ \\ 
S13 & $>28.23$ & $>28.93$ & 28.50 & $>28.33$ & $>28.24$ & $>27.22$ & $>26.18$ & $>25.31$ & 25.22 & 22.55 & 22.73 & 22.33 & $>22.81$ & $>18.56$ & $>14.35$ & $>14.55$ & $>14.15$ \\ 
S14 & $>28.23$ & 25.96 & 26.09 & 25.68 & $>28.24$ & $>27.22$ & $>26.18$ & $>25.31$ & $>25.94$ & $>25.27$ & $>25.16$ & $>22.85$ & $>22.81$ & $>18.56$ & $>14.35$ & $>14.55$ & $>14.15$ \\ 
S15 & 23.87 & 23.49 & 23.40 & 23.08 & 22.84 & 22.42 & 21.72 & 21.25 & 21.16 & 20.27 & 20.12 & 20.32 & 19.89 & $>18.56$ & $>14.35$ & $>14.55$ & $>14.15$ \\  \hline\hline
\multicolumn{18}{c}{Abell 1689}  \\ 
ID  & \nodata & \nodata & $g_{475}$   & $r_{625}$ & $i_{775}$ & $z_{850}$ & $J_{110}$ & $H_{160}$ & \nodata & 3.6$\mu$m & 4.5$\mu$m & 5.8$\mu$m & 8.0$\mu$m & 24$\mu$m & 250$\mu$m & 350$\mu$m & 500$\mu$m \\ \hline
A1 & $\cdots$ & $\cdots$ & 26.61 & 25.78 & 24.79 & 25.36 & 24.29 & 23.55 & $\cdots$ & 18.25 & 18.45 & 18.72 & 19.10 & 16.94 & $>14.55$ & $>14.75$ & $>14.35$ \\ 
A2 & $\cdots$ & $\cdots$ & 27.35 & 26.28 & 25.84 & 25.59 & $>26.75$ & 24.36 & $\cdots$ & $>25.35$ & $>24.85$ & $>23.95$ & $>23.45$ & $>20.05$ & $>14.55$ & $>14.75$ & $>14.35$
\enddata 
\tablecomments{The $u^*$ magnitudes are {\tt MAG\_AUTO} values given by our {\sc SExtractor} photometry.
Similarly, $BVR_c i^\prime z^\prime$ magnitudes are {\tt MAG\_AUTO} values listed in the \citet{furusawa2008}'s catalogs.
The {\it Spitzer} photometry values are total magnitudes estimated from the SEIP aperture flux densities with
the aperture correction, while the {\it HST} photometry values (for A1 and A2) are the total magnitudes calculated 
from {\tt MAG\_APER} values of our {\sc SExtractor} measurements. 
The aperture corrections for the {\it Spitzer} and {\it HST} photometry
are provided by the {\it Spitzer}
and {\it HST} websites, respectively (see text).
The lower limits correspond to the $3\sigma$ levels. 
The S1 and S2 objects are not observed in $u^*$ band and $JHK$ bands. 
}
\tablenotetext{$\dagger$}{The secure photometry cannot be carried out due to the blending with a bright neighboring objects for the B-500 $\mu$m bands. See Figure \ref{fig:postage_stamps}.}
\tablenotetext{$\dagger\dagger$}{The positions of the optical-NIR counterpart are far from the ALMA source center by $\sim1.\!^{\prime\prime}2$. }
\end{deluxetable*} 
\end{turnpage}

\subsection{Lensed ALMA Sources in A1689}
\label{sec:lensed_a1689}

\begin{figure}
\begin{center}
\includegraphics[trim=0cm 0cm 0cm 0cm, clip, angle=0,width=0.48\textwidth]{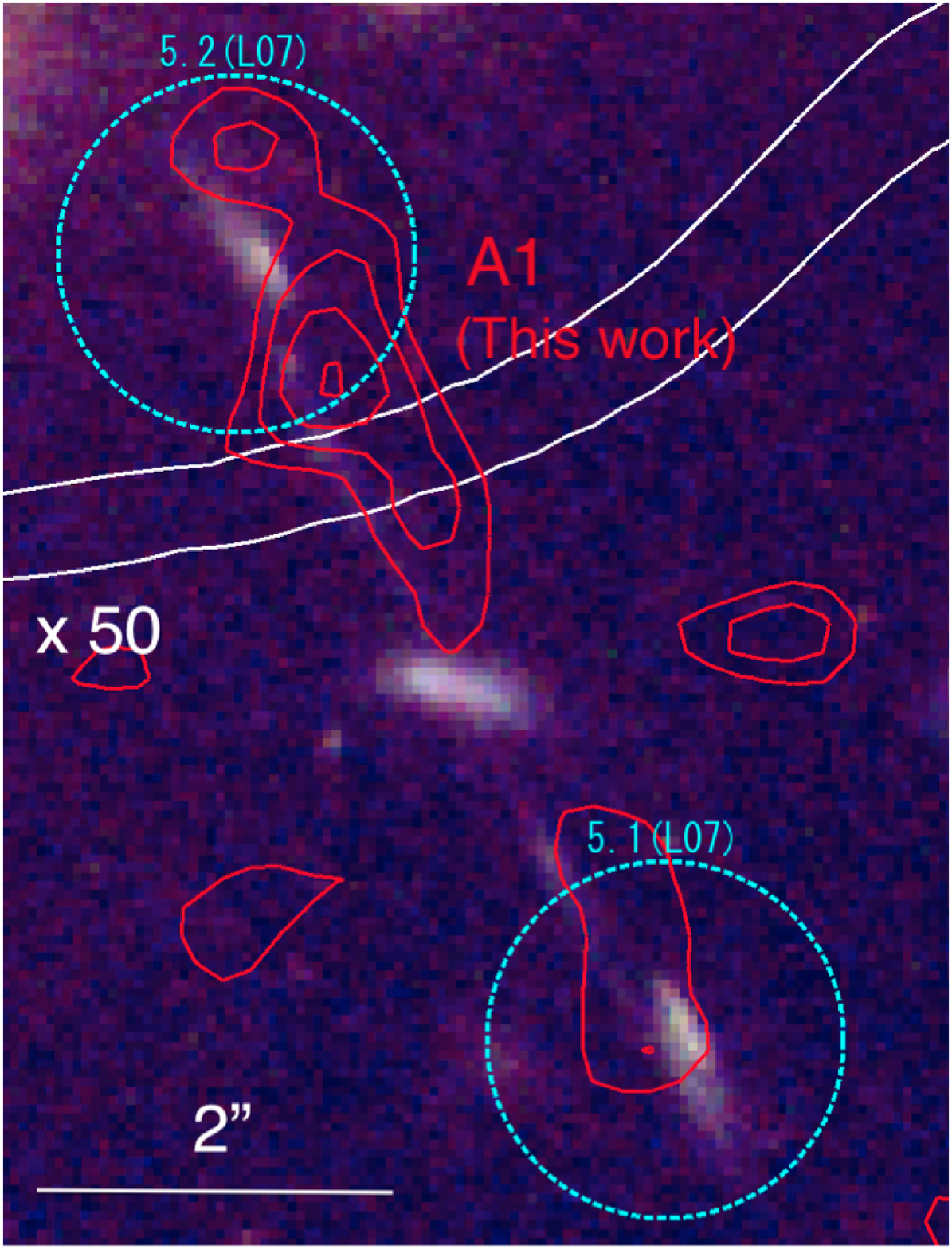}
 \caption[]
{False-color image of A1. The used images and color assignment are the same as Figure \ref{fig:a1689}. 
The center of the cyan circles pinpoint the positions of the 5.1 and  5.2 sources, which are two of the multiple images of the galaxy at $z_{\rm spec}=2.60$ \citep{limousin2007, coe2010}. 
The radius of the cyan circle is $1\farcs0$ that corresponds to the search radius for the counterpart identifications. 
The red contours show the ALMA 1.3 mm emission at the 2, 3, 4, and 5 $\sigma$ levels. The white curves represent the lines for the lensing magnification of $\mu=50$.  
\label{fig:A1}}
\end{center}
\end{figure}

We identify two optical counterparts of A1 and A2 that are lensed by the A1689 cluster. 
The optical image at the position of A1 
is presented in Figure \ref{fig:A1}. There is a small positional offset
of $\sim0\farcs8$ between the optical and mm emission, but 
this small difference
can be explained by the positional uncertainty of the ALMA image
(Section \ref{sec:counterpart_id})
\footnote{
There remains a possibility that the offset would be a real positional difference
between the optical and mm emitting regions. However, it requires 
ALMA data analysis for a positional accuracy better than this study.
}.

This optical counterpart of A1 is one of three known multiple images that is dubbed 5.2, while
the rest of two multiple images are called 5.1 and 5.3 \citep{limousin2007}.
We investigate our ALMA image at the optical positions of 5.1 and 5.3 sources. 
In the ALMA image, the 5.1 source is marginally detected at the $3\sigma$ level, while 
the 5.3 source has no signals above the $3\sigma$ level.
Table \ref{tab:multiple_photometry} summarizes the physical properties of these three multiple images.
The difference of mm emissivities between these multiple images
is discussed in Section \ref{sec:counterpart_id}.

Although the 5.2 source, i.e. A1, does not have a spectroscopic redshift,
the redshift of the 5.1 and 5.3 sources is $z=2.60$ determined by spectroscopy (Richard et al. in preparation, see \citealt{limousin2007}).
We thus regard that the redshift of A1 is $z=2.60$.
We calculate the intrinsic 1.2 mm flux of A1 from our 1.3 mm data
with the flux scaling factor from 1.3 mm to 1.2 mm (Section \ref{sec:data_analysis}), 
and obtain $0.020\pm0.008$ mJy that is corrected for 
the gravitational lensing magnification of $\mu=22.8$ and the primary beam sensitivity.
\citet{chen2014} report that 
the 5.1 and 5.3 sources are not detected above the $3\sigma$ level by their 
Submillimeter Array (SMA) $870\ \mu$m observations, but only the 5.2 source.
\citet{chen2014} measure the flux of the 5.2 source, and find
the intrinsic flux of 5.2 source is 0.085 $\pm0.035$ mJy at $870\ \mu$m, 
which corresponds to 
$0.033\pm0.014$ mJy at 1.2 mm 
with the $k$-correction from $870\ \mu$m to 1.2 mm (Section \ref{sec:data_analysis}). 
This 1.2 mm flux of $0.033\pm0.014$ mJy is consistent with our measurement of 
$0.020 \pm 0.008$ mJy within the 1$\sigma$ uncertainties.

The optical-NIR counterpart of A2 is identified in the optical bands with a lensing magnification factor of $\mu=4.9$. 
The intrinsic 1.2 mm flux of A2 is estimated to be $0.077\pm0.035$ mJy that are corrected 
for the gravitational lensing magnification and the primary beam attenuation. 
Neither $z_{\rm spec}$ nor $z_{\rm photo}$ of A2 are available in the literature. 

We compare these A1689 sources with optical-NIR counterparts (Table \ref{tab:multiple_photometry}) 
and the previously-known lensed sources,
Cosmic Eylash \citep[e.g.,][]{swinbank2010} and SMMJ16358 \citep[e.g.,][]{kneib2004}, 
identified by single-dish observations.
Cosmic Eyelash and SMMJ16358 have intrinsic SFRs of $\sim 210$ and $500 M_{\odot}{\rm yr}^{-1}$, respectively. 
These SFRs correspond to $0.5$ and $1.3$ mJy at 1.2 mm under the assumption of the $SFR-L_{\rm IR}$ relation \citep{kennicutt1998}. 
On the other hand, our optical-NIR counterparts have intrinsic fluxes of
$\sim0.02-0.08$ mJy at 1.2 mm. 
Thus, our ALMA study has identified sources with intrinsic fluxes $\gtrsim 10$ times fainter than 
these lensed sources.

\begin{figure}
\begin{center}
\center{\includegraphics[trim=0.0cm 0.0cm 0cm 0cm, clip,angle=0,width=0.48 \textwidth]{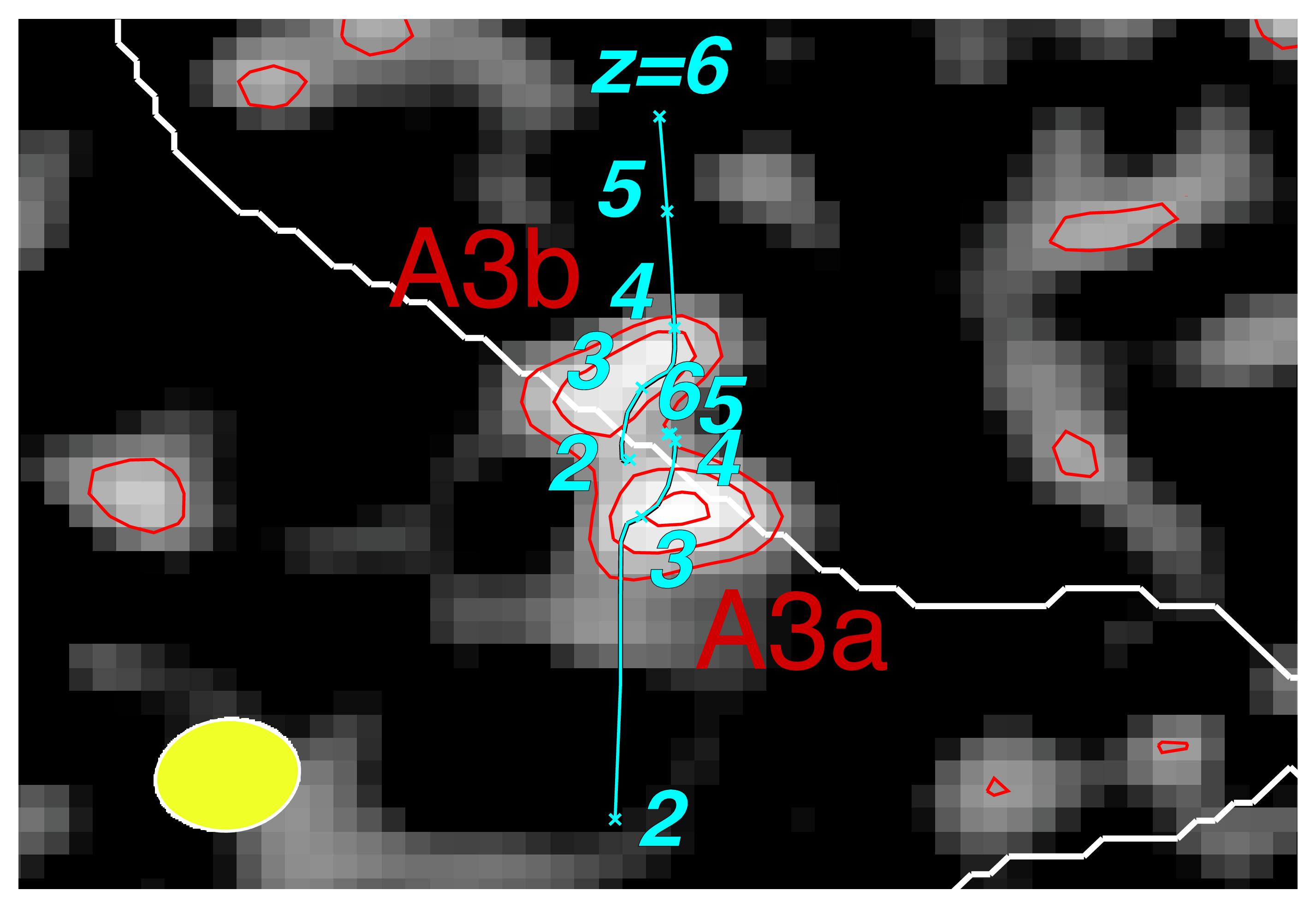}}
\vspace{-0.8cm}
\center{\includegraphics[trim=0.0cm 0.0cm -0.1cm 0cm, clip,angle=0,width=0.48\textwidth]{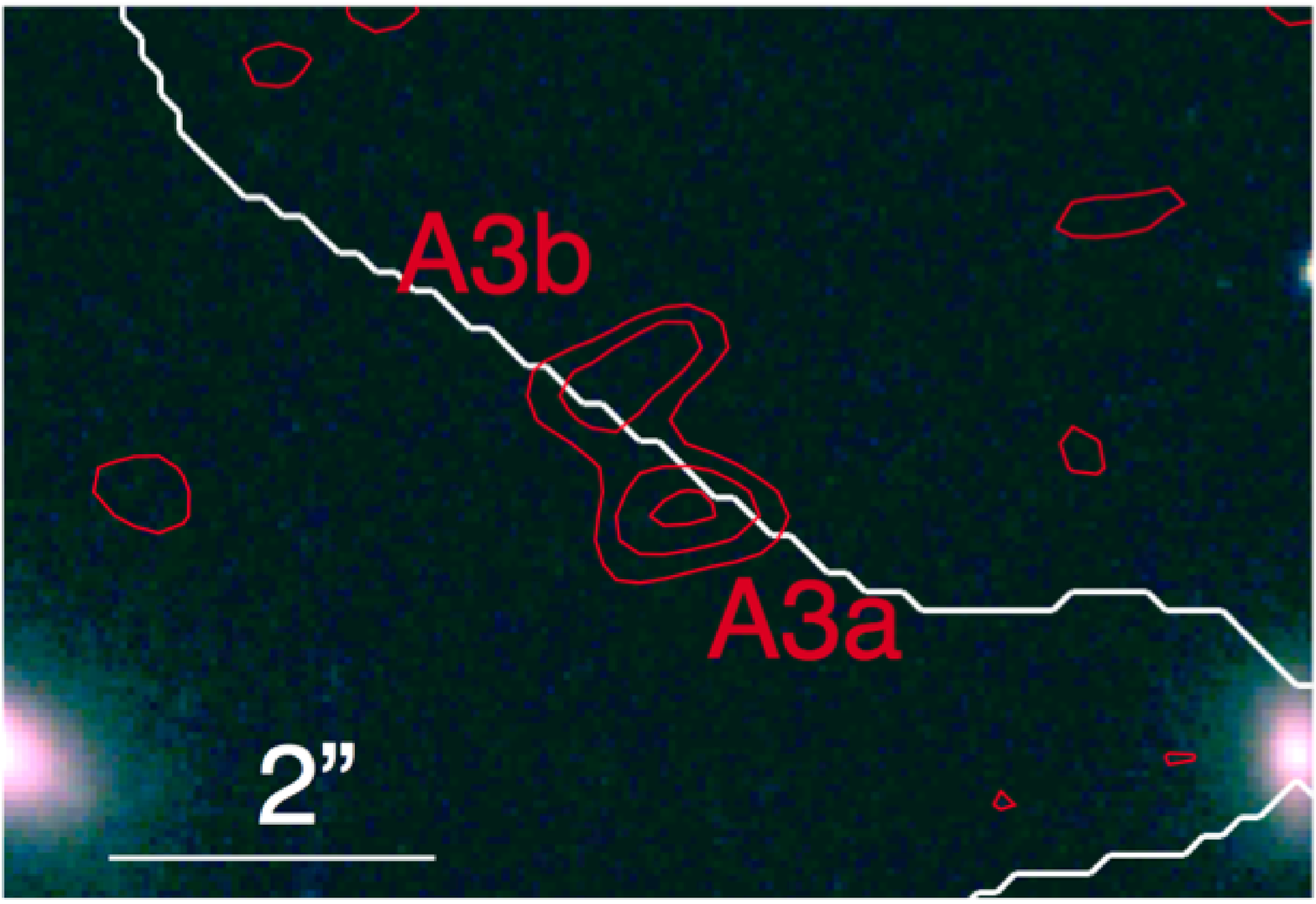}}
\caption{
Very strongly lensed source candidate, A3, found in A1689. 
The red contours show the ALMA 1.3 mm emission at the 2, 3, 4 $\sigma$ levels. 
The white curves denote the critical lines for a $z=3$ source 
that are estimated with our mass model. 
{\it Top}: ALMA continuum image of A3. 
The cyan tracks show the positions of multiple images for lensed sources at $z=2, 3, 4 , 5,$ and 6
that are predicted with our mass model. 
The ALMA multiple image positions 
(red contours; A3a and A3b) 
are well reproduced by
a lensed source at $z\sim 3$.
{\it Bottom}:
False-color image of HST optical data showing the same region as the top panel.
The used images and color assignment are the same as Figure \ref{fig:a1689}.
No optical counterparts of A3 are identified. 
\label{fig:A3}}
\end{center}
\end{figure}

Although there are no optical-NIR counterparts, we find a very strongly 
lensed source candidate, A3, that shows ALMA flux double peaks dubbed A3a and A3b.
The upper panel of Figure \ref{fig:A3} shows 
the ALMA image of A3. The A3 source is very close to the
critical line for a $z=3$ source (Figure \ref{fig:A3}), and the critical line
is located between the double peaks of A3. More quantitatively, we investigate
whether these double peaks can be explained by multiple images
of a single lensed source behind A1689. Using our mass model,
we calculate positions of multiple images made by a lensed source
at $z=2-6$ (cyan lines in the upper panel of Figure \ref{fig:A3}).
We find that multiple images at $z\sim 3$ reproduce the positions
of the double peaks very well. The best-fit redshift is $z=3.07$,
and the magnification factor is very high, $159$. 
The bottom panel of Figure \ref{fig:A3} is the HST optical images.
No optical counterparts of A3 are found. 
We summarize the properties of A3 in Table \ref{tab:A3}.
%

\begin{deluxetable*}{cccccccccccc}
\tablecolumns{12}
\tablewidth{0pt}
\tablecaption{Multiple Images at $z=2.60$ with the ALMA Counterpart\\
\label{tab:multiple_photometry}}
\tablehead{
\colhead{ID}
    & \colhead{ID}
    & \colhead{R.A.}
    & \colhead{Decl.}
    & \colhead{$S_{\rm 1.3mm}^{\rm obs}$}
    & \colhead{$S_{\rm 1.3mm}^{\rm corr}$}
    & \colhead{$S_{\rm 1.2mm}^{\rm corr}$}
    & \colhead{$g_{475}$}
    & \colhead{$r_{625}$}
    & \colhead{$i_{775}$}
    & \colhead{$z_{850}$}
    & \colhead{$\mu$} \\
(L07) & (This work) & (J2000)& (J2000)& (mJy) & (mJy) & (mJy) & \multicolumn{4}{c}{(mag$_{\rm AB}$)} & \\
    (1) & (2) &  \multicolumn{2}{c}{(3)} & (4) & (5) & (6) & \multicolumn{4}{c}{(7)} & (8)
}
\startdata
5.1 & \nodata & 13:11:29.064 & $-$1:20:48.64 & 0.18 $\pm$ 0.06\tablenotemark{$\dagger$} & 0.049 $\pm$ 0.021 & 0.067 $\pm$ 0.029 & 26.29 & 25.53 & 25.37 & 25.17 & 5.7 \\
5.2 & A1 & 13:11:29.224 & $-$1:20:44.24 & 0.30 $\pm$ 0.06 &  0.015 $\pm$ 0.006 & 0.020 $\pm$ 0.008 & 26.61 & 25.78 & 25.58 & 25.36 & 22.8 \\
5.3 & \nodata & 13:11:34.120 & $-$1:20:20.96 & $<$  0.28 & $<$ 0.093 & $<$ 0.13 & 26.61 & 25.85 & 25.85 & 25.60 &  3.0
\enddata
\tablecomments{(1) and (2) ID names given in \citet{limousin2007} (L07) and this work, respectively.
(3) R.A. and Decl. in the optical data.
(4) Peak flux density SExtractor measurement (Section \ref{sec:source_extraction}) with the primary beam correction. 
If no significant $1.32$ mm fluxes are detected at the optical source positions, three sigma upper limits are presented. 
(5) Observed flux density at 1.32 mm on the uv-taper image that are measured 
by 2D Gaussian-fitting routine of $imfit$ in CASA with the primary beam correction.
The magnification uncertainty (Section \ref{sec:mass_model}) is included in the error. 
(6) Source flux density at 1.2 mm with the flux scaling from 1.32 mm to 1.2 mm (Section \ref{sec:data_analysis}).
(7) Total magnitudes in the {\it HST}/ACS bands.
(8) Magnification factor at the peak of the optical flux.
}
\tablenotetext{$\dagger$}{
Marginal detection of the $3 \sigma$ level that is below our source selection limit for the cluster data ($>3.9 \sigma$; see Section \ref{sec:spurious_rate}).
}
\end{deluxetable*}

\begin{deluxetable*}{ccccccccccc}
\tablecolumns{11}
\tablewidth{0pt}
\tablecaption{Very Strongly Lensed Source Candidate\\
\label{tab:A3}}
\tablehead{
\colhead{ID}
    & \colhead{R.A}
    & \colhead{Decl.}
    & \colhead{$S_{\rm 1.3mm}^{\rm obs}$}
    & \colhead{$S_{\rm 1.3mm}^{corr}$}
    & \colhead{$S_{\rm 1.2mm}^{corr}$}
    & \colhead{$g_{475}$}
    & \colhead{$r_{625}$}
    & \colhead{$i_{775}$}
    & \colhead{$z_{850}$}
    & \colhead{$\mu$} \\
    &  (J2000)& (J2000)& (mJy) & (mJy) & (mJy) & \multicolumn{4}{c}{(mag$_{\rm AB}$)} & \\
    (1) &  \multicolumn{2}{c}{(2)} & (3) & (4) & (5) & \multicolumn{4}{c}{(6)} & (7)
}
\startdata
A3 & 13:11:26.897 & $-$1:20:32.02 & 0.24$\pm$0.06 & $0.0015\pm0.0007$ & $0.0020\pm0.0010$ & $>28.56$ & $>28.33$ & $>28.24$ & $>27.22$ & 159.5
\enddata
\tablecomments{
(1) $-$ (7) correspond to (2) $-$ (8) in Table \ref{tab:multiple_photometry}.
These values are for the A3a component (Figure \ref{fig:A3}). 
Because A3a and A3b are confused in the uv-taper image, these fluxes are obtained
on the image without uv-taper.
}
\end{deluxetable*}

\subsection{Color and Luminosity Properties of the Optical-NIR Counterparts}
\label{sec:color_luminosity_properties}

We investigate color (i.e. SED) and luminosity properties of the optical-NIR counterparts, S1-15 and A1-2,
to understand the population of the faint ALMA sources with the detectable optical-NIR continuum emission.
First of all, we refer the catalog of photo-$z$ estimated from the optical-NIR SEDs \citep{williams2009}, 
and find that S11, S12, S15 have photo-$z$ values of $1.54^{+0.08}_{-0.02}$, $1.57^{+0.07}_{-0.02}$, and $1.45^{+0.03}_{-0.02}$, respectively.
However, the photo-$z$ values of the S4 and S6 sources are $0.77^{+0.23}_{-0.03}$ and $0.67^{+0.01}_{-0.02}$, respectively.
We thus regard that the S4 and S6 sources reside at low redshift ($z<1$),
and find that $\sim 10$\% $(=2/17)$ optical-NIR counterparts are low-$z$ objects.
Because the majority of mm sources are
thought to be high-$z$ galaxies and AGNs \citep{casey2014},
the rest of 15 ($=17-2$) optical-NIR counterparts are candidates
of high-$z$ galaxies.
We thus compare colors and luminosities of these high-$z$ galaxy candidates 
with those of high-$z$ sources.
The major high-$z$ populations are SMGs, LBGs (including BX/BM), DRGs, BzKs, 
LAEs, and AGNs. Since the SMGs, DRGs, and LAEs
generally have weak optical continua that cannot be reliably
compared with our optical-NIR counterparts, we focus
on the comparisons with LBGs (BX/BM) and BzKs.

\subsubsection{Comparison with the LBG BX/BM Populations}
\label{sec:bxbm_color}

The BX and BM galaxies, generally included in the LBG population,
reside in the redshift ranges of $2.0\lesssim z\lesssim 2.5$ and 
$1.5\lesssim z\lesssim2.0$, respectively \citep{steidel2004}.
Similarly, there are LBGs at $z\sim 3$.
These galaxies are defined by the selection criteria of 
$U_{n}-G$ and $G-R$ colors. We apply these color selection criteria
to our optical-NIR counterparts, but the photometry bands of our data
are different from the $U_{n}$, $G$ and $R$ bands.
Following the procedures of \cite{ly2011}, we alternatively use a color set of $U-BV$ and $BV-R_{c}i'$.
The $BV$ and $R_{c}i'$ photometry are defined by
\begin{eqnarray}
BV = -2.5{\rm log}\left[\frac{x_{1}f_{B}+(1-x_{1}f_{V})}{3630\ \mu Jy}\right], \nonumber \\
R_{c}i' = -2.5{\rm log} \left[\frac{x_{2}f_{R}+(1-x_{2}f_{i'})}{3630\ \mu Jy}\right],
\end{eqnarray}
where $f_{X}$ is the flux density per unit frequency 
(erg s$^{-1}$ cm$^{-2}$ Hz$^{-1}$) in a band 'X' corresponding to
$B$, $V$, $R_{c}$, and $i'$. The coefficients of 
$x_{1}$ and $x_{2}$ are $0.314$ and $0.207$,
respectively. With these photometric relations
and the selection colors \citep{steidel2004}, 
the color criteria of BX objects are re-written as
\begin{eqnarray}
&BV& - R_{c}i' \ \geq\  -0.2, \nonumber \\
&U&- BV \ \geq\  BV-R_{c}i' + 0.23, \nonumber \\
&BV& - R_{c}i' \ \leq\  0.2(u - BV) + 0.4, \nonumber \\ 
&U& - BV \ \leq\  BV - R_{c}i' + 1.0.
\end{eqnarray}
Similarly, the criteria of BM objects are
\begin{eqnarray}
&BV& - R_{c}i' \ \geq\  -0.2, \nonumber \\
&U& - BV \ \geq\  -0.1, \nonumber \\
&BV& - R_{c}i' \ \leq\  -0.382(U -BV)+0.853, \nonumber \\
&BV& - R_{c}i' \ \leq\  0.70(U- BV) + 0.280,\  \nonumber \\
&U& - BV \ \geq\  BV - R_{c}i' + 0.23.
\end{eqnarray}
%
\begin{figure}
\begin{center}
\includegraphics[trim=0cm 0.5cm 0.5cm 1.0cm, clip, angle=0,width=0.5\textwidth]{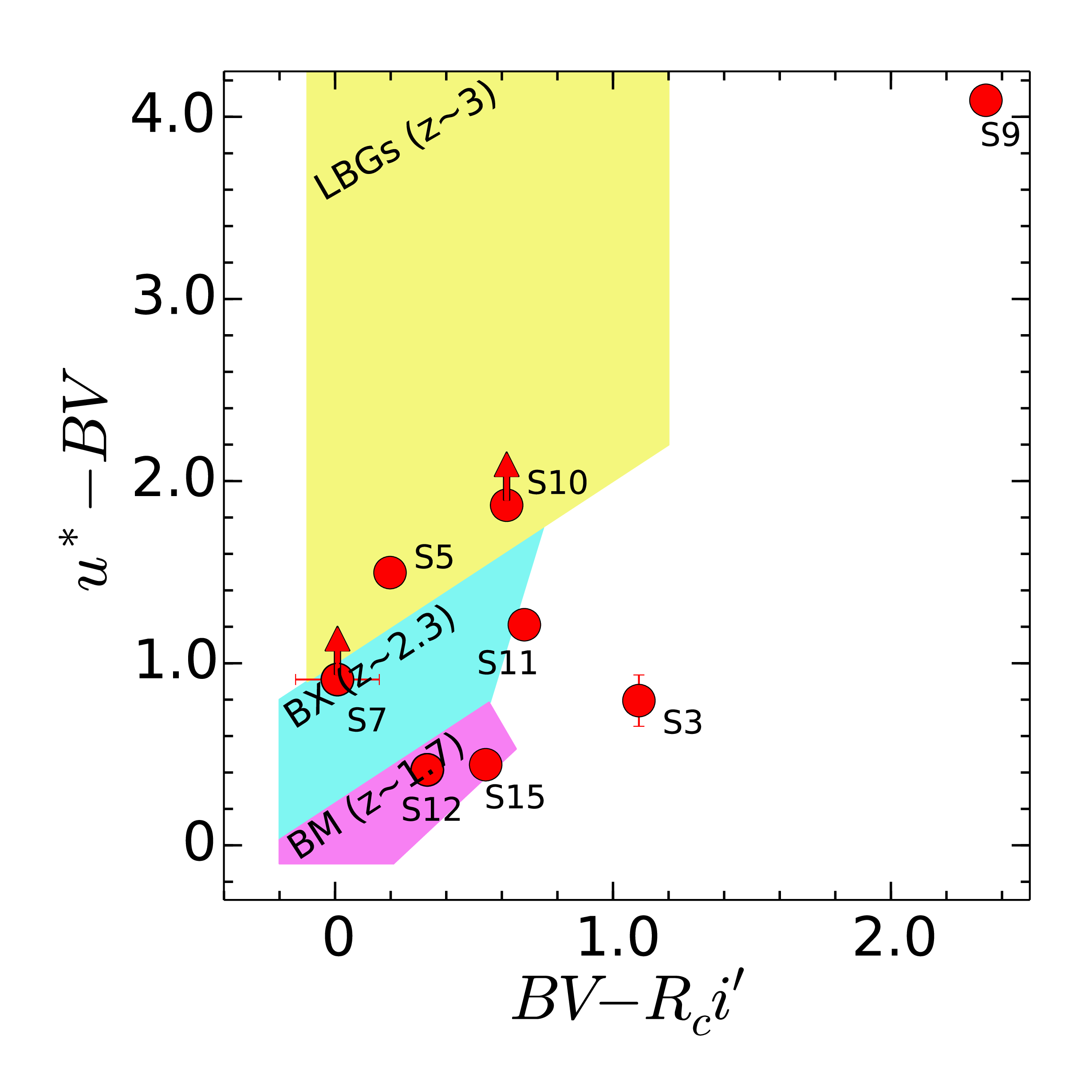}
 \caption[]
{Two-color 
diagram for our optical-NIR counterparts. 
The cyan and magenta-shaded regions are the BX and BM selection windows 
that are defined in the $BV-R_{c}i'$ and $u^{*}-BV$ color plane \citep{ly2011}. 
The BX and BM galaxies reside at $z\simeq2-2.5$ and $z\simeq1.5-2.0$, respectively. 
The yellow shades indicate the approximate region of the selection window for 
$z\sim 3$ LBGs \citep{steidel2003}, where we assume that the original photometry
of $U_{n}$, $G$ and $R$ bands are roughly the same as those of 
our $u^{*}$, $B$, and $R_{c}$ bands, respectively. 
\label{fig:bxbm}}
\end{center}
\end{figure}

We compare these color criteria with our
optical-NIR counterparts. Among the 17 optical-NIR counterparts, 
10 are observed with the photometric system of 
$u^{*}BVR_{c}i'$ necessary for the comparison.
Removing the 2 obvious low-$z$ objects of S4 and S6,
we use 8 optical-NIR counterparts.
Figure \ref{fig:bxbm} displays the two color diagram of
these 8 sources with the color selection criteria. 
Six sources fall in or near the selection windows of
$z\sim 2$ BX/BM or $z\sim 3$ LBGs, where we include
S11 source at the border of the selection window.
The S9 source is placed at the upper right corner of
Figure \ref{fig:bxbm}. These colors of the S9 source
are similar to $z\gtrsim 4$ LBGs (see, e.g., \citealt{steidel2003}).
It is likely that S9 source is a LBG at $z\gtrsim 4$.
On the other hand, S3 clearly escapes from the selection windows.
We thus find that 7 out of 8 optical-NIR counterparts 
are have colors consistent with either $z\sim 2$ BX/BM or 
$z\gtrsim 3$ LBGs. Including the two low-$z$ objects of S4 and S6 sources, 
70\% $(=7/[8+2])$ of the optical-NIR counterparts
meet the LBG (BX/BM) color criteria.
We then compare magnitudes of the 10 optical-NIR counterparts
with the known $z\sim 2$ BX/BM and $z\gtrsim 3$ LBGs,
omitting $u^{*}$ magnitudes that would include a Lyman break
whose central wavelength is very sensitive to a source redshift.
We find that
all of the 10 optical-NIR sources have $R_c$ magnitudes in the range of $23-27$ mag (Table \ref{tab:multiband_photometry})
that is typical brightness of $z\sim 2$ BX/BM and $z\gtrsim 3$ LBGs \citep{steidel2003,steidel2004}. 
We thus conclude that a majority ($\sim 70$\%) of the 10 optical-NIR counterparts 
belong to LBG (BX/BM) population.

\subsubsection{Comparison with the BzK Populations}
\label{sec:bzk_color}
In contrast with BX/BM galaxies and LBGs,
the BzK galaxy population includes not only
blue star-forming galaxies, but also moderately
reddened dust-poor star-forming galaxies (a.k.a sBzK) 
and passive galaxies (a.k.a pBzK) 
at $1.4 \lesssim z \lesssim 2.5$ \citep{daddi2004}. 
We compare the color criteria of the BzK population with our optical-NIR counterparts.
There are 10 out of the 17 optical-NIR counterparts that are located in the SXDS region
with the photometric measurements of $B$, $z'$, and $K$ bands (Table \ref{tab:multiband_photometry}).\footnote{
Note that the A1689 region misses $K$ band photometry 
critical for the BzK selection.
}
Removing the 2 obvious low-$z$ objects of S4 and S6,
we use 8 optical-NIR counterparts for the comparison.

Because the photometry system of ours (SXDS) is different from that of the BzK studies
\citep{daddi2004}, we again apply corrections to the color selection criteria for BzK galaxies,
which are suggested by \cite{yuma2012},
\begin{eqnarray}
(B-z)_{\rm Daddi+04} = (B-z)_{\rm SXDS} +0.3, \nonumber \\
(z-K)_{\rm Daddi+04} = (z-K)_{\rm SXDS} +0.1.
\end{eqnarray}

\begin{figure}
\begin{center}
\includegraphics[trim=0cm 0.5cm 0.5cm 1.0cm, clip, angle=0,width=0.5\textwidth]{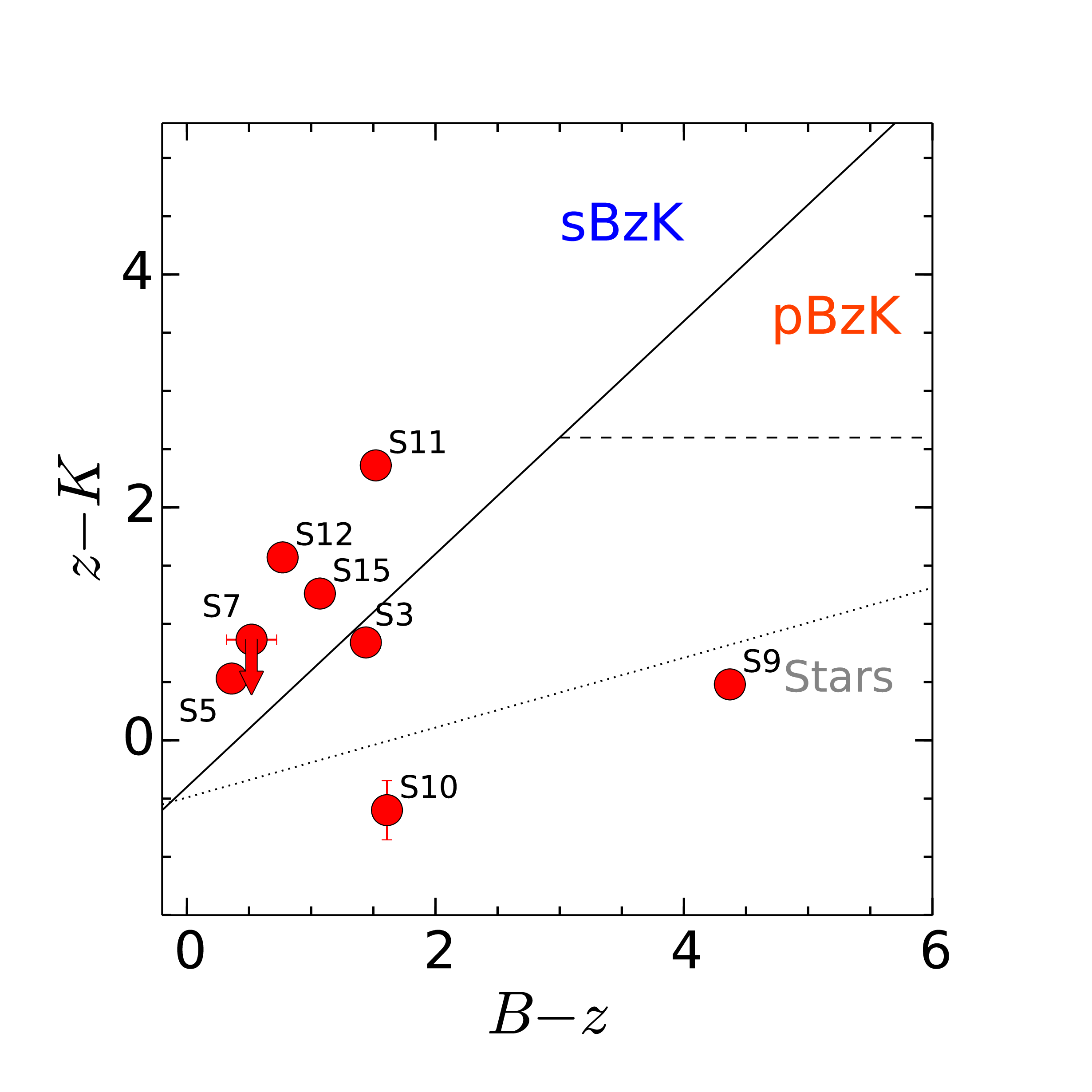}
 \caption[]
{Two-color 
diagram of the BzK galaxy selection
in the SXDS photometry system 
The red circles denote the 7 optical-NIR counterparts.
Star-forming galaxies at $z\simeq 1.4-2.5$, called sBzK galaxies,
are selected in the region beyond the solid line 
defined by 
$(z-K)-(B-z)\ \geq\ -0.2$. 
The pBzK galaxies, passively evolving galaxies at $z\simeq 1.4-2.5$,
are selected in the triangular region defined with the
sold and dashed ($z-K>2.5$) lines at the upper right corner.
Galactic stars fall in the region below the dotted line (See also \citealt{daddi2004,yuma2012}).
\label{fig:bzk}}
\end{center}
\end{figure}

Figure \ref{fig:bzk} presents the $B-z$ and $z-K$ colors
of the 8 optical-NIR counterparts. 
We find that 5 out of the 8 optical-NIR counterparts meet the color criteria of sBzK, 
while the rest of the three sources, S3, S9 and S10, have colors different from sBzK or pBzK galaxies
whose redshift range is $z\simeq 1.4-2.5$. 
Because S3 also shows the colors different from those of BX/BM galaxies and LBGs (Section \ref{sec:bxbm_color}), 
S3 would reside at $z\lesssim1.4$ that is out of the BzK redshift window.
Although the colors of S9 and S10 are similar to those of stars, 
in Section \ref{sec:bxbm_color}, the BX/BM galaxy and LBG selections 
suggest these two sources have
colors consistent with $z\gtrsim 3$ galaxies.
Thus, S9 and S10 would be galaxies at $z\gtrsim 3$ that escape from
the BzK redshift window. In this way, we
confirm that the results of the BzK selection are consistent with those of BX/BM galaxies and LBGs.
In summary, including the two low-$z$ objects of S4 and S6 sources,
we find that  
$50\%$(=5/[8+2])
of the optical-NIR counterparts
are sBzK galaxies at $z\simeq 1.4-2.5$.
We then compare magnitudes of these optical-NIR counterparts (Table \ref{tab:multiband_photometry}),
and find that the optical and NIR magnitudes are comparable with
the sBzK galaxies \citep{daddi2004}.

\subsubsection{Comparison with the AGN Population}
\label{sec:agn_color}
\begin{figure}
\begin{center}
\includegraphics[trim=0cm 0.5cm 0.5cm 1.0cm, clip, angle=0,width=0.5\textwidth]{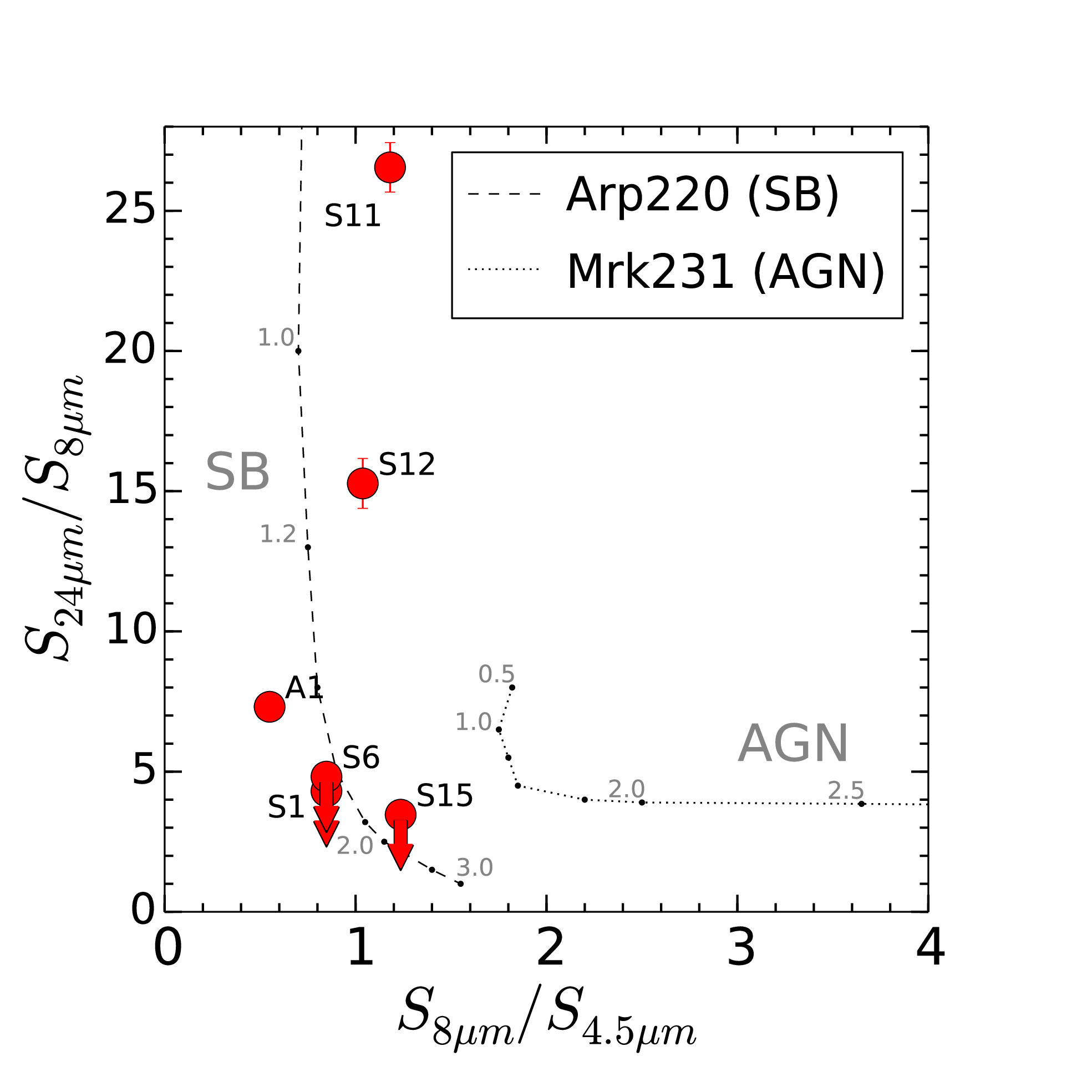}
 \caption[]
{
Color-color diagram for the AGN diagnostics.
The red circles and upper limits are our five mid-IR bright counterparts.
The dotted and dashed lines present the color tracks of Mrk 231 (AGN)
and Arp 220 (SB), respectively, that are redshifted from $z\simeq 0.5$ to $3.0$.
The numbers labeled to the lines represent the redshifts. 
\label{fig:agn_diagram}}
\end{center}
\end{figure}

Six counterparts of S1, S6, S11, S12, S15, and A1 are detected in
the mid IR of $8$ and $24\ \mu$m bands. Because mid-IR bright 
objects are prominent candidates of AGN that have 
strong hot-dust emission, we compare colors of the six mid-IR bright counterparts
and a typical AGN (Mrk 231) as well as a starburst (Arp 220).
Figure \ref{fig:agn_diagram} shows the colors of the five mid-IR bright counterparts,
Mrk 231, and Arp 220 in the diagnostic color diagram
of $S_{8.0 \mu\rm m}$/$S_{4.5 \mu\rm m}$ vs. $S_{24 \mu\rm m}$/$S_{8.0 \mu\rm m}$ 
that separates AGN and starburst (SB) populations \citep{ivison2004}.
In Figure \ref{fig:agn_diagram}, the dashed and dotted lines 
indicate the color tracks of Mrk 231 and Arp 220 
in the redshift range of $0.5-3.0$
that is taken from Figure 4 of \cite{ivison2004}. 
All of our five mid-IR bright counterparts are placed near the color track of Arp 220,
and far from the one of Mrk 231, which suggest that
these six counterparts are not AGN but starbursts.

\begin{deluxetable*}{lccccccccccc} 
\tablecolumns{12} 
\tablewidth{0pt} 
\tablecaption{Properties of our Optical-NIR Counterparts\label{tab:properties_id}}
\tablehead{
ID   & R.A.       & Decl.     & $\lambda_{\rm obs}$ & $S^{\rm obs}_{\lambda_{\rm obs}}$ & $S^{\rm corr}_{\rm 1.2mm}$ & $z_{\rm photo}$ & $z_{\rm spec}$ & sBzK & pBzK & BX$/$BM$+$LBG & AGN \\
      &  (J2000) & (J2000)  &  (mm) &   (mJy) & (mJy)                                              &     &  &   &                       &    \\
      &                &               &   (1)    &    (2)    &     (3)                                                                                                           &      (4)             &        (5)           &  (6)    &  (7)     &   (8)                        & (9)
}
\startdata 
\renewcommand\thefootnote{\alph{footnote}}
S1 & 34.612686 & $-$4.583063 &1.22 &  0.36 $\pm$ 0.07 &  0.43 $\pm$ 0.09 & $\cdots$     & $\cdots$ & $\cdots$ & $\cdots$ & $\cdots$ & N     \\
S2 & 34.613739 & $-$4.588505 &1.22 &  0.31 $\pm$ 0.09 &  0.31 $\pm$ 0.10 &$\cdots$     & $\cdots$  & $\cdots$ & $\cdots$ & $\cdots$ &  $\cdots$\\
S3 & 34.437805 & $-$5.494339 &1.03 &  0.056 $\pm$ 0.015 &  0.032 $\pm$ 0.010 &$\cdots$ & $\cdots$    &  N  & N            & N          &  $\cdots$ \\
S4 & 34.446648 & $-$5.007715 &1.26 &  0.047 $\pm$ 0.014 &  0.044 $\pm$ 0.016 & 0.77$^{+0.23}_{-0.03}$ &$\cdots$    &  $\cdots$  & $\cdots$  & $\cdots$  &  $\cdots$ \\
S5 & 34.446007 & $-$5.008159 &1.26 &  0.049 $\pm$ 0.014 &  0.050 $\pm$ 0.017 & $\cdots$ &$\cdots$    & Y        &  N      & Y          &  $\cdots$ \\
S6 & 34.387825 & $-$5.220457 &1.25 &  0.41 $\pm$ 0.11 &  0.42 $\pm$ 0.13 & $0.67^{+0.01}_{-0.02}$ &$\cdots$    & $\cdots$ & $\cdots$   &$\cdots$     & N             \\
S7 & 34.411190 & $-$4.745051 &1.23 &  0.30 $\pm$ 0.08 &  0.14 $\pm$ 0.09}\tablenotemark{$\dagger$ & $\cdots$ &$\cdots$    &   Y$'$ & N& Y$'$ & $\cdots$ \\
S8\footnotemark[1] & 34.305553 & $-$5.067293 &1.30 &  0.29 $\pm$ 0.07 &  0.22 $\pm$ 0.09 & $\cdots$ &$\cdots$    &   $\cdots$ & $\cdots$ &  $\cdots$ & $\cdots$ \\
S9 & 34.194702 & $-$5.056259 & 1.25 &  0.43 $\pm$ 0.12 &  0.39 $\pm$ 0.15 & $\cdots$ &$\cdots$    &  N          & N         & Y$'$            & $\cdots$ \\
S10 & 34.197121 & $-$5.059778 &1.25 &  0.31 $\pm$ 0.08 & 0.37 $\pm$ 0.10 & $\cdots$ &$\cdots$   &   N & N         &  Y$'$            & $\cdots$ \\
S11\footnotemark[2] & 34.442795 & $-$4.911066 &1.23 &  0.53 $\pm$ 0.10 & 0.56 $\pm$ 0.11 & $1.54^{+0.08}_{-0.02}$ &$\cdots$   & Y           & N          &  Y$'$            & N         \\
S12 & 34.441357 & $-$4.912045 &1.23 &  0.32 $\pm$ 0.09 & 0.29 $\pm$ 0.09 & $1.57^{+0.07}_{-0.02}$ &$\cdots$   &   Y         &  N         & Y            & N         \\
S13 & 34.747295 & $-$4.858103 &1.25 &  0.37 $\pm$ 0.10 & 0.26 $\pm$ 0.11 &  $\cdots$ &$\cdots$  &  $\cdots$ & $\cdots$ & $\cdots$ & $\cdots$ \\
 S14  & 34.274231 &$-$4.860201 & 1.30 &  0.32 $\pm$ 0.09 &  0.43 $\pm$ 0.12 & $\cdots$  & $\cdots$ & $\cdots$& $\cdots$ & $\cdots$ & $\cdots$  \\
 S15  & 34.587593 &$-$5.318766 & 1.25 &  0.32 $\pm$ 0.09 &  0.21 $\pm$ 0.10 & $1.45^{+0.03}_{-0.02}$  & $\cdots$ & Y     & N     & Y        &  N     \\
A1\footnotemark[3]  & 197.871590 & $-$1.345783 &1.32 &  0.30 $\pm$ 0.06 &  0.020 $\pm$ 0.008\tablenotemark{$\dagger\dagger$} & $\cdots$ &2.60  & $\cdots$ & $\cdots$  &$\cdots$ &N  \\
 A2 & 197.869086 &$-$1.337143 & 1.32 &  0.31 $\pm$ 0.08 &  0.077 $\pm$ 0.033\tablenotemark{$\dagger\dagger$} & $\cdots$ & $\cdots$ & $\cdots$ & $\cdots$ & $\cdots$ & $\cdots$ 
\enddata 
\tablecomments{
(1): Wavelength in the observed frame.
(2): 
Peak flux density of the SExtractor measurement (Section \ref{sec:source_extraction}) 
with the primary beam correction at the observed wavelength. 
(3): 
The best-estimate source flux density at 1.2 mm.
The source flux is estimated with  the 2D Gaussian-fitting routine of {\it imfit} in CASA (Section \ref{sec:flux_measure})
with the primary beam correction and the flux scaling to 1.2 mm (Section \ref{sec:data_analysis}). 
For A1 and A2 sources, the lensing magnification corrections are also applied.
The lensing magnification factors are estimated at the ALMA flux peak positions of A1 and A2.
(4): Photometric redshift estimated by \cite{williams2009}. 
(5): Spectroscopic redshift obtained by \cite{limousin2007}.
(6), (7), (8), and (9): ''Y'' indicates the objects that meet the color selection criteria of 
sBzK, pBzK, BX/BM, and AGN populations, respectively. 
"N" represents the sources escaping from the color space of the selection criteria. 
"Y$'$" is presented for the sources having colors consistent with the populations.
}
\tablenotetext{$\dagger$}{
The photometry of {\it imfit} may be biased, due to the systematic noise.
The SExtractor peak flux measurement suggests $0.32\pm0.08$ mJy. 
}
\tablenotetext{$\dagger\dagger$}{ 
The magnification uncertainty (Section \ref{sec:mass_model}) is also included in the error. 
}
\footnotetext[1]{Same as the AS2 source in \cite{hatsukade2015b}}
\footnotetext[2]{Same as the AS4 source in \cite{hatsukade2015b}}
\footnotetext[3]{Same as the Chen-5 source in \cite{chen2014}}
\end{deluxetable*} 

\subsubsection{IRX-$\beta$ Relation}
\label{sec:irx_beta}

\begin{figure}
\begin{center}
\includegraphics[trim=0cm 0cm 0cm 0cm, clip, angle=0,width=0.5\textwidth]{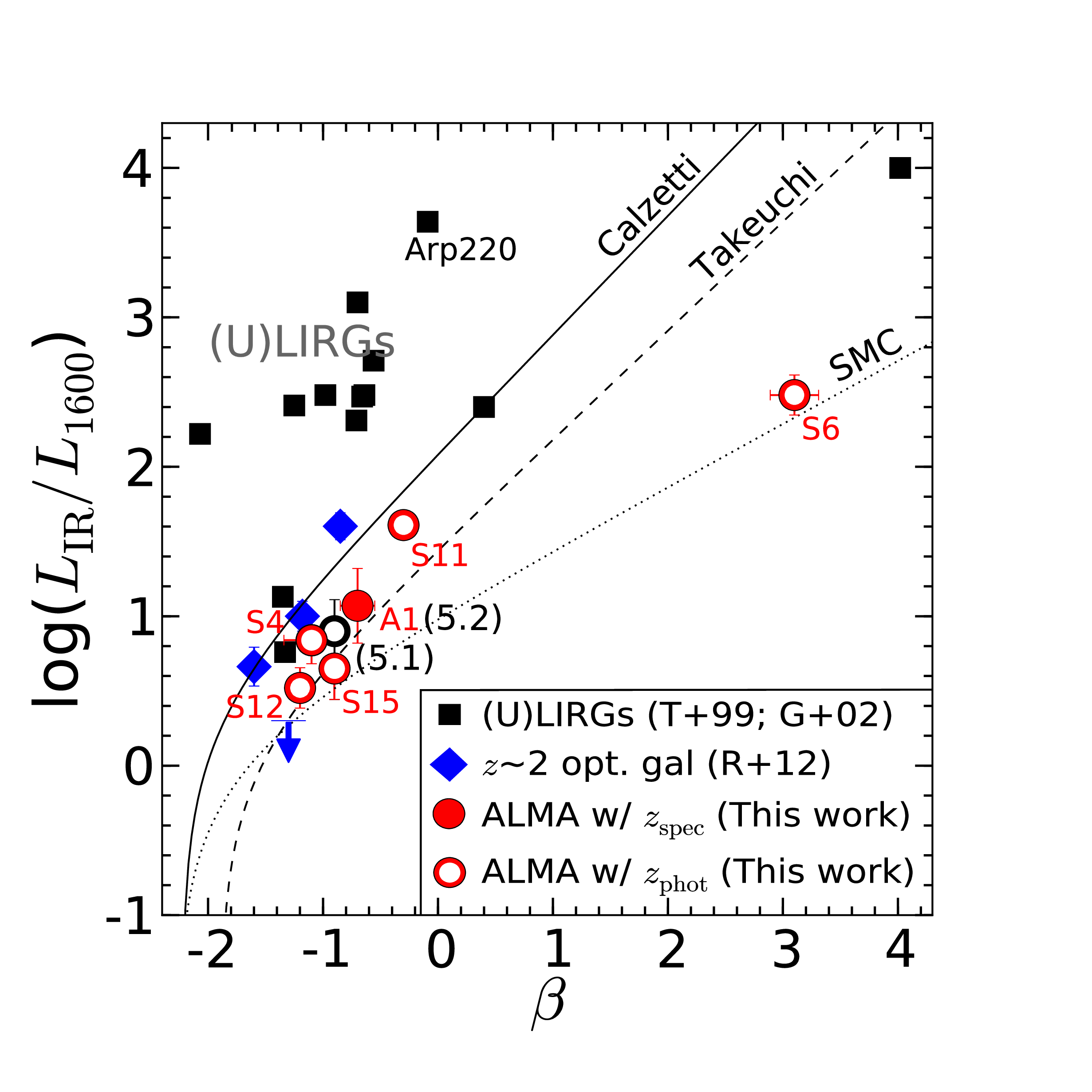}
 \caption[]{ 
Infrared-to-UV luminosity ratio IRX as a function of the UV slope $\beta$. 
The red filled circle is the spectroscopically-confirmed A1 (5.2) source at $z_{\rm spec}=2.60$. 
The red open circles denote the faint ALMA sources with the photometric redshifts. 
The open black circle is the 5.1 source of the A1 multiple image system 
that is marginally detected in the ALMA data (Table \ref{tab:multiple_photometry}).
The blue diamonds (arrow) represent(s) the  
measurements (the upper limit) of the $z\sim 2$ optically-selected galaxies 
obtained by the stacking analysis \citep{reddy2012}.
The black squares indicate nearby (U)LIRGs \citep{trentham1999, goldader2002}. 
The solid, dashed, and dotted curves denote the IRX-$\beta$ relation of the extinction curves
for the Calzetti law \citep{meurer1999}, the updated Calzetti law \citep{takeuchi2012}, and 
the SMC law \citep{bouchet1985}. 
For the Calzetti law curve, we shift the original relation of the Calzetti law
\citep{meurer1999} by $+0.24$ dex, because \citet{meurer1999}
define their infrared luminosities at $40-120\mu$m that is different
from our definition ($8-1000\mu$m).
\label{fig:irx_beta}}
\end{center}
\end{figure}

\begin{figure*}
\begin{center}
 \includegraphics[trim=0cm 0cm 0cm 0.5cm, angle=0,width=1.0\textwidth]{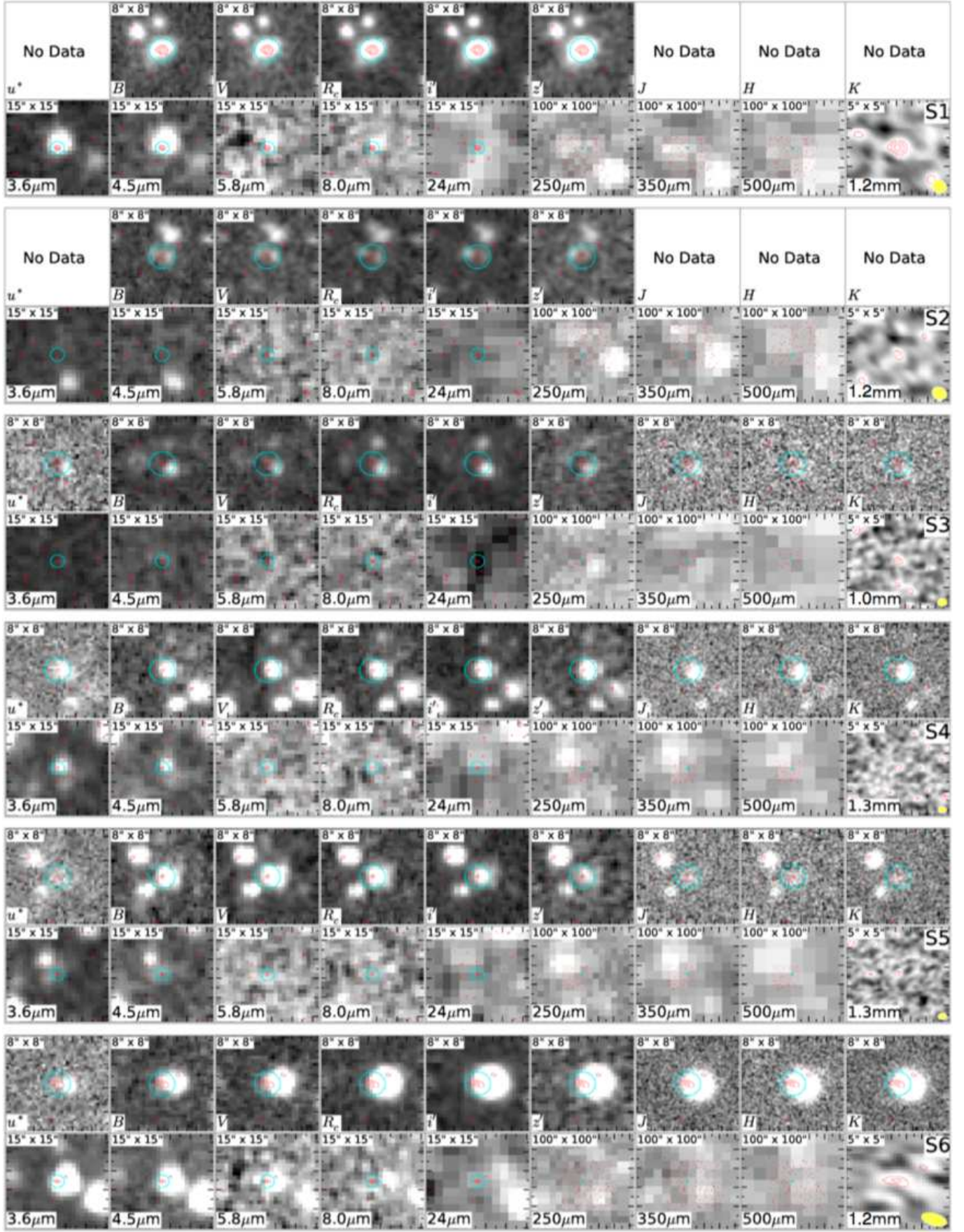}
 \caption[]
{The snapshot images of the optical-NIR counterparts. 
The red contours indicate the ALMA mm band intensity at 2.4, 3.4, 4.4, and 5.4 $\sigma$ levels. 
The cyan circle denotes the $1\farcs$0 search radius for the counterpart identifications.
The wavelengths of the images range from 
$0.4$ to $500\ \mu$m, which are shown with the labels in the
images. The bottom right panels present the ALMA images with the
beam sizes (yellow ellipses).
In each panel, the size of the image is indicated.
North is up, and east is to the left.
\label{fig:postage_stamps}}
\end{center}
\end{figure*}

To evaluate the SED properties quantitatively,
we calculate the infrared-to-UV luminosity ratio IRX($\equiv L_{\rm IR}/L_{1600}$) and 
the rest-frame UV continuum slope $\beta$ for 
6 faint ALMA sources of S4, S6, S11, S12, S15, and A1 that have spectroscopic or photometric redshifts.
Here $L_{\rm IR}$ is the integrated infrared flux density defined in a range of rest-frame 8-1000 $\mu$m. 
The $\beta$ values are estimated with two broadband magnitudes, $m_{1}$ and $m_{2}$:
\begin{eqnarray}
\beta=-\frac{m_{1}-m_{2}}{2.5\log(\lambda^{1}_{\rm c}/\lambda^{2}_{\rm c})} -2,
\end{eqnarray}
where $\lambda^{1}_{\rm c}$ and $\lambda^{2}_{\rm c}$ are the central wavelengths of the two broadband filters. 
We use $(m_{1}, m_{2})$ = ($u^{*}$, $B$), ($u^{*}$, $B$), ($V$, $i'$), ($V$, $i'$), ($V$, $i'$), ($r_{625}$, $z_{850}$) 
for S4, S6, S11, S12, S15, and A1, respectively. 
These IRX and $\beta$ values are shown in Figure \ref{fig:irx_beta},
together with those of nearby (U)LIRGs and optically selected star-forming galaxies at $z\sim2$.
In Figure \ref{fig:irx_beta}, we also plot the 5.1 source of the multiple image (Table \ref{tab:multiple_photometry}) that
is marginally detected ($3\sigma$). For the lensed sources of A1 and 5.1, we assume that the positional offsets between
the submm and optical sources are real, and use the lensing magnification values estimated at the submm (optical) source positions 
to derive the intrinsic submm (optical) luminosities.

In Figure \ref{fig:irx_beta}, all of the six faint ALMA sources fall in the region 
nearly-enclosed by the Calzetti and SMC law curves.
The six faint ALMA sources are placed clearly below the (U)LIRGs that have
large IRX values at a given $\beta$.
The clear difference between the faint ALMA sources and (U)LIRGs suggests that
the faint ALMA sources are not miniature (U)LIRGs 
with infrared luminosities fainter than (U)LIRGs by $1-2$ order(s) of magnitude. 
Note that these results are based on the 59\% of the faint ALMA sources with the optical-NIR counterparts.
The rest of the 41\% of the faint ALMA sources that have no optical-NIR counterparts 
could have IRX values as high as those of (U)LIRGs, because these sources have
low $L_{1600}$ values that are suggested by the fact of no optical-NIR counterparts.

These 6 faint ALMA sources are composed of 4 high-$z$ sources (S11, S12, S15, and A1) 
and 2 low-$z$ sources (S4 and S6), where we define high-$z$ and low-$z$ by $z\geq1$ and $z<1$, respectively.
Four high-$z$ sources have the IRX-$\beta$ relation similar to those of optically selected star-forming galaxies at $z=2$ \citep{reddy2012}. 
One low-$z$ source of S4 is also placed at the region of $z=2$ optically selected galaxies in Figure \ref{fig:irx_beta}, while
the other low-$z$ source of S6 has large IRX and $\beta$ values. 
It implies that majority of high-$z$ sources are similar to $z=2$ optically selected galaxies,
but that low-$z$ sources would be widely distributed in the IRX-$\beta$ plot.

Figure \ref{fig:irx_beta} shows that the multiple images of 5.2 (i.e. A1) and 5.1
have similar IRX-$\beta$ values. This fact confirms that these sources are 
multiple images of a single high-$z$ source.

\subsubsection{What are the Optical-NIR Counterparts of the Faint ALMA Sources?}
\label{sec:summary_optnir_counterpart}

We summarize the results of Sections \ref{sec:bxbm_color}-\ref{sec:agn_color}
in Table \ref{tab:properties_id}.
We find, in Sections \ref{sec:bxbm_color} and \ref{sec:bzk_color},
that 7 out of the 10 optical-NIR counterparts (including the low-$z$ galaxies) 
meet the color criteria of BX/BM LBG and sBzK galaxy populations (see Table \ref{tab:properties_id}),
and that the magnitudes of the counterparts are consistent with these populations.
The rest of the three counterparts are low-$z$ objects (2 objects) and 
probably the foreground/background object of the chance projection 
(1 object; Section \ref{sec:counterpart_id}).
These results suggest that not all but the majority of the optical-NIR counterparts of our faint ALMA sources 
belong to the sBzKs and BX/BM LBGs populations.
Moreover, the optical-NIR counterparts show neither AGN nor (U)LIRG signatures
(Sections \ref{sec:agn_color} and \ref{sec:irx_beta}).

We thus conclude that the majority of our faint ALMA sources with a detectable
optical continuum ($\sim 25$ mag) are UV bright star-forming galaxies of BX/BM LBGs and sBzKs
with no AGN. 
This conclusion is supported by the 
clustering analysis results in Section \ref{sec:clustering}
that presents a weak clustering signal of $b_g<3.5$ comparable
to those of LBGs and sBzKs.

Note that we find the optical-NIR counterparts in 
59\% of our faint ALMA sources (Section \ref{sec:counterpart_id}).
In other words, we understand that 
roughly about a half of the faint ALMA sources are LBGs and sBzKs.
For the remaining half of the faint ALMA sources,
the physical origin is unclear.
Our study identifies a new issue of faint ALMA sources
that should be addressed in the
future studies.

\section{Summary} \label{sec:summary}
In this paper, we identify 133 faint 1.2-mm continuum sources with a flux density of $0.02-1$ mJy
in the fifty blank-field and one cluster, Abell 1689, maps taken by $\sim 120$ deep ALMA Band 6/7 pointings.
This is the large faint source sample reaching $0.02$ mJy,
which is drawn from our deep ALMA projects and the complete archive search
for the deep ALMA data that are open for public by 2015 June. 
Using this large sample of the faint 1.2 mm sources,
we derive the 1.2 mm number counts and discuss the faint-source contribution to the EBL.
We then investigate the physical origin of these faint 1.2 mm sources
by the statistical technique of clustering analysis 
and the individual-source approach of optical-NIR counterparts.
The major results of this paper are summarized below.

\begin{enumerate}
\item 
We derive the 1.2-mm number counts in a flux density range of $0.02-1$ mJy.
Our number counts are consistent with the previous ALMA and single-dish lensing
results, and largely push the flux density limit down to $0.02$ mJy.
We find that the number counts are well represented by the Schechter function at $\gtrsim 0.02$ mJy, 
combining the faint and bright number counts from
our and the previous studies, respectively.

\item
We estimate the total integrated 1.2 mm flux density 
with the derived number counts, and obtain 
$22.9^{+6.7}_{-5.6}$ Jy deg$^{-2}$ down to
the flux limit of $S_{\rm limit}$.
This integrated 1.2 mm flux density
corresponds to $104^{+31}_{-25}$\% of the EBL
measured by COBE observations.
Because we reach $\sim 100$\% of the EBL-resolved fraction 
at $\sim 0.02$ mJy within the errors that are mainly originated from the 
COBE EBL estimates, the major 1.2 mm EBL contributors are sources 
with $\gtrsim 0.02$ mJy. In other words, 1.2 mm sources with a flux density below $\sim 0.02$ mJy
make a negligibly small contribution to the EBL. These results suggest a possibility 
that there is a flattening and/or truncation in the number counts at the
very faint flux density regime of $\lesssim 0.02$ mJy.

\item 
We perform the counts-in-cells analysis for clustering properties of our faint ALMA sources,
assuming that these sources reside at $z\sim 1-4$. We obtain an upper limit of 
the galaxy bias $b_{g}<3.5$ that is smaller than those of massive galaxy
populations of DRGs and SMGs. This upper limit is comparable with
the galaxy bias values of UV-bright sBzK and LBG populations.

\item 
We search for optical-NIR counterparts of our faint ALMA sources
located in two regions of SXDS and Abell 1689 that have 
deep optical-NIR images and the various  multi-wavelength data.
We identify 17 optical-NIR counterparts, and find that $59$\% of
our faint ALMA sources have the counterparts with the
detectable ($m\lesssim 28$ mag) optical-NIR continuum. 
Neither X-ray nor radio counterparts are found in any of our faint ALMA sources,
and the mid-IR diagnostics indicates that there are no AGN activities
in these sources.
We apply the color selection criteria of BX/BM LBGs, sBzKs, and pBzKs
for our optical-NIR counterparts with a photometric data set
necessary for the color selections. We find that a majority
of these optical-NIR counterparts have colors meeting the 
BX/BM LBG and sBzK criteria, and that their magnitudes
are comparable to the BX/BM LBGs and sBzKs. 
Because $59$\% of our
faint ALMA sources have the detectable optical-NIR counterparts,
we conclude that about a half of faint ALMA sources are
BX/BM LBGs and sBzKs that are high-$z$ galaxies
with a star-formation rate and dust extinction 
much smaller than SMGs.
Moreover, the IRX-$\beta$ relation plot indicates that 
the optical-NIR counterparts of our faint ALMA sources
are clearly different from (U)LIRGs,
and that faint ALMA sources are not miniature (U)LIRGs
simply scaled down with the infrared brightness.

\end{enumerate}

\section*{Acknowledgements}
We are grateful to Akifumi Seko and Koji Ohta for providing us their reduced data, and Bunyo Hatsukade for giving us helpful advice on analyzing the data and sending their results. 
We also thank Ryohei Kawabe, Kotaro Kohno, Soh Ikarashi, Kirsten K. Knudsen, Chian-Chou Chen, and Ian Smail for useful information, comments, and discussions. 
We appreciate the support of the staff at the ALMA Regional Center, especially Kazuya Saigo and Bunyo Hatsukade. This paper makes use of the following ALMA data: 
ADS/JAO.ALMA{\#}2011.0.00115.S,  
{\#}2011.0.00232.S,
{\#}2011.0.00243.S,
{\#}2011.0.00319.S,
{\#}2011.0.00648.S,
{\#}2011.0.00767.S,
{\#}2012.1.00076.S, 
{\#}2012.1.00602.S,  
{\#}2012.1.00676.S,
{\#}2012.1.00719.S,
{\#}2012.1.00323.S,
{\#}2012.1.00536.S,
{\#}2012.1.00610.S,
{\#}2012.1.00934.S,
and {\#}2012.1.00953.S.
ALMA is a partnership of ESO (representing its member states), 
NSF (USA) and NINS (Japan), together with NRC (Canada) and NSC and ASIAA (Taiwan), 
in cooperation with the Republic of Chile. 
The Joint ALMA Observatory is operated by ESO, AUI/NRAO and NAOJ.
This work was supported by World Premier International Research Center Initiative
(WPI Initiative), MEXT, Japan, and KAKENHI (23244025) 
and (15H02064)
Grant-in-Aid 
for Scientific Research (A) through Japan Society for the Promotion of Science (JSPS).

\begin{figure*}
\begin{center}
 \includegraphics[trim=0cm 0cm 0cm 0.5cm, angle=0,width=1.0\textwidth]{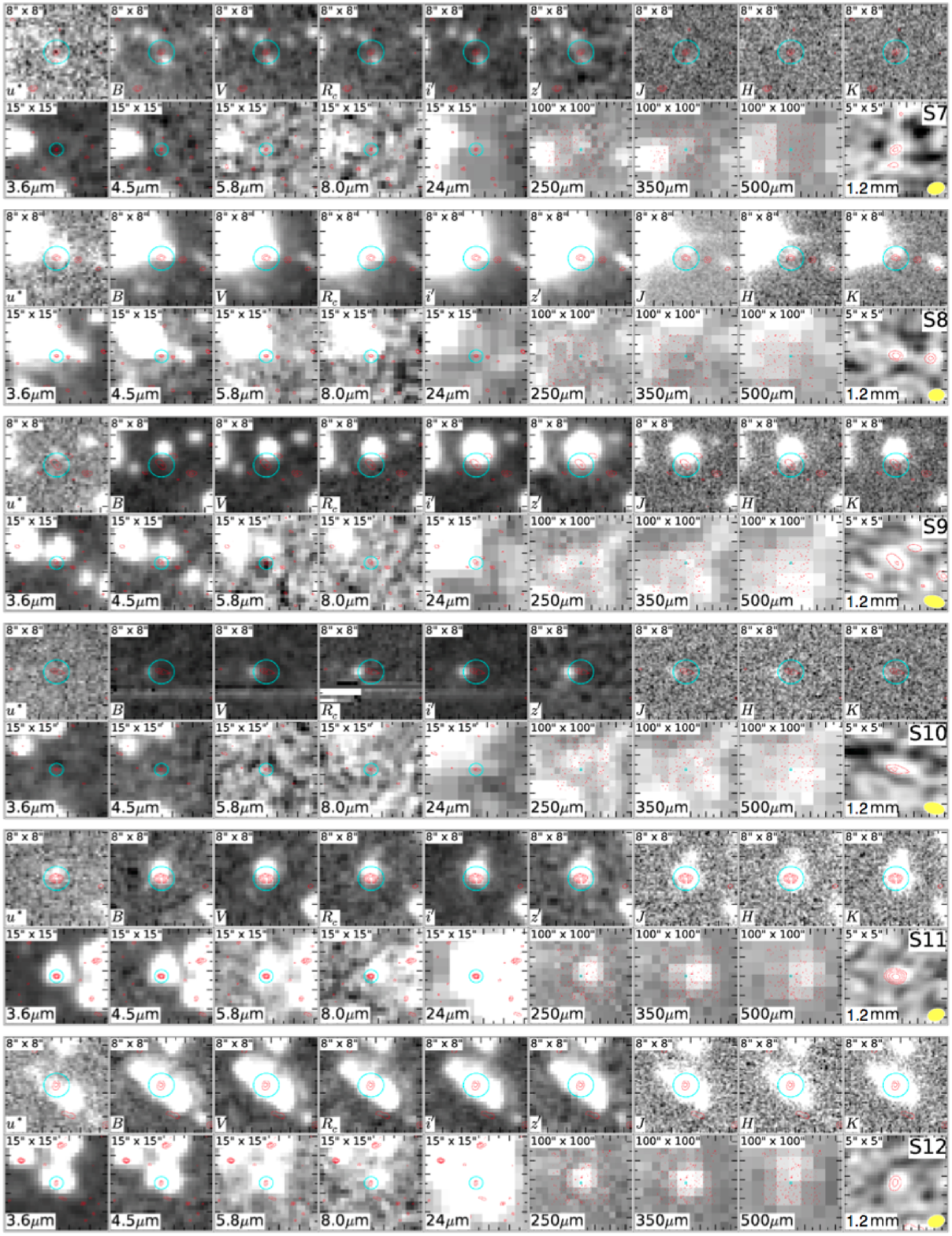}
Fig. \ref{fig:postage_stamps} (continued)
\end{center}
\end{figure*}

\begin{figure*}
\begin{center}
\includegraphics[trim=0cm 0cm 0cm 0.5cm, angle=0,width=1.0\textwidth]{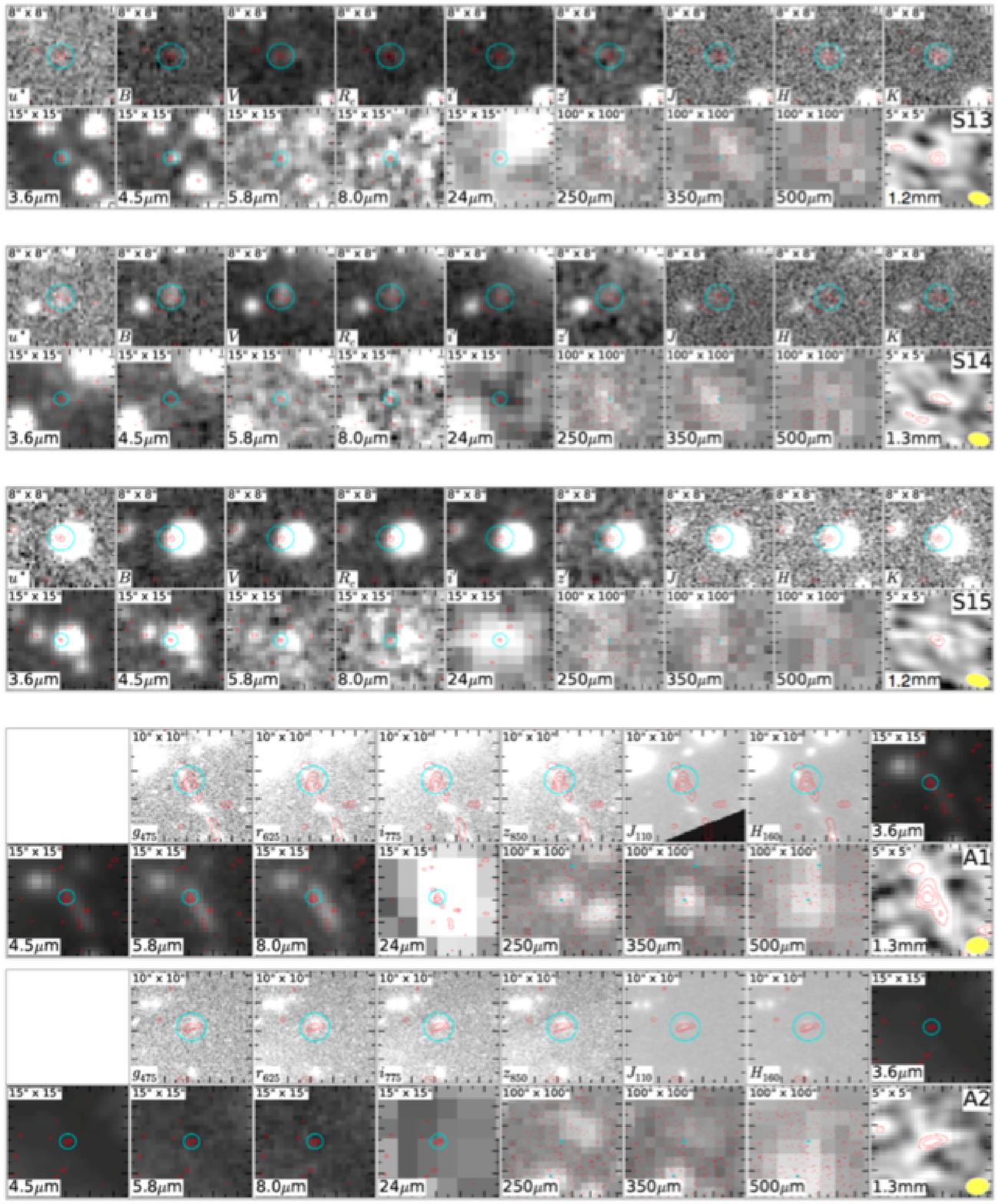}
Fig. \ref{fig:postage_stamps} (continued)
\end{center}
\end{figure*}


\bibliographystyle{apj}
\bibliography{apj-jour,reference}

\end{document}